\documentclass[accepted]{uai2025} % after acceptance, for a revised version; 
\usepackage[american]{babel}

\usepackage{mathtools} % amsmath with fixes and additions
\usepackage{booktabs} % commands to create good-looking tables
\usepackage{tikz} % nice language for creating drawings and diagrams

\usepackage{ascmac}
\usepackage{float}
\usepackage{perpage}
\MakeSorted{figure}
\MakeSorted{table}

\usepackage{url}
\usepackage{natbib}
\usepackage{chapterbib}

\usepackage{color}
\usepackage{tikz}
\tikzset{%
mynode/.style={circle,minimum width=.5ex, fill=none,draw}, % no filling
myfillnode/.style={circle,minimum width=.5ex, fill=lightgray,draw}, % fill with black
}
\usepackage{amssymb}
\usepackage{natbib}

\newcommand{\indep}{\perp \!\!\! \perp}
\usepackage{amsmath}               
\usepackage{lscape}
\usepackage{algorithm}
\usepackage{color}
\usepackage{tikz}
\usepackage{amsmath,amsthm}
\newtheorem{theorem}{Theorem}[section]
\newtheorem{definition}{Definition}[section]
\newtheorem{assumption}{Assumption}[section]
\newtheorem{lemma}{Lemma}[section]
\newtheorem{proposition}{Proposition}[section]

\usepackage{comment}
\usepackage{here}
\allowdisplaybreaks[4]
\usepackage{bbm}
\usepackage{caption}
\usepackage{bbding}
\usepackage{arydshln}
\usepackage{afterpage}

\usepackage{mathrsfs}

\usepackage{soul}

\usepackage{listings}
\floatstyle{ruled}
\newfloat{listing}{tb}{lst}{}
\floatname{listing}{Listing}
\usepackage[utf8]{inputenc} % allow utf-8 input
\usepackage[T1]{fontenc}    % use 8-bit T1 fonts
\usepackage{hyperref}       % hyperlinks
\usepackage{url}            % simple {\boldsymbol U}RL typesetting
\usepackage{booktabs}       % professional-quality tables
\usepackage{amsfonts}       % blackboard math symbols
\usepackage{nicefrac}       % compact symbols for 1/2, etc.
\usepackage{microtype}      % microtypography
\usepackage{xcolor}         % colors

\usepackage{stackengine}
\def\defeq{\mathrel{\ensurestackMath{\stackon[1pt]{=}{\scriptscriptstyle\Delta}}}}
\usepackage{natbib} % has a nice set of citation styles and commands
\usepackage{mathtools} % amsmath with fixes and additions
\usepackage{booktabs} % commands to create good-looking tables
\usepackage{tikz} % nice language for creating drawings and diagrams

\title{Decomposition of Probabilities of Causation with Two Mediators}

\author[1]{\href{mailto:<Yuta.Kawakami@mbzuai.ac.ae}{Yuta Kawakami}}
\author[1]{\href{mailto:<Jin.Tian@mbzuai.ac.ae}{Jin Tian}}
\affil[1]{%
Mohamed bin Zayed University of Artificial Intelligence, UAE
}
  
\begin{document}
\maketitle

%We investigate the path-specific probability of necessity and sufficiency (PNS) to decompose the total PNS into path-specific components, incorporating two mediators.

\begin{abstract}
Mediation analysis for probabilities of causation (PoC) provides a fundamental framework for evaluating the necessity and sufficiency of treatment in provoking an event through different causal pathways. 
One of the primary objectives of causal mediation analysis is to decompose the total effect into path-specific components.
In this study, we investigate the path-specific probability of necessity and sufficiency (PNS) to decompose the total PNS into path-specific components along distinct causal pathways between treatment and outcome, incorporating two mediators.
We define the path-specific PNS for decomposition and provide an identification theorem.
Furthermore, we conduct numerical experiments to assess the properties of the proposed estimators from finite samples and demonstrate their practical application using a real-world educational dataset.
\end{abstract}

\section{Introduction}

%Mediation analysis for probabilities of causation (PoC) is an essential framework for revealing the necessity and sufficiency of treatment to provoke an event via different pathways between treatment and outcome through a mediator \citep{Kawakami2024med}.
%It is a member of the PoC framework \citep{Robins1989,Tian2000,Pearl09,Kuroki2011,Dawid2014,Murtas2017,Shingaki2021,Kawakami2023b,Rubinstein2024} and also the causal mediation analyses \citep{Wright1921,Wright1934,Baron1986,Robins1992,Imai2010a,Imai2010b,TchetgenTchetgen2012,Rubinstein2024}. Both PoC and causal mediation analysis are valuable techniques for explainable artificial intelligence (XAI) \citep{Shin2021}.

Probabilities of causation (PoC) \citep{Robins1989,Tian2000,Pearl09,Kuroki2011,Dawid2014,Murtas2017,Shingaki2021,Kawakami2023b,Rubinstein2024} and causal mediation analysis \citep{Wright1921,Wright1934,Baron1986,Robins1992,Imai2010a,Imai2010b,TchetgenTchetgen2012,Rubinstein2024} are valuable tools for decision-making and  enhancing explainability in AI (XAI) \citep{Shin2021}. 
One of the primary objectives of causal mediation analysis is to decompose the total effect into path-specific components. 
%Recently, \citep{Kawakami2024med} provided PoC-based mediation analysis for measuring the necessity and sufficiency of the treatment ($X$) in causing the outcome ($Y$) via a mediator ($M$).  They introduced the natural direct and indirect probability of necessity and sufficiency, ND-PNS and NI-PNS, that can be used to  answer the following causal questions:
Recently, \citep{Kawakami2024med} provided  mediation analysis for the probability of necessity and sufficiency (PNS) of the treatment ($X$) in causing the outcome ($Y$) via a mediator ($M$). They introduced the natural direct PNS (ND-PNS) and natural indirect PNS (NI-PNS) such that PNS can be decomposed into its direct component ND-PNS and indirect component NI-PNS. ND-PNS and NI-PNS can be used to  answer the following causal questions:
\begin{center}
\vspace{0cm}
%({\bf Q1}). {\it Would the treatment still be necessary and sufficient had the value of the mediator been fixed to a certain value?}\\\vspace{0.2cm}
({\bf Q-a}). {\it Would the treatment still be necessary and sufficient had there been no influence via the mediator $M$?}\\\vspace{0.1cm}
({\bf Q-b}). {\it Would the treatment still be necessary and sufficient had %there only existed the influence 
the influence only existed via the mediator $M$?}
\vspace{0cm}
\end{center}
%\yuta{Their definitions enable the decomposition of T-PNS into its direct and indirect components through $M$.}
%as $\text{\normalfont T-PNS}=\text{\normalfont ND-PNS}+\text{\normalfont NI-PNS}$.}

In many studies, researchers often face multiple mediators  and study the causal effects along a specific pathway.
{Causal mediation analysis of the total effect $\mathbb{E}[Y_{x}-Y_{x'}]$ with two or more mediators has been studied across various fields, including statistics \citep{Lin2017,Miles2017,Miles2019,Zhou2022}, AI \citep{Avin2005,Shpitser2008}, medicine \citep{Albert2011,VanderWeele2014m,Daniel2015,Vansteelandt2017}, political science \citep{Zhou2023}, and cognitive science \citep{Shpitser2013}.
For example, \citet{Daniel2015} studied the effect of heavy drinking in the previous year on systolic blood pressure, mediated through body mass index (BMI) and gamma-glutamyl transpeptidase (GGT).}
%For example,
{\citet{Daniel2015} examined the decomposition of the total effect $\mathbb{E}[Y_1-Y_0]$ into path-specific influences.}
%\citep{Kawakami2024med} have not considered the path-specific influence of multiple mediators.
In this paper, we study the decomposition of PNS into path-specific components with two mediators and aim to answer the following four additional path-specific causal questions, incorporating an additional mediator $N$ downstream of $M$:
%two causally ordered mediators ($M$) and  ($N$):
\begin{center}
\vspace{0cm}
({\bf Q-a1}). {\it Would the treatment still be necessary and sufficient had the influence via neither of the two mediators $M$ and $N$  existed?}\\\vspace{0.1cm}
({\bf Q-a2}). {\it Would the treatment still be necessary and sufficient had the influence only existed via the second mediator $N$ but not via the first mediator $M$?
}\\\vspace{0.1cm}
({\bf Q-b1}). {\it Would the treatment still be necessary and sufficient had the influence only existed via both the first mediator $M$ and the second mediator $N$?}\\\vspace{0cm}
({\bf Q-b2}). {\it Would the treatment still be necessary and sufficient had the influence only existed via the first mediator $M$ but not via the second mediator $N$?}
\vspace{0.1cm}
\end{center}
We provide definitions of path-specific PNS %for decomposition.
%that satisfy the decomposition relationships.
%: (i) $\text{\normalfont T-PNS}=\text{\normalfont PNS}^{X \rightarrow Y}+\text{\normalfont PNS}^{X \rightarrow {N} \rightarrow  Y}+\text{\normalfont PNS}^{X \rightarrow {M} \rightarrow {N} \rightarrow  Y}+\text{\normalfont PNS}^{X \rightarrow {M}  \rightarrow  Y}$, (ii) $\text{\normalfont ND-PNS}^{{M}}=\text{\normalfont PNS}^{X \rightarrow Y}+\text{\normalfont PNS}^{X \rightarrow {N} \rightarrow  Y}$, and (iii) $\text{\normalfont NI-PNS}^{{M}}=\text{\normalfont PNS}^{X \rightarrow {M} \rightarrow {N} \rightarrow  Y}+\text{\normalfont PNS}^{X \rightarrow {M}  \rightarrow  Y}$.
%Our definitions 
that enable the decomposition of the PNS into path-specific components. 
%In this paper, we first define the path-specific PNS to address the four causal questions outlined above.
We then present an identification theorem for the path-specific PNS. 
Finally, we conduct numerical experiments to evaluate the finite-sample properties of the estimators and demonstrate their application using a real-world educational dataset.

\section{Backgrounds and Notations}

We represent a single or vector variable with a capital letter $(X)$ and its realized value with a small letter $(x)$.
Let $\mathbb{I}(\cdot)$ be an indicator function that takes $1$ if the statement in $(\cdot)$ is true and $0$ otherwise, and $\mathbbm{1}(\cdot)$ be a delta function.
Denote $\Omega_Y$ be the domain of variable $Y$,
$\mathbb{E}[Y]$ be the expectation of $Y$, 
$\mathbb{P}(Y\prec y)$ be the cumulative distribution function (CDF) of continuous variable $Y$, and $\mathfrak{p}_Y(y)$ be the probability density function (PDF) of continuous variable $Y$.
We use $X \indep Y|C$ to denote that $X$ and $Y$ are conditionally independent given $C$.
We use $\preceq$ to denote a total order. {In the univariate case, the total order $\preceq$ reduces to the standard order $\leq$.}
A formal definition of total order is given in Appendix \ref{appA}.

{\bf Structural causal models (SCM).}
We use the language of SCMs as our basic %semantic and inferential 
framework and follow the standard definition in the following \citep{Pearl09}. 
An SCM ${\cal M}$ is a tuple $\left<{\boldsymbol V},{\boldsymbol U}, {\cal F}, \mathbb{P}_{\boldsymbol U} \right>$, where ${\boldsymbol U}$ is a set of exogenous (unobserved) variables following a distribution $\mathbb{P}_{\boldsymbol U}$, and ${\boldsymbol V}$ is a set of endogenous (observable) variables whose values are determined by structural functions ${\cal F}=\{f_{V_i}\}_{V_i \in {\boldsymbol V}}$ such that $v_i:= f_{V_i}({\mathbf{pa}}_{V_i},{\boldsymbol u}_{V_i})$ where ${\mathbf{PA}}_{V_i} \subseteq {\boldsymbol V}$ and $\boldsymbol{U}_{V_i} \subseteq {\boldsymbol U}$. 
Each SCM ${\cal M}$ induces an observational distribution $\mathbb{P}_{\boldsymbol V}$ over ${\boldsymbol V}$, and a causal graph $G({\cal M})$ %over ${\boldsymbol V}$ 
in which there exists a directed edge from every variable in ${\mathbf{PA}}_{V_i}$ and $\boldsymbol{U}_{V_i}$ to $V_i$. 
An intervention of setting a set of endogenous variables ${\boldsymbol X}$ to constants ${\boldsymbol x}$, denoted by $do({\boldsymbol x})$, replaces the original equations of ${\boldsymbol X}$
 by the constants ${\boldsymbol x}$ and induces a \textit{sub-model}  ${\cal M}_{{\boldsymbol x}}$.
We denote the potential outcome $Y$ under intervention $do({\boldsymbol x})$ by $Y_{{\boldsymbol x}}({\boldsymbol u})$, which is the solution of $Y$ in the sub-model ${\cal M}_{{\boldsymbol x}}$ given ${\boldsymbol U}={\boldsymbol u}$.

{\bf Probabilities of causation (PoC) and mediation analysis for PoC.}
\citet{Kawakami2024} defined the (multivariate  conditional) PoC for vectors of continuous or discrete variables as follows:
\begin{definition}[PNS with Evidence] \citep{Kawakami2024}
\label{def41}
%For any $x',x \in \Omega_X$, $y \in \Omega_Y$, and $c \in \Omega_C$, 
The 
%(multivariate  conditional) 
PNS with evidence is defined as 
%\begin{equation}
$\text{\normalfont PNS}(y;x',x,{\cal E},c)\defeq\mathbb{P}(Y_{x'} \prec y \preceq Y_{x}|{\cal E},C=c)$,
%\end{equation}
%\begin{equation}
%\text{\normalfont PN}(y;x',x,{\cal E},c)\defeq\mathbb{P}(Y_{x'} \prec y |y \preceq Y,X=x,{\cal E},C=c),
%\end{equation}
%\begin{equation}
%\text{\normalfont PS}(y;x',x,{\cal E},c)\defeq\mathbb{P}(y \preceq Y_{x} |Y \prec y,X=x',{\cal E},C=c).
%\end{equation}
\end{definition}
{$\text{\normalfont PNS}(y;x',x,{\cal E},c)$ provides a measure of the necessity and sufficiency of $x$ w.r.t. $x'$ to produce $Y \succeq y$ given $C= c$ and evidence ${\cal E}$, that is, when $X$ is set to $X=x$, the event $Y \succeq y$ occurs; when $X$ is set to $X=x'$, the event $Y \succeq y$ does not occur.}
{Note that PNS with ${\cal E}=(y \leq Y, X=x)$ coincides with PN, and PNS with ${\cal E}=(Y < y, X=x')$ coincides with PS \citep{Kawakami2024med}.}

%\noindent $\text{\normalfont PNS}(y;x',x,{\cal E},c)$ provides a measure of the necessity and sufficiency of $x$ w.r.t. $x'$ to produce $Y\succeq y$ given $C=c$ and ${\cal E}$.
%$\text{\normalfont PNS}(y;x',x,{\cal E},c)$ include PN and PS, which is defined as $\text{\normalfont PN}(y;x',x,c)\defeq\mathbb{P}(Y_{x'} \prec y |y \preceq Y,X=x,C=c)$ and $\text{\normalfont PS}(y;x',x,c)\defeq\mathbb{P}(y \preceq Y_{x} |Y \prec y,X=x',C=c)$.
%$\text{\normalfont PN}(y;x',x,c)$ and $\text{\normalfont PS}(y;x',x,c)$ provide a measure of the necessity and sufficiency, respectively, of $x$ w.r.t. $x'$ to produce $Y\succeq y$ given $C=c$.
%$\text{\normalfont PS}(y;x',x,c)$ provides a measure of the sufficiency of $x$ w.r.t. $x'$ to produce $Y\succeq y$ given $C=c$.
We will often call $\text{\normalfont PNS}$ \textit{total PNS (T-PNS)} and denote it by $\text{\normalfont T-PNS}(y;x',x,{\cal E},c)$  for convenience. 
%Figure \ref{fig2} (a) shows the situation of potential outcomes in T-PNS.
%When treatment $X$ and outcome $Y$ are binary, 
%PNS, PS, and PS become (setting $y=1$) 
%$\text{\normalfont PNS}(c)=\mathbb{P}(Y_{0}=0,Y_{1}=1|C=c)$,
%$\text{\normalfont PN}(c)=\mathbb{P}(Y_{0}=0|Y=1,X=1,C=c)$, and 
%$\text{\normalfont PS}(c)=\mathbb{P}(Y_{1}=1|Y=0,X=0,C=c)$
%for any $c \in \Omega_C$, which reduce to Pearl's (1999) original definition when $C=\emptyset$.
%In the studies \citep{Dawid2017,Cuellar2020}, PN is called PoC.

\begin{figure}[tb]
%\vspace{-0.5cm}
   % \hspace{0.3cm}
    \centering
    \scalebox{1}{
\begin{tikzpicture}
    % x node set with absolute coordinates
    \node[mynode] (x) at (0,0) {$X$};
    \node[mynode] (y) at (4,0) {$Y$};
    \node[mynode] (u) at (2,1.25) {$C$};
    \node[mynode] (m) at (2,-1.25) {${M}$};
    
    %\node[mynode] (m2) at (3,-1.5) {${N}$};

    % Directed edge
    \path (x) edge[->] (y);
    %\path (x) edge[dotted,<->,bend right] (y);
%    \path (z) edge[->] (x);
    \path (u) edge[->] (y);
%    \path (u) edge[dotted,<->,bend left] (y);
    \path (u) edge[->]  (x);
%\path (x) edge[dotted,<->,bend left] (u);

\path (x) edge[->] (m);
\path (m) edge[->] (y);
\path (u) edge[->] (m);

%\path (x) edge[->] (m2);
%\path (m) edge[->] (m2);
%\path (u) edge[->] (m2);
%\path (m2) edge[->] (y);
\end{tikzpicture}
}
\vspace{-0cm}
    \caption{A causal graph representing SCM ${\cal M}_1$.}
    \label{DAG0}
    \end{figure}
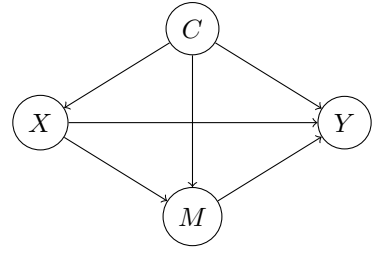

Recently, \citet{Kawakami2024med} considered the following SCM ${\cal M}_1${, corresponding to the causal graph in Figure~\ref{DAG0}}:
\begin{equation}
\begin{gathered}
Y:=f_Y(X,{M},C,U_Y), {M}:=f_{{M}}(X,C,U_{{M}}), \\
X:=f_X(C,U_X), C:=f_C(U_C),
\end{gathered}
\end{equation}
where all variables can be vectors, and $U_X$, $U_C$, $U_Y$, and $U_{{M}}$ are latent exogenous variables.

\citet{Kawakami2024med} defined the (conditional) controlled direct, natural direct, and natural indirect probabilities of necessity and sufficiency with evidence.
%to answer the questions (Q-a) and (Q-b)
%as below.
\begin{definition}[CD-PNS, ND-PNS, and NI-PNS with Evidence]
\label{def3}
%For each $x',x \in \Omega_X$, $m \in \Omega_M$, $y \in \Omega_Y$, and $c \in \Omega_C$, 
The controlled direct, natural direct, and natural indirect %probabilities of necessity and sufficiency 
PNS (CD-PNS, ND-PNS, and NI-PNS) with evidence w.r.t. $M$ are defined by
%\begin{align}
$\text{\normalfont CD-PNS}(y;x',x,m,{\cal E},c)\defeq\mathbb{P}(Y_{x',m} \prec y \preceq Y_{x,m}|{\cal E},C=c)$, 
$\text{\normalfont ND-PNS}(y;x',x,{\cal E},c)\defeq\mathbb{P}(Y_{x'} \prec y \preceq Y_{x}, Y_{x',M_{x}} \prec y|{\cal E},C=c)$, and
$\text{\normalfont NI-PNS}(y;x',x,{\cal E},c)\defeq\mathbb{P}(Y_{x'} \prec y \preceq Y_{x},y \preceq Y_{x',M_{x}}|{\cal E},C=c)$.
%\end{align}
\end{definition}
$C$ and ${\cal E}$ represent the information used to characterize a specific targeted subpopulation.
$C$ consists of subjects’ pre-treatment covariates, and ${\cal E}$ contains information about post-treatment variables, commonly referred to as evidence.
%, including the evidence of PN and PS.
ND-PNS and NI-PNS can answer the causal questions (Q-a) and (Q-b), respectively.
$\text{\normalfont T-PNS}$ is decomposed as $\text{\normalfont ND-PNS}+\text{\normalfont NI-PNS}$.
However, their applicability is restricted to the single mediator case, limiting their ability to capture more complex mediation pathways involving additional mediators.

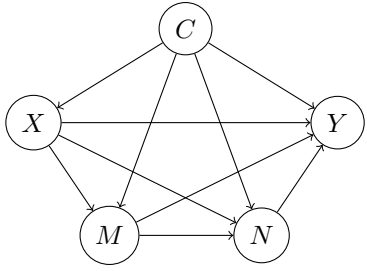
\begin{figure}[tb]
%\vspace{-0.5cm}
   % \hspace{0.3cm}
    \centering
    \scalebox{1}{
\begin{tikzpicture}
    % x node set with absolute coordinates
    \node[mynode] (x) at (0,0) {$X$};
    \node[mynode] (y) at (4,0) {$Y$};
    \node[mynode] (u) at (2,1.25) {$C$};
    \node[mynode] (m) at (1,-1.5) {${M}$};
    
    \node[mynode] (m2) at (3,-1.5) {${N}$};

    % Directed edge
    \path (x) edge[->] (y);
    %\path (x) edge[dotted,<->,bend right] (y);
%    \path (z) edge[->] (x);
    \path (u) edge[->] (y);
%    \path (u) edge[dotted,<->,bend left] (y);
    \path (u) edge[->]  (x);
%\path (x) edge[dotted,<->,bend left] (u);

\path (x) edge[->] (m);
\path (m) edge[->] (y);
\path (u) edge[->] (m);

\path (x) edge[->] (m2);
\path (m) edge[->] (m2);
\path (u) edge[->] (m2);
\path (m2) edge[->] (y);
\end{tikzpicture}
}
\vspace{-0cm}
    \caption{A causal graph representing SCM ${\cal M}_2$.}
    \label{DAG1}
    \end{figure}

{\bf Causal mediation analysis for two mediators.}
%\citet{Daniel2015} and \citet{Zhou2023} have studied the causal mediation analysis with multiple mediators.
Researchers often consider the following SCM ${\cal M}_2$ with two mediators, corresponding to the causal graph in Figure~\ref{DAG1}:
\begin{equation}
\begin{gathered}
Y:=f_Y(X,{M},{N},C,U_Y), {N}:=f_{{N}}(X,{M},C,U_{{N}}),\\ {M}:=f_{{M}}(X,C,U_{{M}}),X:=f_X(C,U_X),C:=f_C(U_C),
\end{gathered}
\end{equation}
where all variables can be vectors, 
and $U_X$, $U_C$, $U_Y$, $U_{{M}}$, and $U_{{N}}$ are latent exogenous variables.
We assume that the domains $\Omega_Y$ and $\Omega_{U_Y} \times \Omega_{U_{{M}}} \times \Omega_{U_{{N}}}$ are totally ordered sets with $\preceq$.
SCM ${\cal M}_2$ means that two mediators are causally ordered, or ${M}$ is the cause of ${N}$.

One of the most widely used models in mediation analysis with multiple mediators is a linear SCM ${\cal M}^{L2}$ %with normal distribution 
%\citep{Baron1986} 
consisting of 
%Especially, the linear SCM with normal distribution ${\cal M}^L$ consists of 
$Y:=\alpha_0+\alpha_1 X+\alpha_2 {M}+\alpha_3 {N}+\alpha_4 C+U_Y$, ${N}:=\beta_0+\beta_1 X+\beta_2 {M}+\beta_3 C+U_{{N}}$, ${M}:=\gamma_0+\gamma_1 X+\beta_3 C+U_{{M}}$, where $U_C\sim {\cal N}(0,\sigma_C)$, $U_X\sim {\cal N}(0,\sigma_X)$, $U_Y\sim {\cal N}(0,\sigma_Y)$, $U_{{M}} \sim {\cal N}(0,\sigma_{{M}})$, $U_{{N}} \sim {\cal N}(0,\sigma_{{N}})$, and they are mutually independent normal distributions.
${\cal N}(\mu,\sigma)$ means a normal distribution whose mean is $\mu$ and standard deviation is $\sigma$.
%${\cal M}^L$ is the most widely used model in the mediation analysis \citep{Baron1986}.
%Under SCM ${\cal M}^L$, the total effect of $X$ on $Y$ is $\alpha_1+\beta_1\alpha_2$, the indirect effect is $\beta_1\alpha_2$, and the direct effect is $\alpha_1$.

Then, \citet{Daniel2015} defined the 
%\hl{56}\jin{??} 
natural path-specific causal effects for binary treatment using the expectation of the counterfactuals, e.g.,
$\mathbb{E}[Y_{1,{M}_{1},{N}_{1,{M}_{0}}}]-\mathbb{E}[Y_{1,{M}_{1},{N}_{0,{M}_{0}}}]$.
They impose the following assumption to identify the path-specific causal effects.
%$\mathbb{P}\Big(Y_{x,{M}_{x'},{N}_{x'',{M}_{x'''}}}>y\Big)$.
\begin{assumption}%[Assumptions for Path-Specific Effects]
\label{SCAS}
The following conditional independence statements hold:
%\begin{equation}
%\begin{gathered}
${\normalfont (1)} \{Y_{x,{m},{n}},{M}_x,{N}_{x,{m}}\} \indep X|C=c$,
${\normalfont (2)} \{Y_{x,{m},{n}},{N}_{x,{m}}\} \indep {M}|C=c,X=x$, and
${\normalfont (3)} Y_{x,{m},{n}} \indep {N}|C=c,X=x$,
%\end{gathered}
%\end{equation}
for any ${m} \in \Omega_{{M}}$, ${n} \in \Omega_{{N}}$, $x \in \Omega_X$, and $c \in \Omega_C$, where $\mathfrak{p}_{X|C}(x|c)>0$, $\mathfrak{p}_{{M}|C,X}({m}|c,x)>0$, $\mathfrak{p}_{{N}|C,X,{M}}({n}|c,x'',{{m}}')>0$, and $\mathfrak{p}_{{M}_{x'''}|C,{M}_{x'}}({{m}}'|c,{m})>0$ for any ${m}, {{m}}' \in \Omega_{{M}}$, ${n} \in \Omega_{{N}}$, $x \in \Omega_X$, and $c \in \Omega_C$.
\end{assumption}
These independence conditions hold when there are no unmeasured confounders (or bidirected edges) between $\{X,{M},{N}\}$$\rightarrow$$Y$, $\{X,{M}\}$$\rightarrow$${N}$, and $X$$\rightarrow$${M}$.
The consistency conditions $\{X,{M},{N}\}$ on $Y$, $X$ on ${M}$, and $\{X,{M}\}$ on ${N}$ assumed in \citep{Daniel2015} hold under SCM ${\cal M}_2$.

\begin{lemma}%[$\mathbb{P}\Big(Y_{x,{M}_{x'},{N}_{x'',{M}_{x'''}}}\prec y\Big)$] 
\citep{Daniel2015}
\label{lem2}
Under SCM ${\cal M}_2$  and Assumption \ref{SCAS}, 
the conditional CDF of potential outcome $\mathbb{P}(Y_{x,{M}_{x'},{N}_{x'',{M}_{x'''}}}\prec y|C=c)$ is given by
\begin{align}
\label{eq3}
&\mathbb{P}(Y_{x,{M}_{x'},{N}_{x'',{M}_{x'''}}}\prec y|C=c)=\int_{\Omega_{{M}}}\int_{\Omega_{{M}}}\int_{\Omega_{{N}}}\nonumber\\
&\hspace{0.5cm}\mathbb{P}(Y\prec y|X=x,{M}={m},{N}={n},C=c)\nonumber\\
&\hspace{0.5cm}\times\mathfrak{p}_{{N}|C,X,{M}}({n}|c,x'',{{m}}')\ \mathfrak{p}_{{M}_{x'''}|C,{M}_{x'}}({{m}}'|c,{m})\nonumber\\
&\hspace{0.5cm}\times\mathfrak{p}_{{M}|C,X}({m}|c,x')\ d{n}d{m}d{{m}}'
\end{align}
for any $x, x', x'', x''' \in \Omega_X$, $y \in \Omega_Y$, and $c \in \Omega_C$.
\end{lemma}
This lemma does not imply the identification of $\mathbb{P}(Y_{x,{M}_{x'},{N}_{x'',{M}_{x'''}}}\prec y|C=c)$.
Instead, it states that $\mathbb{P}(Y_{x,{M}_{x'},{N}_{x'',{M}_{x'''}}}\prec y|C=c)$ is identifiable if $\mathfrak{p}_{{M}_{x''}|C,{M}_{x'}}({{m}}'|c,{m})$ is known or identifiable.
{Appendix \ref{appA3} presents the explicit form of the distribution of $Y_{x,{M}_{x'},{N}_{x'',{M}_{x'''}}}$ derived under simple SCMs.}

Furthermore, \citep{Daniel2015} showed three special cases in which $\mathfrak{p}_{{M}_{x''}|C,{M}_{x'}}({{m}}'|c,{m})$ is identifiable.
First, if $x'=x'''$, then $\mathfrak{p}_{{M}_{x'''}|C,{M}_{x'}}({{m}}'|c,{m})=\mathbbm{1}({m}={{m}}')$ holds.
Second, if there exists no effect of ${M}$ on ${N}$, then
%{\small
%\begin{align}
$\mathbb{P}(Y_{x,{M}_{x'},{N}_{x'',{M}_{x'''}}}\prec y|C=c)=\int_{\Omega_{{M}}}\int_{\Omega_{{M}}}\int_{\Omega_{{N}}}\mathbb{P}(Y\prec y|X=x,{M}={m},{N}={n},C=c)\mathfrak{p}_{{N}|C,X}({n}|c,x'')\mathfrak{p}_{{M}|C,X}({m}|c,x')d{n}d{m}$
%\end{align}
%}
holds.
Third, if we assume a specific model with Gaussian noise, i.e., ${M}|X,C \sim {\cal N}(f(X,C;\alpha),\sigma^2)$, then we can identify $\mathfrak{p}_{{M}_{x'''}|C,{M}_{x'}}({{m}}'|c,{m})$, where $f(X,C;\alpha)$ represents a parametric model.

\section{Path-specific PNS with Two Mediators}

%In this section, we define and explain the controlled, path-specific PNS with two mediators.
%to answer the causal questions (Q-a1), (Q-a2), (Q-b1), and (Q-b2).

%{\bf Definition of controlled direct PNS with two mediators.} \jin{Why do you want to define this?}
%\begin{definition}[Controlled direct PNS with two mediators]
%\st{We define the controlled direct PNS with two mediators as $\text{\normalfont CD-PNS}(y;x',x,{m},{n},{\cal E},c)\defeq\mathbb{P}(Y_{x',{m},{n}} \prec y \preceq Y_{x,{m},{n}}|{\cal E},C=c)$.}
%\end{definition}
%\st{This does not have the corresponding path-specific PNS like the controlled direct PNS with a single mediator} \citep{Kawakami2024med}, and T-PNS cannot be decomposed through CD-PNS.

%{\bf Definition of path-specific PNS with two mediators.}
%\jin{Before providing the new definition, you should provide a discussion of how to extend \citep{Kawakami2024med}'s definition to two mediator setting. For example, an obvious extension is to simply plug $\{M,N\}$ into $M$'s place. What would they represent? What's wrong with or what's the limitation of this straightforward extension? Another possible extension is to apply \citep{Kawakami2024med}'s definition blindly to the setting in Figure \ref{DAG1} by simply ignoring $N$. However, why would one want to do this since it's clearly not a valid extension?}

We can extend the definition proposed by \citet{Kawakami2024med} to a two-mediator setting by replacing the mediator $M$ in \citet{Kawakami2024med} with the set $\{M,N\}$.
However, this extension accounts for only two aggregated pathways: $X$$\rightarrow$$Y$ and $X$$\rightarrow$$\{M,N\}$$\rightarrow$$Y$.
Alternatively, one may apply the definition proposed by \citet{Kawakami2024med} while ignoring $N$, treating it as an unobserved exogenous variable of $Y$.
This approach accounts for only two marginalized pathways: $X$$\rightarrow$$Y$ and $X$$\rightarrow$$M$$\rightarrow$$Y$.
In Figure \ref{DAG1}, conditioning $C=c$, there exist four pathways between $X$ and $Y$, $X$$\rightarrow$$Y$, $X$$\rightarrow$${N} \rightarrow Y$, $X$$\rightarrow$${M}$$\rightarrow$${N}$$\rightarrow$$Y$, and $X$$\rightarrow$${M}$$\rightarrow$$Y$.

In this paper, we propose a definition of path-specific PNSs that satisfies the key decomposition relationships, ensuring that the sum of all path-specific components equals T-PNS.
%: (i) $\text{\normalfont T-PNS}=\text{\normalfont PNS}^{X \rightarrow Y}+\text{\normalfont PNS}^{X \rightarrow {N} \rightarrow  Y}+\text{\normalfont PNS}^{X \rightarrow {M} \rightarrow {N} \rightarrow  Y}+\text{\normalfont PNS}^{X \rightarrow {M}  \rightarrow  Y}$, (ii) $\text{\normalfont ND-PNS}^{{M}}=\text{\normalfont PNS}^{X \rightarrow Y}+\text{\normalfont PNS}^{X \rightarrow {N} \rightarrow  Y}$, and (iii) $\text{\normalfont NI-PNS}^{{M}}=\text{\normalfont PNS}^{X \rightarrow {M} \rightarrow {N} \rightarrow  Y}+\text{\normalfont PNS}^{X \rightarrow {M}  \rightarrow  Y}$.

%\jin{$\text{\normalfont NI-PNS}^{{M}}$ and $\text{\normalfont NI-PNS}^{{M}}$ are not even defined anywhere in the paper! It looks like (my best guess) they represent applying \citep{Kawakami2024med}'s definition blindly to the setting in Figure \ref{DAG1} by simply ignoring $N$. But (ii) and (iii) can't serve as desired requirements, rather they give an interpretation to those expressions used to define $\text{\normalfont NI-PNS}^{{M}}$ and $\text{\normalfont NI-PNS}^{{M}}$.}

%Then, we define four types of path-specific PNS for decomposition.
\begin{definition}[Path-specific PNS for decomposition]
%For each $x',x \in \Omega_X$, $y \in \Omega_Y$, and $c \in \Omega_C$, 
We define four types of path-specific PNS with two mediators as follows:
\begin{align}
&\text{\normalfont PNS}^{X \rightarrow Y}(y;x',x,{\cal E},c)\defeq\mathbb{P}(Y_{x'} \prec y \preceq Y_{x}, Y_{x',{M}_{x}} \prec y,\nonumber\\
&\hspace{3cm} Y_{x',{M}_{x},{N}_{x,{M}_{x}}} \prec y|{\cal E},C=c),\\
&\text{\normalfont PNS}^{X \rightarrow {N} \rightarrow  Y}(y;x',x,{\cal E},c)\defeq\nonumber\\
&\hspace{1cm}\mathbb{P}(Y_{x'} \prec y \preceq Y_{x}, Y_{x',{M}_{x}} \prec y,\nonumber\\
&\hspace{3cm} y \preceq  Y_{x',{M}_{x},{N}_{x,{M}_{x}}}|{\cal E},C=c),\\
&\text{\normalfont PNS}^{X \rightarrow {M} \rightarrow {N} \rightarrow  Y}(y;x',x,{\cal E},c)\defeq\nonumber\\
&\hspace{1cm}\mathbb{P}(Y_{x'} \prec y \preceq Y_{x},y \preceq Y_{x',{M}_{x}},\nonumber\\
&\hspace{2.5cm}Y_{x',{M}_{x},{N}_{x',{M}_{x'}}} \prec y|{\cal E},C=c),\\
&\text{\normalfont PNS}^{X \rightarrow {M}  \rightarrow  Y}(y;x',x,{\cal E},c)\defeq\nonumber\\
&\hspace{1cm}\mathbb{P}(Y_{x'} \prec y \preceq Y_{x},y \preceq Y_{x',{M}_{x}},\nonumber\\
&\hspace{2.5cm}y \preceq Y_{x',{M}_{x},{N}_{x',{M}_{x'}}}|{\cal E},C=c),
\end{align}
where
%\begin{equation}
${\cal E}\defeq(X=x^e, Y\in {\cal I}_Y)$
%\end{equation}
and ${\cal I}_Y$ is a half-open interval $[y^l,y^u)$ or a closed interval $[y^l,y^u]$ w.r.t. $\prec$. 
\end{definition}
%We refer to the path-specific PNS for decomposition simply as the path-specific PNS.
%All path-specific PNS definitions involve four counterfactual conditions.
If $C=\emptyset$ and ${\cal E}=\emptyset$, they reduce to the path-specific PNS for the entire population.
{We provide pathway representations of $\text{\normalfont PNS}^{X \rightarrow Y}(y;x',x,{\cal E},c)$, $\text{\normalfont PNS}^{X \rightarrow {N} \rightarrow  Y}(y;x',x,{\cal E},c)$, $\text{\normalfont PNS}^{X \rightarrow {M} \rightarrow {N} \rightarrow  Y}(y;x',x,{\cal E},c)$, and $\text{\normalfont PNS}^{X \rightarrow {M}  \rightarrow  Y}(y;x',x,{\cal E},c)$ in Appendix \ref{appA2}.}

{For binary outcome $Y\in\{0,1\}$, the definitions reduce to 
$\text{\normalfont PNS}^{X \rightarrow Y}(y;x',x,{\cal E},c)=\mathbb{P}(Y_{x'}=0,Y_{x}=1,Y_{x',M_x}=0,Y_{x',M_x,N_{x,M_x}}=0|{\cal E},C=c)$,
$\text{\normalfont PNS}^{X \rightarrow {N} \rightarrow  Y}(y;x',x,{\cal E},c)=\mathbb{P}(Y_{x'}=0,Y_{x}=1,Y_{x',{M}_{x}}=0,Y_{x',{M}_{x},{N}_{x,{M}_{x}}}=1|{\cal E},C=c)$,
$\text{\normalfont PNS}^{X \rightarrow {M} \rightarrow {N} \rightarrow  Y}(y;x',x,{\cal E},c)=\mathbb{P}(Y_{x'}=0, Y_{x}=1,Y_{x',{M}_{x}}=1,Y_{x',{M}_{x},{N}_{x',{M}_{x'}}}=0|{\cal E},C=c)$, and
$\text{\normalfont PNS}^{X \rightarrow {M}  \rightarrow  Y}(y;x',x,{\cal E},c)=\mathbb{P}(Y_{x'}=0,Y_{x}=1,Y_{x',{M}_{x}}=1,Y_{x',{M}_{x},{N}_{x',{M}_{x'}}}=1|{\cal E},C=c)$.
}

The following decomposition relationships hold:
\begin{proposition}%[Decompositions]
\label{prop31}
T-PNS 
%, ND-PNS w.r.t. ${M}$, and NI-PNS w.r.t. ${M}$ 
can be decomposed as follows:
\begin{align}
&\text{\normalfont T-PNS}(y;x',x,{\cal E},c)\nonumber\\
&=\text{\normalfont PNS}^{X \rightarrow Y}(y;x',x,{\cal E},c)+\text{\normalfont PNS}^{X \rightarrow {N} \rightarrow  Y}(y;x',x,{\cal E},c)\nonumber\\
&\hspace{1cm}+\text{\normalfont PNS}^{X \rightarrow {M} \rightarrow {N} \rightarrow  Y}(y;x',x,{\cal E},c)\nonumber\\
&\hspace{2cm}+\text{\normalfont PNS}^{X \rightarrow {M}  \rightarrow  Y}(y;x',x,{\cal E},c).
%&\text{\normalfont ND-PNS}^{{M}}(y;x',x,{\cal E},c)\nonumber\\
%&=\text{\normalfont PNS}^{X \rightarrow Y}(y;x',x,{\cal E},c)+\text{\normalfont PNS}^{X \rightarrow {N} \rightarrow  Y}(y;x',x,{\cal E},c),\\
%&\text{\normalfont NI-PNS}^{{M}}(y;x',x,{\cal E},c)=\text{\normalfont PNS}^{X \rightarrow {M} \rightarrow {N} \rightarrow  Y}(y;x',x,{\cal E},c)\nonumber\\
%&+\text{\normalfont PNS}^{X \rightarrow {M}  \rightarrow  Y}(y;x',x,{\cal E},c),
\end{align}
%where $\text{\normalfont ND-PNS}^{{M}}(y;x',x,{\cal E},c)\defeq\mathbb{P}(Y_{x'} \prec y \preceq Y_{x}, Y_{x',{M}_{x}} \prec y|{\cal E},C=c)$ and $\text{\normalfont NI-PNS}^{{M}}(y;x',x,{\cal E},c)\defeq\mathbb{P}(Y_{x'} \prec y \preceq Y_{x},y \preceq Y_{x',{M}_{x}}|{\cal E},C=c)$.
\end{proposition}
This decomposition property is essential for causal mediation analysis.

\begin{proposition}%[Decompositions]
\label{prop32}
We have
\begin{align}
%\label{eq9}
%&\text{\normalfont T-PNS}(y;x',x,{\cal E},c)\nonumber\\
%&=\text{\normalfont PNS}^{X \rightarrow Y}(y;x',x,{\cal E},c)+\text{\normalfont PNS}^{X \rightarrow {N} \rightarrow  Y}(y;x',x,{\cal E},c)\nonumber\\
%&+\text{\normalfont PNS}^{X \rightarrow {M} \rightarrow {N} \rightarrow  Y}(y;x',x,{\cal E},c)%\nonumber\\
%&+\text{\normalfont PNS}^{X \rightarrow {M}  \rightarrow  Y}(y;x',x,{\cal E},c),\\
&\mathbb{P}(Y_{x'} \prec y \preceq Y_{x}, Y_{x',{M}_{x}} \prec y|{\cal E},C=c)\nonumber\\
&=\text{\normalfont PNS}^{X \rightarrow Y}(y;x',x,{\cal E},c)+\text{\normalfont PNS}^{X \rightarrow {N} \rightarrow  Y}(y;x',x,{\cal E},c),\\
%\label{eq10}
&\mathbb{P}(Y_{x'} \prec y \preceq Y_{x},y \preceq Y_{x',{M}_{x}}|{\cal E},C=c)\nonumber\\
&=\text{\normalfont PNS}^{X \rightarrow {M} \rightarrow {N} \rightarrow  Y}(y;x',x,{\cal E},c)\nonumber\\
&\hspace{2cm}+\text{\normalfont PNS}^{X \rightarrow {M}  \rightarrow  Y}(y;x',x,{\cal E},c).
\end{align}
\end{proposition}
The expressions $\mathbb{P}(Y_{x'} \prec y \preceq Y_{x}, Y_{x',{M}_{x}} \prec y|{\cal E},C=c)$ and $\mathbb{P}(Y_{x'} \prec y \preceq Y_{x},y \preceq Y_{x',{M}_{x}}|{\cal E},C=c)$ coincide with the definitions proposed by \citet{Kawakami2024med} by simply ignoring $N$, considering it as one of the unobserved exogenous variables of $Y$.
We denote these quantities as follows: $\text{\normalfont ND-PNS}^{{M}}(y;x',x,{\cal E},c)=\mathbb{P}(Y_{x'} \prec y \preceq Y_{x}, Y_{x',{M}_{x}} \prec y|{\cal E},C=c)$ and $\text{\normalfont NI-PNS}^{{M}}(y;x',x,{\cal E},c)=\mathbb{P}(Y_{x'} \prec y \preceq Y_{x},y \preceq Y_{x',{M}_{x}}|{\cal E},C=c)$.

\begin{figure}[tb]
%\vspace{-0.5cm}
   % \hspace{0.3cm}
    \centering
    %\vspace{0cm}
    \scalebox{1}{
\begin{tikzpicture}
    % x node set with absolute coordinates
    \path (1,0) edge[->] (8,0); 
    \coordinate[label = above:$Y_{x'}$] (A) at (1.5,0);
    \node at (A)[circle,fill,inner sep=1pt]{};
    \coordinate[label = above:$Y_{x',{M}_{x}}$] (B) at (3,0); 
    \node at (B)[circle,fill,inner sep=1pt]{};
    \coordinate[label = above:$Y_{x',{M}_{x},{N}_{x,{M}_{x}}}$] (C) at (4.5,0);
    \node at (C)[circle,fill,inner sep=1pt]{};
    \coordinate[label = above:$y$] (D) at (6,0);
    \node at (D)[circle,fill,inner sep=1pt]{};
    \coordinate[label = above:$Y_{x}$] (D) at (7.5,0);
    \node at (D)[circle,fill,inner sep=1pt]{};
    %\path (B) edge[->,bend right,dotted] (D); 
\end{tikzpicture}
}
\caption*{(a1) Order of potential outcomes in $\text{\normalfont PNS}^{X \rightarrow Y}$.}
%\vspace{0cm}
\scalebox{1}{
\begin{tikzpicture}
    % x node set with absolute coordinates
    \path (1,0) edge[->] (8,0); 
    \coordinate[label = above:$Y_{x'}$] (A) at (1.5,0);
    \node at (A)[circle,fill,inner sep=1pt]{};
    \coordinate[label = above:$Y_{x',{M}_{x}}$] (B) at (3,0); 
    \node at (B)[circle,fill,inner sep=1pt]{};
    \coordinate[label = above:$y$] (C) at (4.5,0);
    \node at (C)[circle,fill,inner sep=1pt]{};
    \coordinate[label = above:$Y_{x',{M}_{x},{N}_{x,{M}_{x}}}$] (D) at (6,0);
    \node at (D)[circle,fill,inner sep=1pt]{};
    \coordinate[label = above:$Y_{x}$] (D) at (7.5,0);
    \node at (D)[circle,fill,inner sep=1pt]{};
    %\path (B) edge[->,bend right,dotted] (D); 
\end{tikzpicture}
}
\caption*{(a2) Order of potential outcomes in $\text{\normalfont PNS}^{X \rightarrow  {N} \rightarrow Y}$.}
    %\vspace{0cm}
    \scalebox{1}{
\begin{tikzpicture}
    % x node set with absolute coordinates
    \path (1,0) edge[->] (8,0); 
    \coordinate[label = above:$Y_{x'}$] (A) at (1.5,0);
    \node at (A)[circle,fill,inner sep=1pt]{};
    \coordinate[label = above:$Y_{x',{M}_{x},{N}_{x',{M}_{x'}}}$] (B) at (3,0); 
    \node at (B)[circle,fill,inner sep=1pt]{};
    \coordinate[label = above:$y$] (C) at (4.5,0);
    \node at (C)[circle,fill,inner sep=1pt]{};
    \coordinate[label = above:$Y_{x',{M}_{x}}$] (D) at (6,0);
    \node at (D)[circle,fill,inner sep=1pt]{};
    \coordinate[label = above:$Y_{x}$] (D) at (7.5,0);
    \node at (D)[circle,fill,inner sep=1pt]{};
    %\path (B) edge[->,bend right,dotted] (D); 
\end{tikzpicture}
}
\caption*{(b1) Order of potential outcomes in $\text{\normalfont PNS}^{X \rightarrow  {M} \rightarrow  {N} \rightarrow Y}$.}

    \scalebox{1}{
\begin{tikzpicture}
    % x node set with absolute coordinates
    \path (1,0) edge[->] (8,0); 
    \coordinate[label = above:$Y_{x'}$] (A) at (1.5,0);
    \node at (A)[circle,fill,inner sep=1pt]{};
    \coordinate[label = above:$y$] (B) at (3,0); 
    \node at (B)[circle,fill,inner sep=1pt]{};
    \coordinate[label = above:$Y_{x',{M}_{x},{N}_{x',{M}_{x'}}}$] (C) at (4.5,0);
    \node at (C)[circle,fill,inner sep=1pt]{};
    \coordinate[label = above:$Y_{x',{M}_{x}}$] (D) at (6,0);
    \node at (D)[circle,fill,inner sep=1pt]{};
    \coordinate[label = above:$Y_{x}$] (D) at (7.5,0);
    \node at (D)[circle,fill,inner sep=1pt]{};
    %\path (B) edge[->,bend right,dotted] (D); 
\end{tikzpicture}
}
\caption*{(b2) Order of potential outcomes in $\text{\normalfont PNS}^{X \rightarrow  {M} \rightarrow Y}$.}
\vspace{-0cm}
\caption{Order of potential outcomes.}% in each path-specific PNS.}
\label{fig2}
\end{figure}

First, $\text{\normalfont PNS}^{X \rightarrow Y}$ has the following four counterfactuals: 
\begin{center}
\vspace{-0.1cm}
({\bf A1}). {\it ``had the treatment been $x'$, the outcome would be $Y\prec y$"} $(Y_{x'}=Y_{x',{M}_{x'},{N}_{x',{M}_{x'}}} \prec y)$,\\\vspace{0.1cm}
({\bf A2}). {\it ``had the treatment been $x'$ but the first mediator was kept at the same value ${M}_{x}$, the outcome would be {$Y \prec y$"}} $(Y_{x',{M}_{x}}=Y_{x',{M}_{x},{N}_{x',{M}_{x}}} \prec y)$, \\\vspace{0.1cm}
({\bf A3}). {\it ``had the treatment been $x'$ but the first mediator was kept at the same value ${M}_{x}$ and the second mediator was kept at the same value ${N}_{x,{M}_{x}}$, the outcome would be {$Y \prec y$"}}
$(Y_{x',{M}_{x},{N}_{x,{M}_{x}}} \prec y)$, and \\\vspace{0.1cm}
({\bf A4}). {\it ``had the treatment been $x$, the outcome would be $y \preceq Y$"}
$(y \preceq Y_{x}=Y_{x,{M}_{x},{N}_{x,{M}_{x}}})$.\vspace{-0.1cm}
\end{center} 
The relative values of the potential outcomes are presented in Figure \ref{fig2} (a1).
Conditions (A1) and (A4) imply that $Y_{x'} \prec y \preceq Y_{x}$, which corresponds to the same condition in T-PNS and represents that the treatment $x$ is necessary and sufficient w.r.t. $x'$ to provoke the event $y \preceq Y$  given $C=c$. 
Conditions (A2) and (A4) imply that $ Y_{x',{M}_{x}} \prec y \preceq Y_{x,{M}_{x}}$, which represents the necessity and sufficiency of $x$ w.r.t. $x'$ to produce $Y\succeq y$ given $C=c$, while keeping the values of the first mediator by the same as ${M}_{x}$.
Conditions (A3) and (A4) imply that $Y_{x',{M}_{x},{N}_{x,{M}_{x}}} \prec y \preceq Y_{x,{M}_{x}}$, which represents the necessity and sufficiency of $x$ w.r.t. $x'$ to produce $Y\succeq y$ given $C=c$, while keeping the values of the first mediator by the same as $M_{x}$ and the second mediator by the same as ${N}_{x,{M}_{x}}$.
Then, these conditions indicate that the treatment would be necessary and sufficient even if the influence only existed neither via the first or second mediators.
$\text{\normalfont PNS}^{X \rightarrow Y}$ answers the question (Q-a1).

Second, $\text{\normalfont PNS}^{X \rightarrow {N} \rightarrow Y}$ has four counterfactuals: 
\begin{center}
\vspace{-0.1cm}
({\bf B1}). {\it ``had the treatment been $x'$, the outcome would be $Y\prec y$"}
$(Y_{x'}=Y_{x',{M}_{x'},{N}_{x',{M}_{x'}}} \prec y)$,\\\vspace{0.1cm}
({\bf B2}). {\it ``had the treatment been $x'$ but the first mediator was kept at the same value ${M}_{x}$, the outcome would be {$Y \prec y$"}}
$(Y_{x',{M}_{x}}=Y_{x',{M}_{x},{N}_{x',{M}_{x}}} \prec y)$, \\\vspace{0.1cm}
({\bf B3}). {\it ``had the treatment been $x'$ but the first mediator was kept at the same value ${M}_{x}$ 
%when the treatment is $x$
and the second mediator was kept at the same value ${N}_{x,{M}_{x}}$, the outcome would be {$y \preceq Y$"}}
$(y \preceq Y_{x',{M}_{x},{N}_{x,{M}_{x}}})$, and \\\vspace{0.1cm}
({\bf B4}). {\it ``had the treatment been $x$, the outcome would be $y \preceq Y$"}
$(y \preceq Y_{x}=Y_{x,{M}_{x},{N}_{x,{M}_{x}}})$.\vspace{-0.1cm}
\end{center} 
The relative values of the potential outcomes are shown in Figure \ref{fig2} (a2).
Conditions (B1) and (B4) imply that $Y_{x'} \prec y \preceq Y_{x}$, which corresponds to  the same condition in T-PNS and represents that the treatment $x$ is necessary and sufficient w.r.t. $x'$ to provoke the event $y \preceq Y$  given $C=c$. 
Conditions (B2) and (B4) imply that $ Y_{x',{M}_{x}} \prec y \preceq Y_{x,{M}_{x}}$, which represents the necessity and sufficiency of $x$ w.r.t. $x'$ to produce $Y\succeq y$ given $C=c$, while keeping the values of the first mediator the same as ${M}_{x}$.
Conditions (B2) and (B3) imply that $Y_{x',{M}_{x}} \prec y \preceq Y_{x',{M}_{x},{N}_{x,{M}_{x}}}$, which represents the necessity and sufficiency of ${N}_{x,{M}_{x}}$ w.r.t. ${N}_{x',{M}_{x}}$ to produce $Y\succeq y$ given $C=c$, while keeping the values of the first mediator the same as ${M}_{x}$.
%when the treatment is $x$.
Then, these conditions indicate that the treatment would be necessary and sufficient even if there existed only the influence via the second mediator, not via the first mediator.
$\text{\normalfont PNS}^{X \rightarrow {N} \rightarrow Y}$ answers the question (Q-a2).

Third, $\text{\normalfont PNS}^{X \rightarrow  {M} \rightarrow  {N} \rightarrow Y}$ has four counterfactuals:
\begin{center}
\vspace{-0.1cm}
({\bf C1}). {\it ``had the treatment been $x'$, the outcome would be $Y\prec y$"}
$(Y_{x'}=Y_{x',{M}_{x'},{N}_{x',{M}_{x'}}} \prec y)$,\\\vspace{0.1cm}
({\bf C2}). {\it ``had the treatment been $x'$ but the first mediator was kept at the same value ${M}_{x}$
%when the treatment is $x$
and the second mediator was kept at the same value ${N}_{x',{M}_{x'}}$, the outcome would be {$Y \prec y$"}}
$(Y_{x',{M}_{x},{N}_{x',{M}_{x'}}} \prec y)$, \\\vspace{0.1cm}
({\bf C3}). {\it ``had the treatment been $x'$ but the first mediator was kept at the same value ${M}_{x}$, the outcome would be $y \preceq  Y$"}
$(y \preceq Y_{x',{M}_{x}}=Y_{x',{M}_{x},{N}_{x',{M}_{x}}})$, and \\\vspace{0.1cm}
({\bf C4}). {\it ``had the treatment been $x$, the outcome would be $y \preceq Y$"}
$(y \preceq  Y_{x}=Y_{x,{M}_{x},{N}_{x,{M}_{x}}})$.\vspace{-0.1cm}
\end{center} 
The relative values of the potential outcomes are shown in Figure \ref{fig2} (b1).
Conditions (C1) and (C4) imply that $Y_{x'} \prec y \preceq Y_{x}$, which corresponds to  the same condition in T-PNS and states that the treatment $x$ is necessary and sufficient w.r.t. $x'$ to provoke the event $y \preceq Y$ given $C=c$. 
Conditions (C1) and (C3) imply that $ Y_{x',{M}_{x'}} \prec y \preceq Y_{x',{M}_{x}}$, which represents the necessity and sufficiency of ${M}_{x}$ w.r.t. ${M}_{x'}$ to produce $Y\succeq y$ given $C=c$, when the treatment is set to $x'$. 
Conditions (C2) and (C3) imply that $Y_{x',{M}_{x},{N}_{x',{M}_{x'}}} \prec y \preceq Y_{x',{M}_{x},{N}_{x',{M}_{x}}}$, which represents the necessity and sufficiency of ${N}_{x',{M}_{x}}$ w.r.t. ${N}_{x',{M}_{x'}}$ to produce $Y\succeq y$ given $C=c$, while keeping the values of the first mediator fixed at ${M}_{x}$ and setting the treatment to $x'$.
Then, these conditions indicate that the treatment would be necessary and sufficient if there existed only the influence both via the first and the second mediators.
$\text{\normalfont PNS}^{X \rightarrow  {M} \rightarrow  {N} \rightarrow Y}$ answers the question (Q-b1).

Fourth, $\text{\normalfont PNS}^{X \rightarrow  {M} \rightarrow Y}$ has four counterfactuals:
\begin{center}
\vspace{-0.1cm}
({\bf D1}). {\it ``had the treatment been $x'$, the outcome would be $Y\prec y$"}
$(Y_{x'}=Y_{x',{M}_{x'},{N}_{x',{M}_{x'}}} \prec y)$,\\\vspace{0.1cm}
({\bf D2}). {\it ``had the treatment been $x'$ but the first mediator was kept at the same value ${M}_{x}$
%when the treatment is $x$
and the second mediator was kept at the same value ${N}_{x',{M}_{x'}}$, the outcome would be {$y \preceq Y$"}}
$(y \preceq Y_{x',{M}_{x},{N}_{x',{M}_{x'}}})$, \\\vspace{0.1cm}
({\bf D3}). {\it ``had the treatment been $x'$ but the first mediator was kept at the same value ${M}_{x}$, the outcome would be $y \preceq  Y$"}
$(y \preceq Y_{x',{M}_{x}}=Y_{x',{M}_{x},{N}_{x',{M}_{x}}})$, and \\\vspace{0.1cm}
({\bf D4}). {\it ``had the treatment been $x$, the outcome would be $y \preceq Y$"}
$(y \preceq  Y_{x}=Y_{x',{M}_{x'},{N}_{x',{M}_{x'}}})$.\vspace{-0.1cm}
\end{center} 
The relative values of the potential outcomes are shown in Figure \ref{fig2} (b2).
Conditions (D1) and (D4) imply that $Y_{x'} \prec y \preceq Y_{x}$, which corresponds to  the same condition in T-PNS and states that the treatment $x$ is necessary and sufficient w.r.t. $x'$ to provoke the event $y \preceq Y$  given $C=c$. 
Conditions (D1) and (D3) imply that $ Y_{x',M_{x'}} \prec y \preceq Y_{x',M_{x}}$, which represents the necessity and sufficiency of $M_{x}$ w.r.t. $M_{x'}$ to produce $Y\succeq y$ given $C=c$, when the treatment is set to $x'$. 
Conditions (D1) and (D2) imply that $Y_{x',{M}_{x'},{N}_{x',{M}_{x'}}} \prec y \preceq Y_{x',{M}_{x},{N}_{x',{M}_{x'}}}$, which represents the necessity and sufficiency of ${M}_{x}$ w.r.t. ${M}_{x'}$ to produce $Y\succeq y$ given $C=c$, while keeping the values of the second mediator fixed at ${N}_{x',{M}_{x'}}$ and setting the treatment to $x'$.
Then, these conditions indicate that the treatment would be necessary and sufficient if there existed only the influence via the first mediator, not via the second mediator.
$\text{\normalfont PNS}^{X \rightarrow  {M} \rightarrow Y}$ answers the question (Q-b2).

{{\bf Remark.}
\citet{Pearl2001} defined the path-specific effects from the perspective of a \emph{path-deactivation process}.
In the two-mediator setting, each path-specific effect, from the perspective of a path-deactivation process, is expressed as:
$Y_{x,{M}_{x'},{N}_{x',{M}_{x'}}}-Y_{x'}$, 
$Y_{x',{M}_{x'},{N}_{x,{M}_{x'}}}-Y_{x'}$, 
$Y_{x',{M}_{x'},{N}_{x',{M}_{x}}}-Y_{x'}$, and 
$Y_{x',{M}_{x},{N}_{x',{M}_{x'}}}-Y_{x'}$, respectively.
Similarly, one may define the four types of path-specific PNS with two mediators as follows:
$\text{\normalfont PNS'}^{X \rightarrow Y}(y;x',x,{\cal E},c)\defeq\mathbb{P}(Y_{x'} \prec y \preceq Y_{x,{M}_{x'},{N}_{x',{M}_{x'}}}|{\cal E},C=c)$, 
$\text{\normalfont PNS'}^{X \rightarrow {N} \rightarrow  Y}(y;x',x,{\cal E},c)\defeq \mathbb{P}(Y_{x'} \prec y \preceq Y_{x',{M}_{x'},{N}_{x,{M}_{x'}}}|{\cal E},C=c)$, 
$\text{\normalfont PNS'}^{X \rightarrow {M} \rightarrow {N} \rightarrow  Y}(y;x',x,{\cal E},c)\defeq\mathbb{P}(Y_{x'} \prec y \preceq Y_{x',{M}_{x'},{N}_{x',{M}_{x}}}|{\cal E},C=c)$, and 
$\text{\normalfont PNS'}^{X \rightarrow {M}  \rightarrow  Y}(y;x',x,{\cal E},c)\defeq\mathbb{P}(Y_{x'} \prec y \preceq Y_{x',{M}_{x},{N}_{x',{M}_{x'}}}|{\cal E},C=c)$. 
However, these definitions do not satisfy the desired decomposition relationships.
%Our approach differs from his approach.
%In terms of a path-deactivation process, the four types of path-specific PNS with two mediators are defined as follows:
%\begin{align}
%$\text{\normalfont PNS'}^{X \rightarrow Y}(y;x',x,{\cal E},c)\defeq\mathbb{P}(Y_{x'} \prec y \preceq Y_{x,{M}_{x'},{N}_{x',{M}_{x'}}}|{\cal E},C=c)$,
%$\text{\normalfont PNS'}^{X \rightarrow {N} \rightarrow  Y}(y;x',x,{\cal E},c)\defeq \mathbb{P}(Y_{x'} \prec y \preceq Y_{x',{M}_{x'},{N}_{x,{M}_{x'}}}|{\cal E},C=c)$,
%$\text{\normalfont PNS'}^{X \rightarrow {M} \rightarrow {N} \rightarrow  Y}(y;x',x,{\cal E},c)\defeq\mathbb{P}(Y_{x'} \prec y \preceq Y_{x',{M}_{x'},{N}_{x',{M}_{x}}}|{\cal E},C=c)$, and
%$\text{\normalfont PNS'}^{X \rightarrow {M}  \rightarrow  Y}(y;x',x,{\cal E},c)\defeq\mathbb{P}(Y_{x'} \prec y \preceq Y_{x',{M}_{x},{N}_{x',{M}_{x'}}}|{\cal E},C=c)$.
%\end{align}
%However, this definition does not satisfy the desired decomposition relationships.
We adopt a different approach from his path-deactivation process in the context of T-PNS decomposition. 
}

%\jin{This needs more detailed discussion, including the following comment. Otherwise, impossible to understand what you are talking about.}

%\yuta{
%[Comment: I need to separately discuss the path-specific PNS from the perspective of a path-deactivation process. In this paper, I call my definition ``path-specific PNS for decomposition". 
%}

\section{Identification of Path-Specific PNS with Two Mediators}

In this section, we provide an identification theorem for the path-specific PNS for decomposition.

{\bf New identification assumption of the conditional PDF of potential outcomes.}
%$\mathfrak{p}_{{M}_{x'''}|C,{M}_{x'}}({{m}}'|c,{m})$.
%for $\mathbb{P}\big(Y_{x,{M}_{x'},{N}_{x'',{M}_{x'''}}}\prec y\big)$}
%$\mathbb{P}\big(Y_{x,{M}_{x'},{N}_{x'',{M}_{x'''}}}\prec y\big)$ is identifiable under a specific model with Gaussian noises, discussed in \citep{Daniel2015}.
%We provide the new identification theorem of $\mathbb{P}\big(Y_{x,{M}_{x'},{N}_{x'',{M}_{x'''}}}\prec y\big|C=c\big)$.
Identifying the conditional PDF of potential outcomes $\mathfrak{p}_{{M}_{x'''}|C,{M}_{x'}}({{m}}'|c,{m})$ remains challenging in Lemma \ref{lem2}, and \citet{Daniel2015} addressed this issue by employing a specific parametric model.

We introduce a new nonparametric identification assumption for the counterfactual conditional PDF $\mathfrak{p}_{{M}_{x'''}|C,{M}_{x'}}({{m}}'|c,{m})$ as follows.
\begin{assumption}[Strictly monotonicity over $f_{{M}}$]
\label{ASM}
The function $f_{{M}}$ is either strictly monotonic increasing on $U_{{M}}$ for all $x \in \Omega_X$ and $c \in \Omega_C$, or strictly monotonic decreasing on $U_{{M}}$ for all $x \in \Omega_X$ and $c \in \Omega_C$, almost surely w.r.t. $\mathbb{P}_{U_{{M}}}$ with $\mathfrak{p}(u_{{M}})>0$.
\end{assumption}
This assumption is similar to Assumption 3.5 (strict monotonicity over $f_Y$) in \citep{Kawakami2024}.
The parametric model used in \citep{Daniel2015} satisfies this assumption.
%; thus, our assumption is weaker than theirs.
Then $\mathfrak{p}_{{M}_{x'''}|C,{M}_{x'}}({{m}}'|c,{m})$ is identifiable as follows.
\begin{lemma}%[Identification of $\mathfrak{p}_{{M}_{x'''}|C,{M}_{x'}}({{m}}'|c,{m})$]
\label{lem41}
Under SCM  ${\cal M}_2$ and Assumption \ref{ASM}, we have
\begin{equation}
\label{eq23}
\begin{aligned}
&\mathfrak{p}_{{M}_{x'''}|C,{M}_{x'}}({{m}}'|c,{m})=\\
&\mathbbm{1}(F^{-1}_{{M}|X=x''',C=c}(\mathbb{P}({M} \preceq {m}|X=x',C=c))={{m}}'),
\end{aligned}
\end{equation}
where $F^{-1}_{{M}|X=x''',C=c}$ is the inverse function of conditional CDF $\mathbb{P}({M} \preceq {m}|X=x''',C=c)$ on ${m}$ given $X=x'''$ and $C=c$.
\end{lemma}
Note that, under Assumption \ref{ASM}, $\mathbb{P}({M} \preceq {m}|X=x''',C=c)$ is monotonically increasing and continuous in ${m}$.
Consequently, an inverse function $F^{-1}_{{M}|X=x''',C=c}$ always exists.
%$F^{-1}_{{M}|X=x''',C=c}\left(\mathbb{P}({M} \preceq {m}|X=x',C=c)\right)$ means the $100\times\mathbb{P}({M} \preceq {m}|X=x',C=c)$ percentile of the conditional distribution $\mathbb{P}({M}|X=x''',C=c)$.
When $x'''=x'$, the expression $\mathbbm{1}(F^{-1}_{{M}|X=x''',C=c}(\mathbb{P}({M} \preceq {m}|X=x',C=c))={{m}}')$ reduces to $\mathbbm{1}({m}={{m}}')$, which aligns with the result presented in \citet{Daniel2015}.
%This result can be applied to the path-specific causal effects defined by \citet{Daniel2015}.

We then have the following theorem:
\begin{theorem}%[Identification of $\mathbb{P}(Y_{x,{M}_{x'},{N}_{x'',{M}_{x'''}}}\prec y|C=c)$]
\label{theo41}
Under SCM  ${\cal M}_2$ and Assumption \ref{ASM}, for any $x,x',x'',x''' \in \Omega_X$ and $c \in \Omega_C$, $\mathbb{P}(Y_{x,{M}_{x'},{N}_{x'',{M}_{x'''}}}\prec y|C=c)$ is identifiable by $\theta(y;x,x',x'',x''',c)$, where
\begin{align}
&\theta(y;x,x',x'',x''',c)=\nonumber\\
&\hspace{0cm}\int_{\Omega_{{M}}}\int_{\Omega_{{N}}}\mathbb{P}(Y\prec y|X=x,{M}={m},{N}={n},C=c)\nonumber\\
&\hspace{0cm}\times\mathfrak{p}_{{N}|C,X,{M}}\big({n}|c,x'',F^{-1}_{{M}|X=x''',C=c}\nonumber\\
&\hspace{3cm}(\mathbb{P}({M} \preceq {m}|X=x',C=c))\big)\ \nonumber\\
&\hspace{0cm}\times\mathfrak{p}_{{M}|C,X}({m}|c,x')\ d{n}d{m}.
\end{align}
\end{theorem}

{\bf Identification of path-specific PNS with two mediators.}
{The identification of PoC relies on monotonicity assumptions, as discussed in the literature \citep{Tian2000, Kawakami2024, Kawakami2024med}. 
We adopt similar assumptions to those in \citep{Kawakami2024med}.} 
%\begin{assumption}
%\label{SUP2}
%Potential outcome $Y_{x,m}$ has conditional PDF $\mathfrak{p}_{Y_{x,m}|C=c}$ for each $x \in \Omega_{X}$, $m \in \Omega_M$, and $c \in \Omega_C$, and its support $\{y \in \Omega_Y: \mathfrak{p}_{Y_{x,m}|C=c}(y) \ne0 \}$ is
%the same
%$[-\infty,\infty]$ 
%for each $x \in \Omega_{X}$, $m \in \Omega_M$, and $c \in \Omega_C$.
%\end{assumption}
\begin{assumption}
\label{SUP1}
Potential outcome $Y_{x,{M}_{x'},{N}_{x'',{M}_{x'''}}}$ has conditional PDF $\mathfrak{p}_{Y_{x,{M}_{x'},{N}_{x'',{M}_{x'''}}}|C=c}$ for each $x, x', x'',x''' \in \Omega_{X}$ and $c \in \Omega_C$, and its support $\{y \in \Omega_Y: \mathfrak{p}_{Y_{x,{M}_{x'},{N}_{x'',{M}_{x'''}}}|C=c}(y) \ne 0 \}$ is
the same
%$[-\infty,\infty]$ 
for each $x, x', x'',x''' \in \Omega_{X}$ and $c \in \Omega_C$.
\end{assumption}
\noindent Assumption~\ref{SUP1} is reasonable for continuous variables. 
For example, potential outcomes $Y_{x,{M}_{x'},{N}_{x'',{M}_{x'''}}}$ often have $[-\infty,\infty]$ support, %for each $x \in \Omega_{X}$, $m \in \Omega_M$, and $c \in \Omega_C$, 
such as in linear SCM ${\cal M}^{L2}$.

Let a compounded function $g$ be
\begin{align}
&g(x,x',x'',x''',c,\tilde{U})\defeq f_Y(x,f_{{M}}(x',c,U_{{M}}),\nonumber\\
&\hspace{1cm} f_{{N}}(x'',f_{{M}}(x''',c,U_{{M}}),c,U_{{N}}),c,U_Y)
\end{align}
for all $x, x', x'',x''' \in \Omega_X$ and $c \in \Omega_C$, where $\tilde{U}=(U_Y,U_{{M}},U_{{N}})$. 
We assume the following %two conditions similar to those in \citep{Kawakami2024} 
for identifying path-specific PNS: 
\begin{assumption}[Monotonicity over $g$]
\label{AS1}
The function $g(x,x',x'',x''',c,\tilde{U})$ is either monotonic increasing on $\tilde{U}$ for all $x,x',x'',x''' \in \Omega_X$ and $c \in \Omega_C$, or monotonic decreasing on $\tilde{U}$ for all $x,x',x'',x''' \in \Omega_X$ and $c \in \Omega_C$, almost surely w.r.t. $\mathbb{P}_{\tilde{U}}$.
\end{assumption}
\noindent Or, alternatively, we assume the following:

\noindent{\bf Assumption 4.3'}
(Conditional monotonicity over $Y_{x,{M}_{x'},{N}_{x'',{M}_{x'''}}}$)
{\it 
The potential outcomes $Y_{x,{M}_{x'},{N}_{x'',{M}_{x'''}}}$ satisfy:  for any $x,x',x'',x''',x^*,x^{**},x^{***},x^{****} \in \Omega_X$, $y \in \Omega_Y$, and $c \in \Omega_C$, 
either $\mathbb{P}(Y_{x,{M}_{x'},{N}_{x'',{M}_{x'''}}}\prec y \preceq Y_{x^{*},{M}_{x^{**}},{N}_{x^{***},{M}_{x^{****}}}}|C=c)=0$ or $\mathbb{P}(Y_{x,{M}_{x'},{N}_{x'',{M}_{x'''}}} \prec y \preceq Y_{x^{*},{M}_{x^{**}},{N}_{x^{***},{M}_{x^{****}}}}|C=c)=0$.
}
\vspace{0.1in}

\noindent %\yuta{Considering a binarized outcome $O=\mathbb{I}(Y \geq y)$, Assumption \ref{AS1}' becomes ``$\mathbb{P}(O_{x',M_{x}}=0,O_{x''',M_{x''}}=1)=0$ or $\mathbb{P}(O_{x',M_{x}}=1,O_{x''',M_{x''}}=0)=0$."}
Similarly to the results in \cite{Kawakami2024med}, Assumptions~\ref{AS1} and 4.3' are equivalent under Assumption~\ref{SUP1}. 
We provide the proof of this statement in Appendix \ref{appB}.
We note that both the linear SCM ${\cal M}^{L2}$ and the nonlinear SCM with normal distribution ${\cal M}^N$ satisfy Assumption \ref{AS1} with $\tilde{U}=U_Y+U_{{M}}+U_{{N}}$. 
%\yuta{[Independence condition is not needed. It is required for Assumption 1.]}
%Furthermore, the nonlinear SCM with normal distribution ${\cal M}^N$
%consists of $Y:=f_Y(X,M,C)+U_Y$ and $M:=f_M(X,C)+U_M$, where $U_Y\sim {\cal N}(0,\sigma^2_Y)$ and $ U_M \sim {\cal N}(\mu_M,\sigma^2_M)$Letting $\tilde{U}=U_Y+U_M$, SCM ${\cal M}^N$, 
%also satisfies both Assumptions \ref{AS1} and \ref{SUP1}.
%Alternatively, we can assume the monotonicity over $Y_x$ or $Y_{x',M_{x}}$, which is equivalent to the monotonicity over $f_Y$ or $f_Y\circ f_M$ (Assumptions \ref{AS2} and \ref{AS1}) under Assumptions \ref{SUP2} and \ref{SUP1}, as in \citep{Kawakami2024}. 

Then, the path-specific PNSs are identifiable as follows.
\begin{theorem}[Identification of path-specific PNS]
\label{theo42}
Let ${\cal I}_Y$ be a half-open interval in evidence ${\cal E}$. 
We have the following two statements.

{\normalfont (1).} 
Under SCM  ${\cal M}_2$ and Assumptions 
%\ref{ASM},
\ref{SCAS}, \ref{SUP1} and \ref{AS1}, for any $x',x \in \Omega_X$, $y \in \Omega_Y$, and $c \in \Omega_C$, we have

{\normalfont (1A).} If $\mathbb{P}(Y \prec y^l|X=x^e,C=c) \ne \mathbb{P}(Y \prec y^u|X=x^e,C=c)$, then
\begin{align}
&\text{\normalfont PNS}^{X \rightarrow Y}(y;x',x,{\cal E},c)=\max\{\gamma^1/\delta,0\},\\
&\text{\normalfont PNS}^{X \rightarrow {N} \rightarrow Y}(y;x',x,{\cal E},c)=\max\{\gamma^2/\delta,0\},
%&\text{\normalfont PNS}^{X \rightarrow {M} \rightarrow {N} \rightarrow  Y}(y;x',x,{\cal E},c)=\max\{\gamma^3/\delta,0\},\\
%&\text{\normalfont PNS}^{X \rightarrow {M} \rightarrow  Y}(y;x',x,{\cal E},c)=\max\{\gamma^4/\delta,0\},
\end{align}
where
\begin{align}
&\gamma^1=\min\big\{\theta(y;x',x',x',x',c),\theta(y;x',x,x',x,c),\nonumber\\
&\theta(y;x',x,x,x,c),\mathbb{P}(Y \prec y^u|X=x^e,C=c)\big\}\nonumber\\
&-\max\big\{\theta(y;x,x,x,x,c),\mathbb{P}(Y \prec y^l|X=x^e,C=c)\big\},\\
&\gamma^2=\min\big\{\theta(y;x',x',x',x',c),\theta(y;x',x,x',x,c),\nonumber\\
&\mathbb{P}(Y \prec y^u|X=x^e,C=c)\big\}-\max\big\{\theta(y;x,x,x,x,c),\nonumber\\
&\mathbb{P}(Y \prec y^l|X=x^e,C=c),\theta(y;x',x,x,x,c)\big\},\\
%&\gamma^3=\min\big\{\theta(y;x',x',x',x',c),\mathbb{P}(Y \prec y^u|X=x^e,C=c),\nonumber\\
%&\theta(y;x',x,x',x',c)\big\}-\max\big\{\theta(y;x,x,x,x,c),\nonumber\\
%&\mathbb{P}(Y \prec y^l|X=x^e,C=c),\theta(y;x',x,x',x,c)\big\},\\
%&\gamma^4=\min\big\{\theta(y;x',x',x',x',c),\mathbb{P}(Y \prec y^u|X=x^e,C=c)\big\}\nonumber\\
%&-\max\big\{\theta(y;x,x,x,x,c),\theta(y;x',x,x',x,c),\nonumber\\
%&\mathbb{P}(Y \prec y^l|X=x^e,C=c),\theta(y;x',x,x',x',c)\big\},\\
&\delta=\mathbb{P}(Y \prec y^u|X=x^e,C=c)-\mathbb{P}(Y \prec y^l|X=x^e,C=c).
\end{align}

{\normalfont (1B).} If $\mathbb{P}(Y \prec y^l|X=x^e,C=c) = \mathbb{P}(Y \prec y^u|X=x^e,C=c)$, then
\begin{align}
&\text{\normalfont PNS}^{X \rightarrow Y}(y;x',x,{\cal E},c)=\mathbb{I}\big(\theta(y;x',x',x',x',c)\nonumber\\
& \leq \mathbb{P}(Y \prec y^u|X=x^e,C=c)< \theta(y;x,x,x,x,c),\nonumber\\
& \theta(y;x',x,x',x,c) \leq \mathbb{P}(Y \prec y^u|X=x^e,C=c),\nonumber\\
& \theta(y;x',x,x,x,c) \leq \mathbb{P}(Y \prec y^u|X=x^e,C=c)\big),\\
&\text{\normalfont PNS}^{X \rightarrow {N} \rightarrow  Y}(y;x',x,{\cal E},c)=\mathbb{I}\big(\theta(y;x',x',x',x',c)\nonumber\\
& \leq \mathbb{P}(Y \prec y^u|X=x^e,C=c)< \theta(y;x,x,x,x,c),\nonumber\\
& \theta(y;x',x,x',x,c) \leq \mathbb{P}(Y \prec y^u|X=x^e,C=c),\nonumber\\
&\mathbb{P}(Y \prec y^u|X=x^e,C=c)<\theta(y;x',x,x,x,c) \big).
%&\text{\normalfont PNS}^{X \rightarrow {M} \rightarrow {N} \rightarrow  Y}(y;x',x,{\cal E},c)\nonumber\\
%&=\mathbb{I}\big(\theta(y;x',x',x',x',c) \leq \mathbb{P}(Y \prec y^u|X=x^e,C=c) \nonumber\\
%&\hspace{4cm}< \theta(y;x,x,x,x,c),\nonumber\\
%&\mathbb{P}(Y \prec y^u|X=x^e,C=c)<\theta(y;x',x,x',x,c),\nonumber\\
%&\theta(y;x',x,x',x',c)\leq \mathbb{P}(Y \prec y^u|X=x^e,C=c) \big),\\
%&\text{\normalfont PNS}^{X \rightarrow {M}  \rightarrow  Y}(y;x',x,{\cal E},c)\nonumber\\
%&=\mathbb{I}\big(\theta(y;x',x',x',x',c) \leq \mathbb{P}(Y \prec y^u|X=x^e,C=c) \nonumber\\
%&\hspace{4cm}< \theta(y;x,x,x,x,c),\nonumber\\
%&\mathbb{P}(Y \prec y^u|X=x^e,C=c)<\theta(y;x',x,x',x,c),\nonumber\\
%&\mathbb{P}(Y \prec y^u|X=x^e,C=c)<\theta(y;x',x,x',x',c)\big).
\end{align}

{\normalfont (2).} Under SCM  ${\cal M}_2$ and Assumptions \ref{SCAS}, \ref{ASM}, \ref{SUP1} and \ref{AS1}, for any $x',x \in \Omega_X$, $y \in \Omega_Y$, and $c \in \Omega_C$, we have

{\normalfont (2A).} If $\mathbb{P}(Y \prec y^l|X=x^e,C=c) \ne \mathbb{P}(Y \prec y^u|X=x^e,C=c)$, then
\begin{align}
%&\text{\normalfont PNS}^{X \rightarrow Y}(y;x',x,{\cal E},c)=\max\{\gamma^1/\delta,0\},\\
%&\text{\normalfont PNS}^{X \rightarrow {N} \rightarrow Y}(y;x',x,{\cal E},c)=\max\{\gamma^2/\delta,0\},\\
&\text{\normalfont PNS}^{X \rightarrow {M} \rightarrow {N} \rightarrow  Y}(y;x',x,{\cal E},c)=\max\{\gamma^3/\delta,0\},\\
&\text{\normalfont PNS}^{X \rightarrow {M} \rightarrow  Y}(y;x',x,{\cal E},c)=\max\{\gamma^4/\delta,0\},
\end{align}
where
\begin{align}
%&\gamma^1=\min\big\{\theta(y;x',x',x',x',c),\theta(y;x',x,x',x,c),\nonumber\\
%&\theta(y;x',x,x,x,c),\mathbb{P}(Y \prec y^u|X=x^e,C=c)\big\}\\
%&-\max\big\{\theta(y;x,x,x,x,c),\mathbb{P}(Y \prec y^l|X=x^e,C=c)\big\},\\
%&\gamma^2=\min\big\{\theta(y;x',x',x',x',c),\theta(y;x',x,x',x,c),\nonumber\\
%&\mathbb{P}(Y \prec y^u|X=x^e,C=c)\big\}-\max\big\{\theta(y;x,x,x,x,c),\nonumber\\
%&\mathbb{P}(Y \prec y^l|X=x^e,C=c),\theta(y;x',x,x,x,c)\big\},\\
&\gamma^3=\min\big\{\theta(y;x',x',x',x',c),\mathbb{P}(Y \prec y^u|X=x^e,C=c),\nonumber\\
&\theta(y;x',x,x',x',c)\big\}-\max\big\{\theta(y;x,x,x,x,c),\nonumber\\
&\mathbb{P}(Y \prec y^l|X=x^e,C=c),\theta(y;x',x,x',x,c)\big\},\\
&\gamma^4=\min\big\{\theta(y;x',x',x',x',c),\mathbb{P}(Y \prec y^u|X=x^e,C=c)\big\}\nonumber\\
&-\max\big\{\theta(y;x,x,x,x,c),\theta(y;x',x,x',x,c),\nonumber\\
&\mathbb{P}(Y \prec y^l|X=x^e,C=c),\theta(y;x',x,x',x',c)\big\}.
%&\delta=\mathbb{P}(Y \prec y^u|X=x^e,C=c)-\mathbb{P}(Y \prec y^l|X=x^e,C=c).
\end{align}

{\normalfont (2B).} If $\mathbb{P}(Y \prec y^l|X=x^e,C=c) = \mathbb{P}(Y \prec y^u|X=x^e,C=c)$, then
\begin{align}
%&\text{\normalfont PNS}^{X \rightarrow Y}(y;x',x,{\cal E},c)\nonumber\\
%&=\mathbb{I}\big(\theta(y;x',x',x',x',c) \leq \mathbb{P}(Y \prec y^u|X=x^e,C=c) \nonumber\\
%&\hspace{4cm}< \theta(y;x,x,x,x,c),\nonumber\\
%& \theta(y;x',x,x',x,c) \leq \mathbb{P}(Y \prec y^u|X=x^e,C=c),\nonumber\\
%& \theta(y;x',x,x,x,c) \leq \mathbb{P}(Y \prec y^u|X=x^e,C=c)\big),\\
%&\text{\normalfont PNS}^{X \rightarrow {N} \rightarrow  Y}(y;x',x,{\cal E},c)\nonumber\\
%&=\mathbb{I}\big(\theta(y;x',x',x',x',c) \leq \mathbb{P}(Y \prec y^u|X=x^e,C=c)\nonumber \\
%&\hspace{4cm}< \theta(y;x,x,x,x,c),\nonumber\\
%& \theta(y;x',x,x',x,c) \leq \mathbb{P}(Y \prec y^u|X=x^e,C=c),\nonumber\\
%&\mathbb{P}(Y \prec y^u|X=x^e,C=c)<\theta(y;x',x,x,x,c) \big),\\
&\text{\normalfont PNS}^{X \rightarrow {M} \rightarrow {N} \rightarrow  Y}(y;x',x,{\cal E},c)=\mathbb{I}\big(\theta(y;x',x',x',x',c)\nonumber\\
& \leq \mathbb{P}(Y \prec y^u|X=x^e,C=c) < \theta(y;x,x,x,x,c),\nonumber\\
&\mathbb{P}(Y \prec y^u|X=x^e,C=c)<\theta(y;x',x,x',x,c),\nonumber\\
&\theta(y;x',x,x',x',c)\leq \mathbb{P}(Y \prec y^u|X=x^e,C=c) \big),\\
&\text{\normalfont PNS}^{X \rightarrow {M}  \rightarrow  Y}(y;x',x,{\cal E},c)=\mathbb{I}\big(\theta(y;x',x',x',x',c)\nonumber\\
& \leq \mathbb{P}(Y \prec y^u|X=x^e,C=c)< \theta(y;x,x,x,x,c),\nonumber\\
&\mathbb{P}(Y \prec y^u|X=x^e,C=c)<\theta(y;x',x,x',x,c),\nonumber\\
&\mathbb{P}(Y \prec y^u|X=x^e,C=c)<\theta(y;x',x,x',x',c)\big).
\end{align}
\end{theorem}

Only the identifications of $\text{\normalfont PNS}^{X \rightarrow {M} \rightarrow {N} \rightarrow  Y}$ and $\text{\normalfont PNS}^{X \rightarrow {M}  \rightarrow  Y}$ require Assumption \ref{ASM} since all counterfactuals in $\text{\normalfont PNS}^{X \rightarrow {N} \rightarrow Y}$ and $\text{\normalfont PNS}^{X \rightarrow Y}$ satisfy $x'=x'''$, which corresponds to a special case in \citet{Daniel2015}.

%Denoting the combination of the causal paths $\text{\normalfont PNS}^{X \rightarrow {M} \rightsquigarrow Y}=\text{\normalfont PNS}^{X \rightarrow {M} \rightarrow {N} \rightarrow  Y}+\text{\normalfont PNS}^{X \rightarrow {M} \rightarrow  Y}$, we can decompose and identify T-PNS into three parts, i.e., $\text{\normalfont T-PNS}=\text{\normalfont PNS}^{X \rightarrow Y}+\text{\normalfont PNS}^{X \rightarrow {N} \rightarrow  Y}+\text{\normalfont PNS}^{X \rightarrow {M} \rightsquigarrow  Y}$, without Assumption \ref{ASM}.
%We present an identification theorem for the general definition in Appendix \ref{appB2}.

{\bf Remark.} 
When ${\cal I}_Y$ is a closed interval $[y^l,y^u]$ in evidence ${\cal E}$, the identification results are obtained by replacing “$Y \prec y^u$” with “$Y \preceq y^u$” in Theorem \ref{theo42}.
By setting $y^u=\infty$ and $y^l=-\infty$, the identification theorem for the path-specific PNS can be derived for the entire population in the absence of evidence (${\cal E}=\emptyset$).

\section{Numerical Experiments}

We conduct numerical experiments to illustrate the finite-sample properties of estimators for the path-specific PNS.

\begin{table*}[!tb]
\centering
\caption{Results of numerical experiments}
\label{tab:f}
\vspace{-0.2cm}
\scalebox{1}{
\begin{tabular}{c|cccc}
\hline
Estimators & $N=20$ & $N=100$ & $N=10000$ &  Ground Truth \\
\hline
\hline
$\text{\normalfont T-PNS}$  & $0.447$ ($[0.286,0.625]$) & $0.448$ ($[0.385,0.519]$) & $0.449$ ($[0.443,0.455]$) &$0.449$ \\
$\text{\normalfont ND-PNS}^{{M}}$  & $0.153$ ($[0.040,0.343]$) & $0.155$ ($[0.103,0.214]$) & $0.156$ ($[0.150,0.161]$) &$0.156$ \\
$\text{\normalfont NI-PNS}^{{M}}$  & $0.296$ ($[0.154,0.443]$) & $0.293$ ($[0.231,0.355]$) & $0.293$ ($[0.282,0.299]$) &$0.293$ \\
$\text{\normalfont PNS}^{X \rightarrow Y}$  & $0.057$ ($[0.003,0.153]$) & $0.059$ ($[0.032,0.093]$) & $0.059$ ($[0.056,0.062]$) &$0.059$ \\
$\text{\normalfont PNS}^{X \rightarrow {N}  \rightarrow Y}$   & $0.095$ ($[0.015,0.251]$) &$0.097$ ($[0.059,0.144]$) &  $0.097$ ($[0.093,0.101]$) &$0.097$ \\
$\text{\normalfont PNS}^{X \rightarrow {M} \rightarrow {N}  \rightarrow Y}$   & $0.133$ ($[0.037,0.287]$) &$0.134$ ($[0.096,0.180]$)  &  $0.134$ ($[0.130,0.139]$) &$0.135$ \\
$\text{\normalfont PNS}^{X \rightarrow {M}  \rightarrow Y}$   & $0.163$ ($[0.031,0.319]$)&$0.160$ ($[0.108,0.215]$)  &  $0.158$ ($[0.153,0.164]$) &$0.158$ \\
\hline
\end{tabular}
}
\end{table*}

{\bf Estimation methods.}
The path-specific PNSs for decomposition under SCM ${\cal M}^{L2}$ are estimable using simple linear regressions.
%$Y:=\alpha_0+\alpha_1 X+\alpha_2 {M}+\alpha_3 {N}+\alpha_4 C+U_Y$, ${N}:=\beta_0+\beta_1 X+\beta_2 {M}+\beta_3 C+U_{{N}}$, ${M}:=\gamma_0+\gamma_1 X+\beta_3 C+U_{{M}}$, where $U_Y\sim {\cal N}(0,\sigma^2_Y)$, $U_{{M}} \sim {\cal N}(0,\sigma^2_{{M}})$, and $U_{{N}} \sim {\cal N}(0,\sigma^2_{{N}})$ are independent normal distributions.
%The counterfactuals $Y_{x,{M}_{x'},{N}_{x'',{M}_{x'''}}}$ follows
$\theta(y;x,x',x'',x''',c)$ in Theorem \ref{theo41} is estimated by $\mathbb{P}(Z<y)$,
%\begin{equation}
%\begin{aligned}
%&\alpha_0+\alpha_1 X+\alpha_2 (\gamma_0+\gamma_1 x'+\beta_3 C+U_{{M}})\\
%&+\alpha_3 (\beta_0+\beta_1 x''+\beta_2 (\gamma_0+\gamma_1 x'''+\beta_3 C+U_{{M}})\\
%&+\beta_3 C+U_{{N}})+\alpha_4 C+U_Y\\
%&\hat{\rho}(y;x,x',x'',x''',c)=\\
%&\mathbb{P}(\alpha_0+\alpha_1 x+\alpha_2 \gamma_0+\alpha_2 \gamma_1 x'+\alpha_2 \beta_3 c+\alpha_3 \beta_0\\
%&+\alpha_3 \beta_1 x'' 
%+\alpha_3 \beta_2 \gamma_0+\alpha_3 \beta_2 \gamma_1 x'''+\alpha_3 \beta_2 \beta_3 c\\
%& + \alpha_3 \beta_2 \beta_3 c+\alpha_4 c+\alpha_2 U_{{M}} +\alpha_3 \beta_2 U_{{M}}\\
%&+\alpha_3 \beta_2 U_{{N}} + U_Y <y),
%\mathbb{P}(Z<y)
%\end{aligned}
%\end{equation}
where $Z \sim {\cal N}(
\hat{\alpha}_0+\hat{\alpha}_1 x+\hat{\alpha}_2 (\hat{\gamma}_0+\hat{\gamma}_1 x'+\hat{\gamma}_2 c)+\hat{\alpha}_3 (\hat{\beta}_0+\hat{\beta}_1 x''+\hat{\beta}_2 (\hat{\gamma}_0+\hat{\gamma}_1 x'''+\hat{\gamma}_2 c)+\hat{\beta}_3 c)+\hat{\alpha}_4 c,
\{(\hat{\alpha}_2+\hat{\alpha}_3 \hat{\beta}_2 )^2 \hat{\sigma}_{{M}}^2+\hat{\alpha}_3^2\hat{\sigma}_{{N}}^2+\hat{\sigma}_Y^2\}^{1/2})$
and
$\{\hat{\alpha}_0,\hat{\alpha}_1,\hat{\alpha}_2,\hat{\alpha}_3,\hat{\beta}_0,\hat{\beta}_1,\hat{\beta}_2,\hat{\gamma}_0,\hat{\gamma}_1,\hat{\sigma}_{Y},\hat{\sigma}_{{M}},\hat{\sigma}_{{N}}\}$ are the estimated parameters of the three linear regressions, $Y\sim X+{M}+{N}$, ${N} \sim X+{M}$, and ${M}\sim X$.

{\bf Setting.}
We consider the following SCM:
\begin{equation}
\begin{gathered}
Y:=X+{M}+ {N}+ C+U_Y,
{N}:=X+ {M}+ C+U_{{N}},\\
{M}:=X+C+U_{{M}}, X:=C+U_X, C:=U_C,
\end{gathered}
\end{equation}
where $U_C\sim {\cal N}(0,1)$, $U_X\sim {\cal N}(0,1)$, $U_Y\sim {\cal N}(0,1)$, $U_{{M}} \sim {\cal N}(0,1)$, $U_{{N}} \sim {\cal N}(0,1)$, which are mutually independent normal distributions.
This SCM satisfies Assumptions \ref{SCAS}, \ref{ASM}, \ref{SUP1}, \ref{AS1}, and 4.3'.
We let $x'=0$, $x=1$, $y=0$, $c=0$, and ${\cal E}=\emptyset$.
We simulate 1000 times with the sample size $N=20$, $N=100$, and $N=10000$.

{\bf Results.}
We present the results of each estimator in Table \ref{tab:f}.
%The ground truth of $\text{\normalfont PNS}^{X \rightarrow Y}$ is $0.059$, with the following estimates:
%\begin{center}
%\textbf{$N=20$}:\, \, \, \, $0.057$ (95\%CI: $[0.003,0.153]$),\\\vspace{0.1cm}
%\textbf{$N=100$}:\, \, \,  $0.059$ (95\%CI: $[0.032,0.093]$),\\\vspace{0.1cm}
%\textbf{$N=10000$}: $0.059$ (95\%CI: $[0.056,0.062]$).
%\end{center}
%The ground truth of $\text{\normalfont PNS}^{X \rightarrow {N}  \rightarrow Y}$ is $0.097$, with the following estimates:
%\begin{center}
%\textbf{$N=20$}:\, \, \, \, $0.095$ (95\%CI: $[0.015,0.251]$),\\\vspace{0.1cm}
%\textbf{$N=100$}:\, \, \,  $0.097$ (95\%CI: $[0.059,0.144]$),\\\vspace{0.1cm}
%\textbf{$N=10000$}: $0.097$ (95\%CI: $[0.093,0.101]$).
%\end{center}
%The ground truth of $\text{\normalfont PNS}^{X \rightarrow {M} \rightarrow {N}  \rightarrow Y}$ is $0.135$, with the following estimates:
%\begin{center}
%\textbf{$N=20$}:\, \, \, \, $0.133$ (95\%CI: $[0.037,0.287]$),\\\vspace{0.1cm}
%\textbf{$N=100$}:\, \, \,  $0.134$ (95\%CI: $[0.096,0.180]$),\\\vspace{0.1cm}
%\textbf{$N=10000$}: $0.134$ (95\%CI: $[0.130,0.139]$).
%\end{center}
%The ground truth of $\text{\normalfont PNS}^{X \rightarrow {M}  \rightarrow Y}$ is $0.158$, with the following estimates:
%\begin{center}
%\textbf{$N=20$}:\, \, \,  \, $0.163$ (95\%CI: $[0.031,0.319]$),\\\vspace{0.1cm}
%\textbf{$N=100$}:\, \, \,  $0.160$ (95\%CI: $[0.108,0.215]$),\\\vspace{0.1cm}
%\textbf{$N=10000$}: $0.158$ (95\%CI: $[0.153,0.164]$).
%\end{center}
All means of the estimates are close to the ground truth. 
However, for small sample sizes, the estimators exhibit large 95 $\%$ CIs, indicating higher variability in estimation.
%All means of the estimators are close to the ground truth. 
%However, estimators for small sample sizes have large 95 $\%$ CIs.

We provide three additional experiments in Appendix \ref{appD1} under the following conditions: (1) no effect between ${M}$ and ${N}$, (2) no effect between $\{{M},{N}\}$ and $Y$, and (3) only effect through $X$$\rightarrow$${M}$$\rightarrow$${N}$$\rightarrow$$Y$.
In the setting (1), $\text{\normalfont PNS}^{X \rightarrow {M} \rightarrow {N}  \rightarrow Y}$ is equal to $0$.
In the setting (2), $\text{\normalfont PNS}^{X \rightarrow {N}  \rightarrow Y}$, $\text{\normalfont PNS}^{X \rightarrow {M} \rightarrow {N}  \rightarrow Y}$, and $\text{\normalfont PNS}^{X \rightarrow {M} \rightarrow Y}$ are all equal to $0$.
In the setting (3), $\text{\normalfont PNS}^{X \rightarrow Y}$, $\text{\normalfont PNS}^{X \rightarrow {N}  \rightarrow Y}$, and $\text{\normalfont PNS}^{X \rightarrow {M} \rightarrow Y}$ are all equal to $0$.
These results offer intuitive decompositions of T-PNS.

{We provide a sensitivity analysis by introducing a non-monotonic term in SCM in \ref{appD2}, i.e., $Y:=X+M+ N+ C+\alpha U_Y+ (1-\alpha) U_Y^4$, where $\alpha \in [0,1]$ controls the degree of the violation of the monotonicity.
The magnitude of bias increases with greater violations of the monotonicity.
We additionally report experimental results using logistic regression for binary outcomes in Appendix \ref{appD3}.
The estimates obtained from logistic regression are reliable when the sample size is large.}

\section{Application to Real-World}

We present an application using a real-world dataset.

{\bf Dataset.}
We use an open dataset from 
%the UC Irvine Machine Learning Repository 
({https://archive.ics.uci.edu/dataset/320/student+performance}) on student performance in mathematics from secondary education in two Portuguese schools.
Secondary education lasts for three years, and students are tested once per year, resulting in a total of three tests.
%This data approaches student achievement in secondary education of two Portuguese schools. 
The dataset includes attributes related to demographics, social factors, school-related features, and student grades. 
%and it was collected by using school reports and questionnaires.
The sample size is 649, with no missing values.
Prior research using this dataset aimed to predict students’ performance based on their attributes \citep{Cortez2008, Helwig2017}.
\citet{Kawakami2024} assess the causal relationship between the students' performance, study time, and extra paid classes via estimating PoC.
%introduced in this paper.
In this paper, we analyze the causal relationship between students’ performance in the final period, study time, and extra paid classes, considering their performance in the first and second periods as mediators.

{\bf Variables.}
We consider the mathematics scores in the first period (${M}$) and the second period (${N}$) as mediators, while the mathematics score in the final period ($Y$) serves as the outcome variable.
They take values from $0$ to $20$, respectively.
%$\{0, 1, \ldots, 20\}$, respectively. 
%We consider these variables as discretized versions of normally distributed variables.
We note that \citet{Kawakami2024} considered all mathematics scores from the first period, the second period, and the final period as the vector of outcome variables.
We consider “study time in a week” ($X^1$) and “extra paid classes within the course subject” ($X^2$) (where yes: $X^2=2$, no: $X^2=1$) as the treatment variables, denoted as $X= (X^1, X^2)$.
%We consider ``\emph{study time in a week}'' ($X^1$) and ``\emph{extra paid classes within the course subject}'' ($X^2$) (yes: $X^2=2$, no: $X^2=1$) as treatment variables $X= (X^1, X^2)$. 
We select “sex,” “failures,” “schoolsup,” “famsup,” and “goout” as the covariates ($C$), following their selection in \citep{Helwig2017} and note that they are also used in \citep{Kawakami2024}.
%We select ``sex'', ``failures'', ``schoolsup'', ``famsup'', and ``goout'' as the covariates ($C$), which were chosen in  \citep{Helwig2017} and used in \citep{Kawakami2024}.
We estimate the path-specific PNS using linear regression models, as described in Section 5.
We conduct 1,000 bootstrap resampling iterations \citep{Efron1979} to examine the distribution of the estimators.

In this dataset, Assumption \ref{AS1}’, for instance, $\mathbb{P}(Y_{x,{M}_{x'},{N}_{x,{M}_{x}}}\ \prec y \preceq Y_{x,{M}_{x'},{N}_{x,{M}_{x'}}}|C=c)=0$ is reasonable.
This is because the scores in the first period, had she studied four hours a week and taken extra classes, appear to be greater than those in the first period had she studied only one hour a week and taken no extra classes. 
Moreover, if her scores in the first had been higher, the scores in the second periods would also have been higher.
Furthermore, if her scores in the second had been higher, the scores in the third periods would also have been higher.
%higher scores in the first period lead to higher scores in the second period, which in turn result in higher scores in the final period.
Assumption \ref{ASM} is also reasonable, as, for example, if $U_{{M}}$ represents a genetic factor influencing mathematics performance, it can exert a strictly monotonic increasing effect on the scores in the first period.
%Assumption \ref{ASM} is also reasonable since, for example, considering $U_{{M}}$ is the genetic factor for math performance, $U_{{M}}$ can have a strictly monotonic increasing influence on the scores in the first period.

{\bf Results.}
We consider the subject whose ID number is 1 and set the values of her covariates as $c_1$.
We define the treatment values as $x'=(1,1)$, $x=(4,2)$, set the outcome threshold to $y=10$, and let the evidence be ${\cal E}=\emptyset$.
The estimates of $\text{\normalfont T-PNS}$,
$\text{\normalfont ND-PNS}^{{M}}$, and $\text{\normalfont NI-PNS}^{{M}}$ 
at $(y;x',x,{\cal E},c_1)$ are $15.259 \%$ $(\text{CI}: [0.000\%,33.022\%])$, $1.032$ $\% (\text{CI}: [0.000\%,7.452\%])$, and $14.226$ $\% (\text{CI}: [0.000\%,2.304\%])$, respectively.
%of $\text{\normalfont T-PNS}$, $\text{\normalfont PNS}^{X \rightarrow Y}$, $\text{\normalfont PNS}^{X \rightarrow {N} \rightarrow Y}$, $\text{\normalfont PNS}^{X \rightarrow {M} \rightarrow {N} \rightarrow Y}$, and $\text{\normalfont PNS}^{X \rightarrow {M} \rightarrow Y}$ given $C=c_1$ are 
%\begin{align}
%&\text{\normalfont T-PNS}: &15.259 \% (\text{CI}: [0.000\%,33.022\%]).\nonumber\\ 
%&\text{\normalfont ND-PNS}^{{M}}: &1.032 \% (\text{CI}: [0.000\%,7.452\%]),\nonumber\\
%&\text{\normalfont NI-PNS}^{{M}}: &14.226 \% (\text{CI}: [0.000\%,29.405\%]).\nonumber
%&\text{\normalfont PNS}^{X \rightarrow Y}: &0.149 \% (\text{CI}: [0.000\%,2.304\%]),\nonumber\\
%&\text{\normalfont PNS}^{X \rightarrow {N} \rightarrow Y}: &0.883 \% (\text{CI}: [0.000\%,6.239\%]),\nonumber\\
%&\text{\normalfont PNS}^{X \rightarrow {M} \rightarrow {N} \rightarrow Y}:  &0.000 \% (\text{CI}: [0.000\%,0.000\%]),\nonumber\\
%&\text{\normalfont PNS}^{X \rightarrow {M} \rightarrow Y}:  &14.226 \% (\text{CI}: [0.000\%,29.405\%]).\nonumber
%\end{align}
The results indicate that studying 4 hours a week and taking extra classes would be necessary and sufficient to achieve a score above 10 in the final period for 15.259 $\%$ of subjects.
Moreover, when ignoring the scores in the second period, studying 4 hours a week and taking extra classes would remain necessary and sufficient at the same level of T-PNS if the influence existed solely through the scores in the first period.
%or through both the scores in the first and second periods.
%However, the results do not account for the path through the score in the second period.

We ask four further causal questions about this dataset, considering the scores in the first and second periods:
\begin{center}
\vspace{-0.2cm}
({\bf Q-a1}'). {\it Would studying 4 hours a week and taking extra classes still be necessary and sufficient to achieve a score above 10 in the final period if the influence through the scores in the first and second periods had not existed?
%, compared to studying 1 hour a week and taking no extra classes?
}\\\vspace{0.2cm}
({\bf Q-a2}'). {\it Would studying 4 hours a week and taking extra classes still be necessary and sufficient to achieve a score above 10 in the final period if the influence occurred only through the scores in the second period, and not through the scores in the first period?
%, compared to studying 1 hour a week and taking no extra classes?
}\\\vspace{0.2cm}
({\bf Q-b1}'). {\it Would studying 4 hours a week and taking extra classes still be necessary and sufficient to achieve a score above 10 in the final period if the influence occurred only through both the scores in the first and second periods?
%, compared to studying 1 hour a week and taking no extra classes?
}\\\vspace{0.2cm}
({\bf Q-b2}'). {\it Would studying 4 hours a week and taking extra classes still be necessary and sufficient to achieve a score above 10 in the final period if the influence occurred only through the scores in the first period, and not through the scores in the second period?
%, compared to studying 1 hour a week and taking no extra classes?
}
\vspace{-0.2cm}
\end{center}
Then, the estimates of $\text{\normalfont PNS}^{X \rightarrow Y}$, $\text{\normalfont PNS}^{X \rightarrow {N} \rightarrow Y}$, $\text{\normalfont PNS}^{X \rightarrow {M} \rightarrow {N} \rightarrow Y}$, and $\text{\normalfont PNS}^{X \rightarrow {M} \rightarrow Y}$ at $(y;x',x,{\cal E},c_1)$ are
%of $\text{\normalfont T-PNS}$, $\text{\normalfont PNS}^{X \rightarrow Y}$, $\text{\normalfont PNS}^{X \rightarrow {N} \rightarrow Y}$, $\text{\normalfont PNS}^{X \rightarrow {M} \rightarrow {N} \rightarrow Y}$, and $\text{\normalfont PNS}^{X \rightarrow {M} \rightarrow Y}$ given $C=c_1$ are 
\vspace{-0.1cm}
\begin{align}
%&\text{\normalfont T-PNS}: &15.259 \% (\text{CI}: [0.000\%,33.022\%]),\nonumber\\ 
%&\text{\normalfont ND-PNS}^{{M}}: &1.032 \% (\text{CI}: [0.000\%,7.452\%]),\nonumber\\
%&\text{\normalfont NI-PNS}^{{M}}: &14.226 \% (\text{CI}: [0.000\%,29.405\%]),\nonumber\\
&\text{\normalfont PNS}^{X \rightarrow Y}:\, \, \, \, \, \, \, \, \, \, \, \, \, \, \, \, \, \,  0.149 \% (\text{CI}: [0.000\%,2.304\%]),\nonumber\\
&\text{\normalfont PNS}^{X \rightarrow {N} \rightarrow Y}:\, \, \, \, \, \, \, \, \, \,  0.883 \% (\text{CI}: [0.000\%,6.239\%]),\nonumber\\
&\text{\normalfont PNS}^{X \rightarrow {M} \rightarrow {N} \rightarrow Y}:  0.000 \% (\text{CI}: [0.000\%,0.000\%]),\nonumber\\
&\text{\normalfont PNS}^{X \rightarrow {M} \rightarrow Y}:\, \, \, \,   14.226 \% (\text{CI}: [0.000\%,29.405\%]).\nonumber
\end{align}
%The results suggest that the answers for (Q-a1'), (Q-a2'), and (Q-b1') could be "no", and the answer for (Q-b2') could be "yes".
The results suggest that the necessity and sufficiency of the treatment are almost entirely attributed to the indirect influence occurring solely through the first mediator, while accounting for the path through the score in the second period $N$.
%Furthermore, studying 4 hours a week and taking extra classes would remain necessary and sufficient at the same level of T-PNS if the influence existed solely through the scores in the first period.

Additionally, we provide the estimates under the evidence condition ${\cal E} = (X=0, 10 \leq Y < 15)$, while maintaining the same settings as in Appendix \ref{appE}. 
%Additionally, we present the estimates, setting the evidence as ${\cal E} = (X=0, 10 \leq Y < 15)$, while keeping the other settings the same as in Appendix \ref{appE}.
Similar results are observed for the subpopulations defined by the specified evidence.

\section{Conclusion}
We study path-specific PNSs with two mediators to answer the causal questions (Q-a1), (Q-a2), (Q-b1), and (Q-b2) by decomposing T-PNS into its path-specific components.
Researchers sometimes analyze causal models with more than three mediators.
We provide the definitions of the path-specific PNS with three mediators in Appendix \ref{appC}.
%\yuta{\st{The definitions of path-specific PNS with more than four mediators, which satisfy decomposition relationships, require complex notation. This will be left for future research.}}
{The definitions of path-specific PNS along an arbitrary path in an arbitrary causal graph that satisfy the decomposition relationships will be left for future research.}

{In the settings where the monotonicity assumption does not hold, we can aim to derive bounds for the path-specific PNS \citep{Tian2000,Li2024}. One approach is to use Fréchet inequalities \citep{Frechet1960}. Deriving bounds for the path-specific PNS will be a future work.}

\section*{Acknowledgements}
The authors thank the anonymous reviewers for their time
and thoughtful comments.

%\newpage
\bibliography{citation}

\newpage
\appendix
\onecolumn

\section*{Appendix to ``Decomposition of Probabilities of Causation with Two Mediators”}

\section{Total orders}
\label{appA}

We show the definition of the total order used in the body of the paper.
The definition of the total order is as follows \citep{Harzheim2005}:
%\jin{This definition could be moved to the Appendix to save space.}
\begin{definition}[Total order]
    A total order on a set $\Omega$ is a relation ``$\preceq $'' on $\Omega$ satisfying the following four conditions for all $a_1, a_2, a_3 \in \Omega$:
    \begin{enumerate}
    \vspace{-0cm}
      \setlength{\parskip}{1pt}
  \setlength{\itemsep}{1pt}
        \item $a_1\preceq a_1$;
        \item if $a_1\preceq a_2$ and $a_2\preceq a_3$ then $a_1\preceq a_3$;
        \item if $a_1\preceq a_2$ and $a_2\preceq a_1$ then $a_1= a_2$;
        \item at least one of $a_1\preceq a_2$ and $a_2\preceq a_1$ holds.
    \end{enumerate}
    \vspace{-0cm}
\end{definition}
\noindent In this case we say that the ordered pair $(\Omega,\preceq)$ is a totally ordered set. 
The inequality $a\preceq  b$ of total order means $a\prec b$ or $a=b$, and the relationship $\lnot (a\preceq  b) \Leftrightarrow a\succ b$ holds for a totally ordered set, where $\lnot$ means the negation.

\section{Marginal Distribution of Counterfactual under Special Cases}
\label{appA3}

{\bf Linear SCM.}
We provide explicit forms of path-specific PNS under linear SCM ${\cal M}^{L2}$, i.e.,
\begin{gather}
Y:=\alpha_0+\alpha_1 X+\alpha_2 {M}+\alpha_3 {N}+\alpha_4 C+U_Y,\nonumber\\
{N}:=\beta_0+\beta_1 X+\beta_2 {M}+\beta_3 C+U_{{N}},
{M}:=\gamma_0+\gamma_1 X+\beta_3 C+U_{{M}}, 
\end{gather}
where $U_C\sim {\cal N}(0,\sigma_C)$, $U_X\sim {\cal N}(0,\sigma_X)$, $U_Y\sim {\cal N}(0,\sigma_Y)$, $U_{{M}} \sim {\cal N}(0,\sigma_{{M}})$, $U_{{N}} \sim {\cal N}(0,\sigma_{{N}})$, and they are mutually independent normal distributions.
${\cal N}(\mu,\sigma)$ means a normal distribution whose mean is $\mu$ and standard deviation is $\sigma$.
The potential outcome $Y_{x,{M}_{x'},{N}_{x'',{M}_{x'''}}}$ given $C=c$ is explicitly expressed as:
\begin{align}
Y_{x,{M}_{x'},{N}_{x'',{M}_{x'''}}}
&=\alpha_0+\alpha_1 x+\alpha_2 (\gamma_0+\gamma_1 x'+\beta_3 c+U_{{M}})\nonumber\\
&\hspace{1cm}+\alpha_3 (\beta_0+\beta_1 x''+\beta_2 (\gamma_0+\gamma_1 x'''+\beta_3 c+U_{{M}})+\beta_3 c+U_{{N}})+\alpha_4 c+U_Y.
\end{align}
Then, using a Gaussian distribution $Z \sim {\cal N}(0,1)$, the counterfactual $Y_{x,{M}_{x'},{N}_{x'',{M}_{x'''}}}$ given $C=c$ is expressed as 
\begin{align}
&(\alpha_0+\alpha_1 x+\alpha_2 (\gamma_0+\gamma_1 x'+\gamma_2 c)+\alpha_3 (\beta_0+\beta_1 x''+\beta_2 (\gamma_0+\gamma_1 x'''+\gamma_2 c)+\beta_3 c)+\alpha_4 c)\nonumber\\
&\hspace{3cm}+\{(\alpha_2+\alpha_3 \beta_2 )^2 \sigma_{{M}}^2+\alpha_3^2\sigma_{{N}}^2+\sigma_Y^2\}^{1/2} Z.
\end{align}
Then, we have
\begin{align}
&\mathbb{P}(Y_{x,{M}_{x'},{N}_{x'',{M}_{x'''}}}<y|C=c)\nonumber\\
&=\mathbb{P}((\alpha_0+\alpha_1 x+\alpha_2 (\gamma_0+\gamma_1 x'+\gamma_2 c)+\alpha_3 (\beta_0+\beta_1 x''+\beta_2 (\gamma_0+\gamma_1 x'''+\gamma_2 c)+\beta_3 c)+\alpha_4 c)\nonumber\\
&\hspace{3cm}+\{(\alpha_2+\alpha_3 \beta_2 )^2 \sigma_{{M}}^2+\alpha_3^2\sigma_{{N}}^2+\sigma_Y^2\}^{1/2} Z<y).
\end{align}
%We denote the transformation of $(\alpha_0+\alpha_1 x+\alpha_2 (\gamma_0+\gamma_1 x'+\gamma_2 c)+\alpha_3 (\beta_0+\beta_1 x''+\beta_2 (\gamma_0+\gamma_1 x'''+\gamma_2 c)+\beta_3 c)+\alpha_4 c)+\{(\alpha_2+\alpha_3 \beta_2 )^2 \sigma_{{M}}^2+\alpha_3^2\sigma_{{N}}^2+\sigma_Y^2\}^{1/2} Z$ as $f(Z;x,x',x'',x''')$.
The parameters $\{\alpha_0,\alpha_1,\alpha_2,\alpha_3,\alpha_4,\beta_0,\beta_1,\beta_2,\beta_3,\gamma_0,\gamma_1,\sigma_{{M}},\sigma_{{N}},\sigma_{{Y}}\}$ are identical through three regressions $M \sim X+C$, $N \sim X+M+C$, and $Y \sim X+M+N+C$.
Since SCM ${\cal M}^{L2}$ satisfies \ref{SCAS}, \ref{ASM}, \ref{SUP1} and \ref{AS1}, the path-specific PNS are identified via Theorem \ref{theo42}, based on the marginal distribution of $Y_{x,{M}_{x'},{N}_{x'',{M}_{x'''}}}$.

%Then, the path-specific PNS (with no evidence) is expressed using a single Gaussian distribution $Z$ as follows:
%\begin{align}
%&\text{\normalfont PNS}^{X \rightarrow Y}(y;x',x,\emptyset,c)\nonumber\\
%&=\mathbb{P}(f(Z;x',x',x',x') < y \leq f(Z;x,x,x,x), f(Z;x',x,x',x) < y,f(Z;x',x,x,x) < y),\\
%&\text{\normalfont PNS}^{X \rightarrow {N} \rightarrow  Y}(y;x',x,\emptyset,c)\nonumber\\
%&=\mathbb{P}(f(Z;x',x',x',x') < y \leq f(Z;x,x,x,x), f(Z;x',x,x',x) < y, y \leq f(Z;x',x,x,x)),\\
%&\text{\normalfont PNS}^{X \rightarrow {M} \rightarrow {N} \rightarrow  Y}(y;x',x,\emptyset,c)\nonumber\\
%&=\mathbb{P}(f(Z;x',x',x',x') < y \leq f(Z;x,x,x,x),y \leq f(Z;x',x,x',x),f(Z;x',x,x',x')< y),\\
%&\text{\normalfont PNS}^{X \rightarrow {M}  \rightarrow  Y}(y;x',x,\emptyset,c)\nonumber\\
%&=\mathbb{P}(f(Z;x',x',x',x') < y \leq f(Z;x,x,x,x),y \leq f(Z;x',x,x',x),y \leq f(Z;x',x,x',x')).
%\end{align}

{\bf Binary Outcome Case with Logistic Model.}
When the outcome is binary and modeled using logistic regression by
\begin{gather}
\mathbb{P}(Y=1)=\frac{1}{1+\exp(-(\alpha_0+\alpha_1 X+\alpha_2 {M}+\alpha_3 {N}+\alpha_4 C))},\\
{N}:=\beta_0+\beta_1 X+\beta_2 {M}+\beta_3 C+U_{{N}},
{M}:=\gamma_0+\gamma_1 X+\beta_3 C+U_{{M}}, 
\end{gather}
then, the distribution of potential outcomes $Y_{x,{M}_{x'},{N}_{x'',{M}_{x'''}}}$ given $C=c$, $U^M=u^M$ and $U^N=u^N$ is explicitly expressed as:
\begin{align}
&\mathbb{P}(Y_{x,{M}_{x'},{N}_{x'',{M}_{x'''}}}=1|C=c,U^M=u^M,U^N=u^N)\\
&=\Big\{1+\exp\Big(-(\alpha_0+\alpha_1 x+\alpha_2 (\gamma_0+\gamma_1 x'+\beta_3 c+u_{{M}})\\
&\hspace{2cm}+\alpha_3 (\beta_0+\beta_1 x''+\beta_2 (\gamma_0+\gamma_1 x'''+\beta_3 c+u_{{M}})+\beta_3 c+u_{{N}})+\alpha_4 c)\Big)\Big\}^{-1}.
\end{align}
The parameters $\{\alpha_0,\alpha_1,\alpha_2,\alpha_3,\alpha_4,\beta_0,\beta_1,\beta_2,\beta_3,\gamma_0,\gamma_1\}$ are identical through regressions $M \sim X+C$ and $N \sim M+X+C$.
Since $U_N$ and $U_M$ are identical as residuals of two regressions $M \sim X+C$ and  $N \sim M+X+C$, $\mathbb{P}(Y_{x,{M}_{x'},{N}_{x'',{M}_{x'''}}}=1|C=c)$ is given by $\mathbb{E}_{(U^M,U^N)}[\mathbb{P}(Y_{x,{M}_{x'},{N}_{x'',{M}_{x'''}}}=1|C=c,U^M,U^N)]$.
Since this SCM satisfies \ref{SCAS}, \ref{ASM}, \ref{SUP1} and \ref{AS1}, the path-specific PNS are identified via Theorem \ref{theo42}, based on the marginal distribution of $Y_{x,{M}_{x'},{N}_{x'',{M}_{x'''}}}$.

\section{Pathway Representations}
\label{appA2}

We provide pathway representations of $\text{\normalfont PNS}^{X \rightarrow Y}(y;x',x,{\cal E},c)$, $\text{\normalfont PNS}^{X \rightarrow {N} \rightarrow  Y}(y;x',x,{\cal E},c)$, $\text{\normalfont PNS}^{X \rightarrow {M} \rightarrow {N} \rightarrow  Y}(y;x',x,{\cal E},c)$, and $\text{\normalfont PNS}^{X \rightarrow {M}  \rightarrow  Y}(y;x',x,{\cal E},c)$ in Figures \ref{DAGA1}, \ref{DAGA2}, \ref{DAGA3}, and \ref{DAGA4}, respectively.

%\begin{align}
%&\text{\normalfont PNS}^{X \rightarrow Y}(y;x',x,{\cal E},c)\defeq\mathbb{P}(Y_{x'} \prec y \preceq Y_{x}, Y_{x',{M}_{x}} \prec y,\nonumber\\
%&\hspace{3cm} Y_{x',{M}_{x},{N}_{x,{M}_{x}}} \prec y|{\cal E},C=c),\\
%&\text{\normalfont PNS}^{X \rightarrow {N} \rightarrow  Y}(y;x',x,{\cal E},c)\defeq\nonumber\\
%&\hspace{1cm}\mathbb{P}(Y_{x'} \prec y \preceq Y_{x}, Y_{x',{M}_{x}} \prec y,\nonumber\\
%&\hspace{3cm} y \preceq  Y_{x',{M}_{x},{N}_{x,{M}_{x}}}|{\cal E},C=c),\\
%&\text{\normalfont PNS}^{X \rightarrow {M} \rightarrow {N} \rightarrow  Y}(y;x',x,{\cal E},c)\defeq\nonumber\\
%&\hspace{1cm}\mathbb{P}(Y_{x'} \prec y \preceq Y_{x},y \preceq Y_{x',{M}_{x}},\nonumber\\
%&\hspace{2.5cm}Y_{x',{M}_{x},{N}_{x',{M}_{x'}}} \prec y|{\cal E},C=c),\\
%&\text{\normalfont PNS}^{X \rightarrow {M}  \rightarrow  Y}(y;x',x,{\cal E},c)\defeq\nonumber\\
%&\hspace{1cm}\mathbb{P}(Y_{x'} \prec y \preceq Y_{x},y \preceq Y_{x',{M}_{x}},\nonumber\\
%&\hspace{2.5cm}y \preceq Y_{x',{M}_{x},{N}_{x',{M}_{x'}}}|{\cal E},C=c),
%\end{align}

\begin{figure}[H]
%\vspace{-0.5cm}
   % \hspace{0.3cm}
    \centering
    \scalebox{1}{
\begin{tikzpicture}
    % x node set with absolute coordinates
    \node[mynode] (x) at (0,0) {$X$};
    \node[mynode] (y) at (4,0) {$Y$};
    \node[mynode] (u) at (2,1.25) {$C$};
    \node[mynode] (m) at (1,-1.5) {${M}$};
    
    \node[mynode] (m2) at (3,-1.5) {${N}$};

    % Directed edge
    \path (x) edge[->, line width =0.7mm] (y);
    %\path (x) edge[dotted,<->,bend right] (y);
%    \path (z) edge[->] (x);
    \path (u) edge[->] (y);
%    \path (u) edge[dotted,<->,bend left] (y);
    \path (u) edge[->]  (x);
%\path (x) edge[dotted,<->,bend left] (u);

\path (x) edge[->] (m);
\path (m) edge[->] (y);
\path (u) edge[->] (m);

\path (x) edge[->] (m2);
\path (m) edge[->] (m2);
\path (u) edge[->] (m2);
\path (m2) edge[->] (y);
\end{tikzpicture}
}
\vspace{-0cm}
    \caption{Pathway representation of $\text{\normalfont PNS}^{X \rightarrow Y}(y;x',x,{\cal E},c)$.}
    \label{DAGA1}
    \end{figure}

\begin{figure}[H]
%\vspace{-0.5cm}
   % \hspace{0.3cm}
    \centering
    \scalebox{1}{
\begin{tikzpicture}
    % x node set with absolute coordinates
    \node[mynode] (x) at (0,0) {$X$};
    \node[mynode] (y) at (4,0) {$Y$};
    \node[mynode] (u) at (2,1.25) {$C$};
    \node[mynode] (m) at (1,-1.5) {${M}$};
    
    \node[mynode] (m2) at (3,-1.5) {${N}$};

    % Directed edge
    \path (x) edge[->] (y);
    %\path (x) edge[dotted,<->,bend right] (y);
%    \path (z) edge[->] (x);
    \path (u) edge[->] (y);
%    \path (u) edge[dotted,<->,bend left] (y);
    \path (u) edge[->]  (x);
%\path (x) edge[dotted,<->,bend left] (u);

\path (x) edge[->] (m);
\path (m) edge[->] (y);
\path (u) edge[->] (m);

\path (x) edge[->, line width =0.7mm] (m2);
\path (m) edge[->] (m2);
\path (u) edge[->] (m2);
\path (m2) edge[->, line width =0.7mm] (y);
\end{tikzpicture}
}
\vspace{-0cm}
    \caption{Pathway representation of $\text{\normalfont PNS}^{X \rightarrow {N} \rightarrow  Y}(y;x',x,{\cal E},c)$.}
    \label{DAGA2}
    \end{figure}

\begin{figure}[H]
%\vspace{-0.5cm}
   % \hspace{0.3cm}
    \centering
    \scalebox{1}{
\begin{tikzpicture}
    % x node set with absolute coordinates
    \node[mynode] (x) at (0,0) {$X$};
    \node[mynode] (y) at (4,0) {$Y$};
    \node[mynode] (u) at (2,1.25) {$C$};
    \node[mynode] (m) at (1,-1.5) {${M}$};
    
    \node[mynode] (m2) at (3,-1.5) {${N}$};

    % Directed edge
    \path (x) edge[->] (y);
    %\path (x) edge[dotted,<->,bend right] (y);
%    \path (z) edge[->] (x);
    \path (u) edge[->] (y);
%    \path (u) edge[dotted,<->,bend left] (y);
    \path (u) edge[->]  (x);
%\path (x) edge[dotted,<->,bend left] (u);

\path (x) edge[->, line width =0.7mm] (m);
\path (m) edge[->] (y);
\path (u) edge[->] (m);

\path (x) edge[->] (m2);
\path (m) edge[->, line width =0.7mm] (m2);
\path (u) edge[->] (m2);
\path (m2) edge[->, line width =0.7mm] (y);
\end{tikzpicture}
}
\vspace{-0cm}
    \caption{Pathway representation of $\text{\normalfont PNS}^{X \rightarrow {M} \rightarrow {N} \rightarrow  Y}(y;x',x,{\cal E},c)$.}
    \label{DAGA3}
    \end{figure}
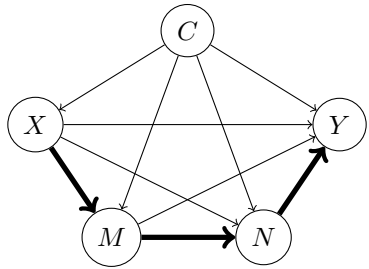

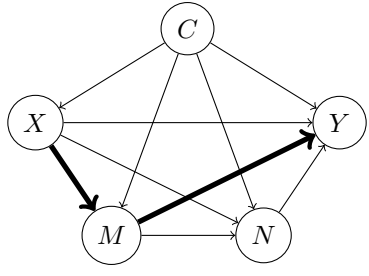
\begin{figure}[H]
%\vspace{-0.5cm}
   % \hspace{0.3cm}
    \centering
    \scalebox{1}{
\begin{tikzpicture}
    % x node set with absolute coordinates
    \node[mynode] (x) at (0,0) {$X$};
    \node[mynode] (y) at (4,0) {$Y$};
    \node[mynode] (u) at (2,1.25) {$C$};
    \node[mynode] (m) at (1,-1.5) {${M}$};
    
    \node[mynode] (m2) at (3,-1.5) {${N}$};

    % Directed edge
    \path (x) edge[->] (y);
    %\path (x) edge[dotted,<->,bend right] (y);
%    \path (z) edge[->] (x);
    \path (u) edge[->] (y);
%    \path (u) edge[dotted,<->,bend left] (y);
    \path (u) edge[->]  (x);
%\path (x) edge[dotted,<->,bend left] (u);

\path (x) edge[->, line width =0.7mm] (m);
\path (m) edge[->, line width =0.7mm] (y);
\path (u) edge[->] (m);

\path (x) edge[->] (m2);
\path (m) edge[->] (m2);
\path (u) edge[->] (m2);
\path (m2) edge[->] (y);
\end{tikzpicture}
}
\vspace{-0cm}
    \caption{Pathway representation of $\text{\normalfont PNS}^{X \rightarrow {M}  \rightarrow  Y}(y;x',x,{\cal E},c)$.}
    \label{DAGA4}
    \end{figure}

\section{Proofs}
\label{appB}

In this section, we provide the proofs of propositions and theorems in the body of the paper.

{\bf Proposition \ref{prop31}.}
{\it
T-PNS 
%, ND-PNS w.r.t. ${M}$, and NI-PNS w.r.t. ${M}$ 
can be decomposed as follows:
\begin{align}
&\text{\normalfont T-PNS}(y;x',x,{\cal E},c)=\text{\normalfont PNS}^{X \rightarrow Y}(y;x',x,{\cal E},c)+\text{\normalfont PNS}^{X \rightarrow {N} \rightarrow  Y}(y;x',x,{\cal E},c)\nonumber\\
&\hspace{4cm}+\text{\normalfont PNS}^{X \rightarrow {M} \rightarrow {N} \rightarrow  Y}(y;x',x,{\cal E},c)+\text{\normalfont PNS}^{X \rightarrow {M}  \rightarrow  Y}(y;x',x,{\cal E},c).
%&\text{\normalfont ND-PNS}^{{M}}(y;x',x,{\cal E},c)\nonumber\\
%&=\text{\normalfont PNS}^{X \rightarrow Y}(y;x',x,{\cal E},c)+\text{\normalfont PNS}^{X \rightarrow {N} \rightarrow  Y}(y;x',x,{\cal E},c),\\
%&\text{\normalfont NI-PNS}^{{M}}(y;x',x,{\cal E},c)=\text{\normalfont PNS}^{X \rightarrow {M} \rightarrow {N} \rightarrow  Y}(y;x',x,{\cal E},c)\nonumber\\
%&+\text{\normalfont PNS}^{X \rightarrow {M}  \rightarrow  Y}(y;x',x,{\cal E},c),
\end{align}
}

\begin{proof}
For any $x',x \in \Omega_X$, $y \in \Omega_Y$, and $c \in \Omega_C$,
T-PNS, ND-PNS, and NI-PNS are decomposed by
\begin{align}
\text{\normalfont T-PNS}(y;x',x,{\cal E},c)
&=\mathbb{P}(Y_{x'} \prec y \preceq Y_{x}|{\cal E},C=c)\nonumber\\
&=\mathbb{P}\Big(Y_{x'} \prec y \preceq Y_{x}, Y_{x',{M}_{x}} \prec y, Y_{x',{M}_{x},{N}_{x,{M}_{x}}} \prec y\Big|{\cal E},C=c\Big)\nonumber\\
&+\mathbb{P}\Big(Y_{x'} \prec y \preceq Y_{x}, Y_{x',{M}_{x}} \prec y, y \preceq  Y_{x',{M}_{x},{N}_{x,{M}_{x}}}\Big|{\cal E},C=c\Big)\nonumber\\
&+\mathbb{P}\Big(Y_{x'} \prec y \preceq Y_{x},y \preceq Y_{x',{M}_{x}},Y_{x',{M}_{x},{N}_{x',{M}_{x'}}} \prec y\Big|{\cal E},C=c\Big)\nonumber\\
&+\mathbb{P}\Big(Y_{x'} \prec y \preceq Y_{x},y \preceq Y_{x',{M}_{x}},y \preceq Y_{x',{M}_{x},{N}_{x',{M}_{x'}}}\Big|{\cal E},C=c\Big)\nonumber\\
&=\text{\normalfont PNS}^{X \rightarrow Y}(y;x',x,{\cal E},c)+\text{\normalfont PNS}^{X \rightarrow {N} \rightarrow  Y}(y;x',x,{\cal E},c)\nonumber\\
&\hspace{1cm}+\text{\normalfont PNS}^{X \rightarrow {M} \rightarrow {N} \rightarrow  Y}(y;x',x,{\cal E},c)+\text{\normalfont PNS}^{X \rightarrow {M}  \rightarrow  Y}(y;x',x,{\cal E},c).
\end{align}
\end{proof}

{\bf Proposition \ref{prop32}.}
{\it 
We have
\begin{align}
%\label{eq9}
%&\text{\normalfont T-PNS}(y;x',x,{\cal E},c)\nonumber\\
%&=\text{\normalfont PNS}^{X \rightarrow Y}(y;x',x,{\cal E},c)+\text{\normalfont PNS}^{X \rightarrow {N} \rightarrow  Y}(y;x',x,{\cal E},c)\nonumber\\
%&+\text{\normalfont PNS}^{X \rightarrow {M} \rightarrow {N} \rightarrow  Y}(y;x',x,{\cal E},c)%\nonumber\\
%&+\text{\normalfont PNS}^{X \rightarrow {M}  \rightarrow  Y}(y;x',x,{\cal E},c),\\
&\mathbb{P}(Y_{x'} \prec y \preceq Y_{x}, Y_{x',{M}_{x}} \prec y|{\cal E},C=c)=\text{\normalfont PNS}^{X \rightarrow Y}(y;x',x,{\cal E},c)+\text{\normalfont PNS}^{X \rightarrow {N} \rightarrow  Y}(y;x',x,{\cal E},c),\\
%\label{eq10}
&\mathbb{P}(Y_{x'} \prec y \preceq Y_{x},y \preceq Y_{x',{M}_{x}}|{\cal E},C=c)=\text{\normalfont PNS}^{X \rightarrow {M} \rightarrow {N} \rightarrow  Y}(y;x',x,{\cal E},c)+\text{\normalfont PNS}^{X \rightarrow {M}  \rightarrow  Y}(y;x',x,{\cal E},c).
\end{align}
}

\begin{proof}
We have
\begin{align}
%\mathbb{P}(Y_{x'} \prec y \preceq Y_{x}, Y_{x',{M}_{x}} \prec y|{\cal E},C=c)
\mathbb{P}(Y_{x'} \prec y \preceq Y_{x}, Y_{x',M_{x}} \prec y|{\cal E},C=c)
&=\mathbb{P}\Big(Y_{x'} \prec y \preceq Y_{x}, Y_{x',{M}_{x}} \prec y, Y_{x',{M}_{x},{N}_{x,{M}_{x}}} \prec y\Big|{\cal E},C=c\Big)\nonumber\\
&+\mathbb{P}\Big(Y_{x'} \prec y \preceq Y_{x}, Y_{x',{M}_{x}} \prec y, y \preceq  Y_{x',{M}_{x},{N}_{x,{M}_{x}}}\Big|{\cal E},C=c\Big)\nonumber\\
&=\text{\normalfont PNS}^{X \rightarrow Y}(y;x',x,{\cal E},c)+\text{\normalfont PNS}^{X \rightarrow {N} \rightarrow  Y}(y;x',x,{\cal E},c),
\end{align}
\begin{align}
%\mathbb{P}(Y_{x'} \prec y \preceq Y_{x},y \preceq Y_{x',{M}_{x}}|{\cal E},C=c)
\mathbb{P}(Y_{x'} \prec y \preceq Y_{x},y \preceq Y_{x',M_{x}}|{\cal E},C=c)
&=\mathbb{P}\Big(Y_{x'} \prec y \preceq Y_{x},y \preceq Y_{x',{M}_{x}},Y_{x',{M}_{x},{N}_{x',{M}_{x'}}} \prec y\Big|{\cal E},C=c\Big)\nonumber\\
&+\mathbb{P}\Big(Y_{x'} \prec y \preceq Y_{x},y \preceq Y_{x',{M}_{x}},y \preceq Y_{x',{M}_{x},{N}_{x',{M}_{x'}}}\Big|{\cal E},C=c\Big)\nonumber\\
&=\text{\normalfont PNS}^{X \rightarrow {M} \rightarrow {N} \rightarrow  Y}(y;x',x,{\cal E},c)+\text{\normalfont PNS}^{X \rightarrow {M}  \rightarrow  Y}(y;x',x,{\cal E},c).
\end{align}
\end{proof}

%{\bf Proposition \ref{prop32}.}
%{\it
%Under SCM ${\cal M}_2$ of no influence from ${M}$ to ${N}$, for each $x',x \in \Omega_X$, $y \in \Omega_Y$, and $c \in \Omega_C$, the natural path-specific PNS become
%\begin{align}
%&\text{\normalfont PNS}^{X \rightarrow Y}(y;x',x,{\cal E},c)=\mathbb{P}\left(Y_{x'} \prec y \preceq Y_{x}, Y_{x',{M}_{x}} \prec y, Y_{x',{M}_{x},{N}_{x}} \prec y\Big|{\cal E},C=c\right),
%\end{align}
%\begin{align}
%&\text{\normalfont PNS}^{X \rightarrow {N} \rightarrow  Y}(y;x',x,{\cal E},c)=\mathbb{P}\left(Y_{x'} \prec y \preceq Y_{x}, Y_{x',{M}_{x}} \prec y,y \preceq  Y_{x',{M}_{x},{N}_{x}}\Big|{\cal E},C=c\right),
%\end{align}
%\begin{align}
%&\text{\normalfont PNS}^{X \rightarrow {M} \rightarrow {N} \rightarrow  Y}(y;x',x,{\cal E},c)=\mathbb{P}\left(Y_{x'} \prec y \preceq Y_{x},y \preceq Y_{x',{M}_{x}},Y_{x',{M}_{x},{N}_{x'}} \prec y\Big|{\cal E},C=c\right)=0,
%\end{align}
%\begin{align}
%&\text{\normalfont PNS}^{X \rightarrow {M}  \rightarrow  Y}(y;x',x,{\cal E},c)=\mathbb{P}\left(Y_{x'} \prec y \preceq Y_{x},y \preceq Y_{x',{M}_{x}},y \preceq Y_{x',{M}_{x},{N}_{x'}}\Big|{\cal E},C=c\right)\\
%&=\mathbb{P}\left(Y_{x'} \prec y \preceq Y_{x},y \preceq Y_{x',{M}_{x}}\Big|{\cal E},C=c\right).
%\end{align}
%}

%\begin{proof}
%This is because we have $Y_{x,{M}_{x'},{N}_{x'',{M}_{x'''}}}=Y_{x,{M}_{x'},{N}_{x''}}$ for any $x,x',x'',x''' \in \Omega_X$ and any subject.
%\end{proof}

{\bf Lemma \ref{lem41}.}
{\it
Under SCM  ${\cal M}_2$ and Assumption \ref{ASM}, we have
\begin{equation}
%\label{eq23}
\begin{aligned}
&\mathfrak{p}_{{M}_{x'''}|C,{M}_{x'}}({{m}}'|c,{m})=\mathbbm{1}(F^{-1}_{{M}|X=x''',C=c}(\mathbb{P}({M} \preceq {m}|X=x',C=c))={{m}}')
\end{aligned}
\end{equation}
where $F^{-1}_{{M}|X=x''',C=c}$ is the inverse function of conditional CDF $\mathbb{P}({M} \preceq {m}|X=x''',C=c)$ on ${m}$ given $X=x'''$ and $C=c$.
}

\begin{proof}
We have
\begin{align}
\mathfrak{p}_{{M}_{x'''}|C,{M}_{x'}}({{m}}'|c,{m})=\frac{d}{d{{{m}}'}}\mathbb{P}({M}_{x'''}\prec{{m}}'|C=c,{M}_{x'} ={m}).
\end{align}
When $f_{{M}}$ is strictly monotonic increasing, we denote $\overline{u}_{{M}_{x'}|C=c}:=\sup\{u_{{M}} \in \Omega_{U_{{M}}};{M}_{x'} \preceq {m}\}$ given $C=c$,
%and $\beta_{x',{m}}=
and we have
$\mathbb{P}(U_{{M}}<\overline{u}_{{M}_{x'}}|C=c)=\mathbb{P}({M}_{x'} \preceq {m}|C=c)=\mathbb{P}({M} \preceq {m}|X=x',C=c)$.
When $f_{{M}}$ is strictly monotonic decreasing, we denote $\overline{u}_{{M}_{x'}|C=c}:=\inf\{u_{{M}_{x'}} \in \Omega_{U_{{M}}};{M}_{x'} \preceq {m}\}$ given $C=c$,
%and $\beta_{x',{m}}=
and we have
$\mathbb{P}(U_{{M}}>\overline{u}_{{M}_{x'}}|C=c)=\mathbb{P}({M}_{x'} \preceq {m}|C=c)=\mathbb{P}({M} \preceq {m}|X=x',C=c)$.
In the both case, we have
\begin{align}
\frac{d}{d{{{m}}'}}\mathbb{P}_{{M}}({M}_{x'''}\prec{{m}}'|C=c,{M}_{x'} ={m})
&=\frac{d}{d{{{m}}'}}\mathbb{P}_{{M}}({M}_{x'''}\prec{{m}}'|C=c,U_{{M}}=\overline{u}_{{M}_{x'}})\nonumber\\
&=\frac{d}{d{{{m}}'}}\mathbb{P}_{{M}}({M}_{x'''}(\overline{u}_{{M}_{x'}})\prec{{m}}'|C=c).
\end{align}
The value of ${M}_{x'''}(\overline{u}_{{M}_{x'}})$, which is a potential outcome ${M}_{x'''}$ for the subject $U_{{M}}=\overline{u}_{{M}_{x'}}$ and is a constant, is equal to $F^{-1}_{{M}|X=x''',C=c}(\mathbb{P}({M} \preceq {m}|X=x',C=c))$ since we have $\mathbb{P}({M}_{x'} \preceq {m}|C=c)=\mathbb{P}({M}_{x'''} \preceq {M}_{x'''}(\overline{u}_{{M}_{x'}})|C=c)$ from the strict monotonicity.
Then, we have
\begin{align}
\frac{d}{d{{{m}}'}}\mathbb{P}_{{M}}({M}_{x'''}\prec{{m}}'|C=c,{M}_{x'} ={m})
&=\frac{d}{d{{{m}}'}}\mathbb{I}(F^{-1}_{{M}|X=x''',C=c}(\mathbb{P}({M} \preceq {m}|X=x',C=c))\prec{{m}}')\nonumber\\
&=\mathbbm{1}(F^{-1}_{{M}|X=x''',C=c}(\mathbb{P}({M} \preceq {m}|X=x',C=c))={{m}}').
\end{align}
\end{proof}

{\bf Theorem \ref{theo41}.}
{\it
Under SCM  ${\cal M}_2$ and Assumption \ref{ASM}, for any $x,x',x'',x''' \in \Omega_X$ and $c \in \Omega_C$, $\mathbb{P}(Y_{x,{M}_{x'},{N}_{x'',{M}_{x'''}}}\prec y|C=c)$ is identifiable by $\theta(y;x,x',x'',x''',c)$, where
\begin{align}
&\theta(y;x,x',x'',x''',c)=\int_{\Omega_{{M}}}\int_{\Omega_{{N}}}\mathbb{P}(Y\prec y|X=x,{M}={m},{N}={n},C=c)\nonumber\\
&\times\mathfrak{p}_{{N}|C,X,{M}}({n}|c,x'',F^{-1}_{{M}|X=x''',C=c}(\mathbb{P}({M} \preceq {m}|X=x',C=c)))\mathfrak{p}_{{M}|C,X}({m}|c,x')\ d{n}d{m}.
\end{align}
}

\begin{proof}
This is because Eq.~\ref{eq3} holds and $\mathfrak{p}_{{M}_{x''}|C,{M}_{x'}}({{m}}'|c,{m})$ is identifiable from Lemma \ref{lem41}.
\end{proof}

{\bf Theorem \ref{theo42}.}
{\it
Let ${\cal I}_Y$ be a half-open interval in evidence ${\cal E}$. We have the following two statements.

(1) Under SCM  ${\cal M}_2$ and Assumptions 
%\ref{ASM},
\ref{SCAS}, \ref{SUP1} and \ref{AS1}, for any $x',x \in \Omega_X$, $y \in \Omega_Y$, and $c \in \Omega_C$, we have

{\normalfont (1A).} If $\mathbb{P}(Y \prec y^l|X=x^e,C=c) \ne \mathbb{P}(Y \prec y^u|X=x^e,C=c)$, then
\begin{align}
&\text{\normalfont PNS}^{X \rightarrow Y}(y;x',x,{\cal E},c)=\max\{\gamma^1/\delta,0\},\text{\normalfont PNS}^{X \rightarrow {N} \rightarrow Y}(y;x',x,{\cal E},c)=\max\{\gamma^2/\delta,0\},
%&\text{\normalfont PNS}^{X \rightarrow {M} \rightarrow {N} \rightarrow  Y}(y;x',x,{\cal E},c)=\max\{\gamma^3/\delta,0\},\\
%&\text{\normalfont PNS}^{X \rightarrow {M} \rightarrow  Y}(y;x',x,{\cal E},c)=\max\{\gamma^4/\delta,0\},
\end{align}
where
\begin{align}
&\gamma^1=\min\Big\{\theta(y;x',x',x',x',c),\theta(y;x',x,x',x,c),\theta(y;x',x,x,x,c),\mathbb{P}(Y \prec y^u|X=x^e,C=c)\Big\}\nonumber\\
&-\max\Big\{\theta(y;x,x,x,x,c),\mathbb{P}(Y \prec y^l|X=x^e,C=c)\Big\},\\
&\gamma^2=\min\Big\{\theta(y;x',x',x',x',c),\theta(y;x',x,x',x,c),\mathbb{P}(Y \prec y^u|X=x^e,C=c)\Big\}\nonumber\\
&-\max\Big\{\theta(y;x,x,x,x,c),\mathbb{P}(Y \prec y^l|X=x^e,C=c),\theta(y;x',x,x,x,c)\Big\},\\
%&\gamma^3=\min\Big\{\theta(y;x',x',x',x',c),\mathbb{P}(Y \prec y^u|X=x^e,C=c),\nonumber\\
%&\theta(y;x',x,x',x',c)\Big\}-\max\Big\{\theta(y;x,x,x,x,c),\nonumber\\
%&\mathbb{P}(Y \prec y^l|X=x^e,C=c),\theta(y;x',x,x',x,c)\Big\},\\
%&\gamma^4=\min\Big\{\theta(y;x',x',x',x',c),\mathbb{P}(Y \prec y^u|X=x^e,C=c)\Big\}\nonumber\\
%&-\max\Big\{\theta(y;x,x,x,x,c),\theta(y;x',x,x',x,c),\nonumber\\
%&\mathbb{P}(Y \prec y^l|X=x^e,C=c),\theta(y;x',x,x',x',c)\Big\},\\
&\delta=\mathbb{P}(Y \prec y^u|X=x^e,C=c)-\mathbb{P}(Y \prec y^l|X=x^e,C=c).
\end{align}

{\normalfont (1B).} If $\mathbb{P}(Y \prec y^l|X=x^e,C=c) = \mathbb{P}(Y \prec y^u|X=x^e,C=c)$, then
\begin{align}
&\text{\normalfont PNS}^{X \rightarrow Y}(y;x',x,{\cal E},c)\mathbb{I}\Big(\theta(y;x',x',x',x',c) \leq \mathbb{P}(Y \prec y^u|X=x^e,C=c)< \theta(y;x,x,x,x,c),\nonumber\\
& \theta(y;x',x,x',x,c) \leq \mathbb{P}(Y \prec y^u|X=x^e,C=c),\theta(y;x',x,x,x,c) \leq \mathbb{P}(Y \prec y^u|X=x^e,C=c)\Big),\nonumber\\
&\text{\normalfont PNS}^{X \rightarrow {N} \rightarrow  Y}(y;x',x,{\cal E},c)=\mathbb{I}\Big(\theta(y;x',x',x',x',c) \leq \mathbb{P}(Y \prec y^u|X=x^e,C=c)< \theta(y;x,x,x,x,c),\nonumber\\
& \theta(y;x',x,x',x,c) \leq \mathbb{P}(Y \prec y^u|X=x^e,C=c),\mathbb{P}(Y \prec y^u|X=x^e,C=c)<\theta(y;x',x,x,x,c) \Big),
%&\text{\normalfont PNS}^{X \rightarrow {M} \rightarrow {N} \rightarrow  Y}(y;x',x,{\cal E},c)\nonumber\\
%&=\mathbb{I}\Big(\theta(y;x',x',x',x',c) \leq \mathbb{P}(Y \prec y^u|X=x^e,C=c) \nonumber\\
%&\hspace{4cm}< \theta(y;x,x,x,x,c),\nonumber\\
%&\mathbb{P}(Y \prec y^u|X=x^e,C=c)<\theta(y;x',x,x',x,c),\nonumber\\
%&\theta(y;x',x,x',x',c)\leq \mathbb{P}(Y \prec y^u|X=x^e,C=c) \Big),\\
%&\text{\normalfont PNS}^{X \rightarrow {M}  \rightarrow  Y}(y;x',x,{\cal E},c)\nonumber\\
%&=\mathbb{I}\Big(\theta(y;x',x',x',x',c) \leq \mathbb{P}(Y \prec y^u|X=x^e,C=c) \nonumber\\
%&\hspace{4cm}< \theta(y;x,x,x,x,c),\nonumber\\
%&\mathbb{P}(Y \prec y^u|X=x^e,C=c)<\theta(y;x',x,x',x,c),\nonumber\\
%&\mathbb{P}(Y \prec y^u|X=x^e,C=c)<\theta(y;x',x,x',x',c)\Big).
\end{align}

(2) Under SCM  ${\cal M}_2$ and Assumptions \ref{SCAS}, \ref{ASM}, \ref{SUP1} and \ref{AS1}, for any $x',x \in \Omega_X$, $y \in \Omega_Y$, and $c \in \Omega_C$, we have

{\normalfont (2A).} If $\mathbb{P}(Y \prec y^l|X=x^e,C=c) \ne \mathbb{P}(Y \prec y^u|X=x^e,C=c)$, then
\begin{align}
%&\text{\normalfont PNS}^{X \rightarrow Y}(y;x',x,{\cal E},c)=\max\{\gamma^1/\delta,0\},\\
%&\text{\normalfont PNS}^{X \rightarrow {N} \rightarrow Y}(y;x',x,{\cal E},c)=\max\{\gamma^2/\delta,0\},\\
&\text{\normalfont PNS}^{X \rightarrow {M} \rightarrow {N} \rightarrow  Y}(y;x',x,{\cal E},c)=\max\{\gamma^3/\delta,0\},\\
&\text{\normalfont PNS}^{X \rightarrow {M} \rightarrow  Y}(y;x',x,{\cal E},c)=\max\{\gamma^4/\delta,0\},
\end{align}
where
\begin{align}
%&\gamma^1=\min\Big\{\theta(y;x',x',x',x',c),\theta(y;x',x,x',x,c),\nonumber\\
%&\theta(y;x',x,x,x,c),\mathbb{P}(Y \prec y^u|X=x^e,C=c)\Big\}\\
%&-\max\Big\{\theta(y;x,x,x,x,c),\mathbb{P}(Y \prec y^l|X=x^e,C=c)\Big\},\\
%&\gamma^2=\min\Big\{\theta(y;x',x',x',x',c),\theta(y;x',x,x',x,c),\nonumber\\
%&\mathbb{P}(Y \prec y^u|X=x^e,C=c)\Big\}-\max\Big\{\theta(y;x,x,x,x,c),\nonumber\\
%&\mathbb{P}(Y \prec y^l|X=x^e,C=c),\theta(y;x',x,x,x,c)\Big\},\\
&\gamma^3=\min\Big\{\theta(y;x',x',x',x',c),\mathbb{P}(Y \prec y^u|X=x^e,C=c),\theta(y;x',x,x',x',c)\Big\}\nonumber\\
&-\max\Big\{\theta(y;x,x,x,x,c),\mathbb{P}(Y \prec y^l|X=x^e,C=c),\theta(y;x',x,x',x,c)\Big\},\\
&\gamma^4=\min\Big\{\theta(y;x',x',x',x',c),\mathbb{P}(Y \prec y^u|X=x^e,C=c)\Big\}\nonumber\\
&-\max\Big\{\theta(y;x,x,x,x,c),\theta(y;x',x,x',x,c),\mathbb{P}(Y \prec y^l|X=x^e,C=c),\theta(y;x',x,x',x',c)\Big\},\\
&\delta=\mathbb{P}(Y \prec y^u|X=x^e,C=c)-\mathbb{P}(Y \prec y^l|X=x^e,C=c).
\end{align}

{\normalfont (2B).} If $\mathbb{P}(Y \prec y^l|X=x^e,C=c) = \mathbb{P}(Y \prec y^u|X=x^e,C=c)$, then
\begin{align}
%&\text{\normalfont PNS}^{X \rightarrow Y}(y;x',x,{\cal E},c)\nonumber\\
%&=\mathbb{I}\Big(\theta(y;x',x',x',x',c) \leq \mathbb{P}(Y \prec y^u|X=x^e,C=c) \nonumber\\
%&\hspace{4cm}< \theta(y;x,x,x,x,c),\nonumber\\
%& \theta(y;x',x,x',x,c) \leq \mathbb{P}(Y \prec y^u|X=x^e,C=c),\nonumber\\
%& \theta(y;x',x,x,x,c) \leq \mathbb{P}(Y \prec y^u|X=x^e,C=c)\Big),\\
%&\text{\normalfont PNS}^{X \rightarrow {N} \rightarrow  Y}(y;x',x,{\cal E},c)\nonumber\\
%&=\mathbb{I}\Big(\theta(y;x',x',x',x',c) \leq \mathbb{P}(Y \prec y^u|X=x^e,C=c)\nonumber \\
%&\hspace{4cm}< \theta(y;x,x,x,x,c),\nonumber\\
%& \theta(y;x',x,x',x,c) \leq \mathbb{P}(Y \prec y^u|X=x^e,C=c),\nonumber\\
%&\mathbb{P}(Y \prec y^u|X=x^e,C=c)<\theta(y;x',x,x,x,c) \Big),\\
&\text{\normalfont PNS}^{X \rightarrow {M} \rightarrow {N} \rightarrow  Y}(y;x',x,{\cal E},c)=\mathbb{I}\Big(\theta(y;x',x',x',x',c) \leq \mathbb{P}(Y \prec y^u|X=x^e,C=c) < \theta(y;x,x,x,x,c),\nonumber\\
&\mathbb{P}(Y \prec y^u|X=x^e,C=c)<\theta(y;x',x,x',x,c),\theta(y;x',x,x',x',c)\leq \mathbb{P}(Y \prec y^u|X=x^e,C=c) \Big),\\
&\text{\normalfont PNS}^{X \rightarrow {M}  \rightarrow  Y}(y;x',x,{\cal E},c)=\mathbb{I}\Big(\theta(y;x',x',x',x',c) \leq \mathbb{P}(Y \prec y^u|X=x^e,C=c)< \theta(y;x,x,x,x,c),\nonumber\\
&\mathbb{P}(Y \prec y^u|X=x^e,C=c)<\theta(y;x',x,x',x,c),\mathbb{P}(Y \prec y^u|X=x^e,C=c)<\theta(y;x',x,x',x',c)\Big).
\end{align}
}

\begin{proof}
    
Let ${\cal I}_Y$ be a half-open interval in the evidence ${\cal E}$.
Under SCM and Assumptions \ref{SCAS}, \ref{SUP1} and \ref{AS1}, for any $x',x \in \Omega_X$, $y \in \Omega_Y$, and $c \in \Omega_C$, we have $\mathbb{P}(Y \prec y^u|X=x^e,C=c)$ and $\mathbb{P}(Y \prec y^l|X=x^e,C=c)$.
Without loss of generality, the function $g(x,x',x'',x''',c,\tilde{U})$ is monotonic increasing on $\tilde{U}$ for all $x,x',x'',x''' \in \Omega_X$ and $c \in \Omega_C$, almost surely w.r.t. $\mathbb{P}_{\tilde{U}}$.
Let $u_{x,y}=\{u \in \Omega_{\tilde{U}};Y_{x}(u) \prec y\}$, $u_{x,x',y}=\{u \in \Omega_{\tilde{U}};Y_{x,{M}_{x'}} \prec y\}$, and $u_{x,x',x'',x'''y}=\{u \in \Omega_{\tilde{U}};Y_{x,{M}_{x'},{N}_{x'',{M}_{x'''}}} \prec y\}$.

(A). If $\mathbb{P}(Y \prec y^l|X=x^e,C=c) \ne \mathbb{P}(Y \prec y^l|X=x^e,C=c)$, then we have
\begin{align}
&\text{\normalfont PNS}^{X \rightarrow Y}(y;x',x,{\cal E},c)\nonumber\\
&=\mathbb{P}\Big(Y_{x'} \prec y \preceq Y_{x}, Y_{x',{M}_{x}} \prec y,Y_{x',{M}_{x},{N}_{x,{M}_{x}}} \prec y\Big|{\cal E},C=c\Big)\nonumber\\
&=\frac{\mathbb{P}\Big(u_{x',y} \leq \tilde{U} < u_{x,y}, u_{x',x,y} \leq \tilde{U}, u_{x',x,x,x,y} \leq \tilde{U},u_{x^e,y^l} \leq \tilde{U} < u_{x^e,y^u}\Big|C=c\Big)}{\mathbb{P}\Big(u_{x^e,y^l} \leq \tilde{U} < u_{x^e,y^u}\Big|C=c\Big)}\nonumber\\
&=\Bigg[\min\Big\{\mathbb{P}\left(Y_{x'} \prec y\Big|C=c\right),\mathbb{P}\left(Y_{x',{M}_{x}} \prec y\Big|C=c\right),\mathbb{P}\left(Y_{x',{M}_{x},{N}_{x,{M}_{x}}} \prec y\Big|C=c\right),\mathbb{P}(Y_{x^e} \prec y^u|C=c)\Big\}\nonumber\\
&-\max\Big\{\mathbb{P}\left(Y_{x} \prec y\Big|C=c\right),\mathbb{P}(Y_{x^e} \prec y^l|C=c)\Big\}\Bigg]\Bigg/ \Big\{\mathbb{P}(Y_{x^e} \prec y^u|C=c)-\mathbb{P}(Y_{x^e} \prec y^l|C=c)\Big\},
\end{align}
\begin{align}
&\text{\normalfont PNS}^{X \rightarrow {N} \rightarrow  Y}(y;x',x,{\cal E},c)\nonumber\\
&=\mathbb{P}\Big(Y_{x'} \prec y \preceq Y_{x}, Y_{x',{M}_{x}} \prec y, y \preceq  Y_{x',{M}_{x},{N}_{x,{M}_{x}}}\Big|{\cal E},C=c\Big)\nonumber\\
&=\frac{\mathbb{P}\Big(u_{x',y} \leq \tilde{U} < u_{x,y}, u_{x',x,y} \leq \tilde{U},\tilde{U}<u_{x',x,x,x,y},u_{x^e,y^l} \leq \tilde{U} < u_{x^e,y^u}\Big|C=c\Big)}{\mathbb{P}\Big(u_{x^e,y^l} \leq \tilde{U} < u_{x^e,y^u}\Big|C=c\Big)}\nonumber\\
&=\Bigg[\min\Big\{\mathbb{P}\left(Y_{x'} \prec y\Big|C=c\right),\mathbb{P}\left(Y_{x',{M}_{x}} \prec y\Big|C=c\right),\mathbb{P}(Y_{x^e} \prec y^u|C=c)\Big\}-\max\Big\{\mathbb{P}\left(Y_{x} \prec y\Big|C=c\right),\nonumber\\
&\mathbb{P}(Y_{x^e} \prec y^l|C=c),\mathbb{P}\left(Y_{x',{M}_{x},{N}_{x,{M}_{x}}} \prec y\Big|C=c\right)\Big\}\Bigg]\Bigg/ \Big\{\mathbb{P}(Y_{x^e} \prec y^u|C=c)-\mathbb{P}(Y_{x^e} \prec y^l|C=c)\Big\},
\end{align}
\begin{align}
&\text{\normalfont PNS}^{X \rightarrow {M} \rightarrow {N} \rightarrow  Y}(y;x',x,{\cal E},c)\nonumber\\
&=\mathbb{P}\Big(Y_{x'} \prec y \preceq Y_{x},y \preceq Y_{x',{M}_{x}},Y_{x',{M}_{x},{N}_{x',{M}_{x'}}} \prec y\Big|{\cal E},C=c\Big)\nonumber\\
&=\frac{\mathbb{P}\Big(u_{x',y} \leq \tilde{U} < u_{x,y}, \tilde{U}<u_{x',x,y},u_{x',x,x',x',y}\leq \tilde{U},u_{x^e,y^l} \leq \tilde{U} < u_{x^e,y^u}\Big|C=c\Big)}{\mathbb{P}\Big(u_{x^e,y^l} \leq \tilde{U} < u_{x^e,y^u}\Big|C=c\Big)}\nonumber\\
&=\Bigg[\min\Big\{\mathbb{P}\left(Y_{x'} \prec y\Big|C=c\right),\mathbb{P}(Y_{x^e} \prec y^u|C=c),\mathbb{P}\left(Y_{x',{M}_{x},{N}_{x',{M}_{x'}}} \prec y\Big|C=c\right)\Big\}-\max\Big\{\mathbb{P}\left(Y_{x} \prec y\Big|C=c\right),\nonumber\\
&\mathbb{P}(Y_{x^e} \prec y^l|C=c),\mathbb{P}\left(Y_{x',{M}_{x}} \prec y\Big|C=c\right)\Big\}\Bigg]\Bigg/ \Big\{\mathbb{P}(Y_{x^e} \prec y^u|C=c)-\mathbb{P}(Y_{x^e} \prec y^l|C=c)\Big\},
\end{align}
\begin{align}
&\text{\normalfont PNS}^{X \rightarrow {M}  \rightarrow  Y}(y;x',x,{\cal E},c)\nonumber\\
&=\mathbb{P}\Big(Y_{x'} \prec y \preceq Y_{x},y \preceq Y_{x',{M}_{x}},y \preceq Y_{x',{M}_{x},{N}_{x',{M}_{x'}}}\Big|{\cal E},C=c\Big)\nonumber\\
&=\frac{\mathbb{P}\Big(u_{x',y} \leq \tilde{U} < u_{x,y}, \tilde{U}<u_{x',x,y},\tilde{U}<u_{x',x,x',x',y},u_{x^e,y^l} \leq \tilde{U} < u_{x^e,y^u}\Big|C=c\Big)}{\mathbb{P}\Big(u_{x^e,y^l} \leq \tilde{U} < u_{x^e,y^u}\Big|C=c\Big)}\nonumber\\
&=\Bigg[\min\Big\{\mathbb{P}\left(Y_{x'} \prec y\Big|C=c\right),\mathbb{P}(Y_{x^e} \prec y^u|C=c)\Big\}-\max\Big\{\mathbb{P}\left(Y_{x} \prec y\Big|C=c\right),\mathbb{P}\left(Y_{x',{M}_{x}} \prec y\Big|C=c\right),\nonumber\\
&\mathbb{P}(Y_{x^e} \prec y^l|C=c),\mathbb{P}\left(Y_{x',{M}_{x},{N}_{x',{M}_{x'}}} \prec y\Big|C=c\right)\Big\}\Bigg]\Bigg/ \Big\{\mathbb{P}(Y_{x^e} \prec y^u|C=c)-\mathbb{P}(Y_{x^e} \prec y^l|C=c)\Big\}.
\end{align}

(B). If $\mathbb{P}(Y \prec y^l|X=x^e,C=c) = \mathbb{P}(Y \prec y^l|X=x^e,C=c)$, then we have

\begin{align}
&\text{\normalfont PNS}^{X \rightarrow Y}(y;x',x,{\cal E},c)\nonumber\\
&=\mathbb{P}\Big(Y_{x'} \prec y \preceq Y_{x}, Y_{x',{M}_{x}} \prec y,Y_{x',{M}_{x},{N}_{x,{M}_{x}}} \prec y\Big|{\cal E},C=c\Big)\nonumber\\
&=\mathbb{P}\Big(u_{x',y} \leq \tilde{U} < u_{x,y}, u_{x',x,y} \leq \tilde{U}, u_{x',x,x,x,y} \leq \tilde{U}\Big|\tilde{U}=u_{x^e,y^u},C=c\Big)\nonumber\\
&=\mathbb{I}\Big(\theta(y;x',x',x',x',c) \leq \mathbb{P}(Y_{x^e} \prec y^u|C=c) < \theta(y;x,x,x,x,c),\nonumber\\
& \hspace{1cm}\theta(y;x',x,x',x,c) \leq \mathbb{P}(Y_{x^e} \prec y^u|_{x^e}C=c), \theta(y;x',x,x,x,c) \leq \mathbb{P}(Y_{x^e} \prec y^u|C=c)\Big),
\end{align}
\begin{align}
&\text{\normalfont PNS}^{X \rightarrow {N} \rightarrow  Y}(y;x',x,{\cal E},c)\nonumber\\
&=\mathbb{P}\Big(Y_{x'} \prec y \preceq Y_{x}, Y_{x',{M}_{x}} \prec y, y \preceq  Y_{x',{M}_{x},{N}_{x,{M}_{x}}}\Big|{\cal E},C=c\Big)\nonumber\\
&=\mathbb{P}\Big(u_{x',y} \leq \tilde{U} < u_{x,y}, u_{x',x,y} \leq \tilde{U},\tilde{U}<u_{x',x,x,x,y}\Big|\tilde{U}=u_{x^e,y^u},C=c\Big)\nonumber\\
&=\mathbb{I}\Big(\theta(y;x',x',x',x',c) \leq \mathbb{P}(Y_{x^e} \prec y^u|C=c)< \theta(y;x,x,x,x,c),\nonumber\\
&\hspace{1cm} \theta(y;x',x,x',x,c) \leq \mathbb{P}(Y_{x^e} \prec y^u|C=c),\mathbb{P}(Y_{x^e} \prec y^u|C=c)<\theta(y;x',x,x,x,c) \Big),
\end{align}
\begin{align}
&\text{\normalfont PNS}^{X \rightarrow {M} \rightarrow {N} \rightarrow  Y}(y;x',x,{\cal E},c)\nonumber\\
&=\mathbb{P}\Big(Y_{x'} \prec y \preceq Y_{x},y \preceq Y_{x',{M}_{x}},Y_{x',{M}_{x},{N}_{x',{M}_{x'}}} \prec y\Big|{\cal E},C=c\Big)\nonumber\\
&=\mathbb{P}\Big(u_{x',y} \leq \tilde{U} < u_{x,y}, \tilde{U}<u_{x',x,y},u_{x',x,x',x',y}\leq \tilde{U}\Big|\tilde{U}=u_{x^e,y^u},C=c\Big)\nonumber\\
&=\mathbb{I}\Big(\theta(y;x',x',x',x',c) \leq \mathbb{P}(Y_{x^e} \prec y^u|C=c)< \theta(y;x,x,x,x,c),\nonumber\\
&\hspace{1cm}\mathbb{P}(Y_{x^e} \prec y^u|C=c)<\theta(y;x',x,x',x,c),\theta(y;x',x,x',x',c)\leq \mathbb{P}(Y_{x^e} \prec y^u|C=c) \Big),
\end{align}
\begin{align}
&\text{\normalfont PNS}^{X \rightarrow {M}  \rightarrow  Y}(y;x',x,{\cal E},c)\nonumber\\
&=\mathbb{P}\Big(Y_{x'} \prec y \preceq Y_{x},y \preceq Y_{x',{M}_{x}},y \preceq Y_{x',{M}_{x},{N}_{x',{M}_{x'}}}\Big|{\cal E},C=c\Big)\nonumber\\
&=\mathbb{P}\Big(u_{x',y} \leq \tilde{U} < u_{x,y}, \tilde{U}<u_{x',x,y},\tilde{U}<u_{x',x,x',x',y}\Big|\tilde{U}=u_{x^e,y^u},C=c\Big)\nonumber\\
&=\mathbb{I}\Big(\theta(y;x',x',x',x',c) \leq \mathbb{P}(Y_{x^e} \prec y^u|C=c)< \theta(y;x,x,x,x,c),\nonumber\\
&\hspace{1cm}\mathbb{P}(Y_{x^e} \prec y^u|C=c)<\theta(y;x',x,x',x,c),\mathbb{P}(Y_{x^e} \prec y^u|C=c)<\theta(y;x',x,x',x',c)\Big).
\end{align}
Furthermore, under Assumption \ref{ASM}, $\theta(y;x',x,x',x',c)$ is identifiable.
\end{proof}

\noindent {\bf Proof of Statement.}
We prove the statement  ``Assumptions~\ref{AS1} and \ref{AS1}' are equivalent under Assumption~\ref{SUP1}" in the body of the paper.

\begin{proof}
We show the proof of equivalence of Assumptions \ref{AS1} and \ref{AS1}’ under Assumption \ref{SUP1}. \vspace{0.2cm}

\noindent (Assumption \ref{AS1} $\Rightarrow$ Assumption \ref{AS1}'.)
For any $c \in \Omega_C$, from Assumption \ref{SUP1}, if we have the negation of Assumption \ref{AS1}'
\begin{center}
there exists a set ${\cal U} \subset \Omega_{\tilde{U}}$ such that $0<\mathbb{P}({\cal U})<1$, and 
\begin{equation}
\begin{aligned}
&g(x,x',x'',x''',c,\tilde{u}_0)\succeq y\succ g(x^{*},x^{**},x^{***},x^{****},c,\tilde{u}_1)\\
&\land g(x,x',x'',x''',c,\tilde{u}_0)\prec y \preceq g(x^{*},x^{**},x^{***},x^{****},c,\tilde{u}_1)        
\end{aligned}
\end{equation} for some $x,x', x'',x''',x^{*},x^{**},x^{***},x^{****} \in \Omega_X$ and $y \in \Omega_Y$ and for any $\tilde{u}_0,\tilde{u}_1 \in {\cal U}$ such that $\tilde{u}_0\preceq \tilde{u}_1$,
\end{center}
then we have
\begin{center}
$g(x,x',x'',x''',c,\tilde{u}_0)\succeq y \succ g(x^{*},x^{**},x^{***},x^{****},c,\tilde{u}_0)$ and $g(x,x',x'',x''',c,\tilde{u}_1)\prec y \preceq g(x^{*},x^{**},x^{***},x^{****},c,\tilde{u}_1)$ for some $x,x', x'',x''',x^{*},x^{**},x^{***},x^{****} \in \Omega_X$ and $y \in \Omega_Y$ and\\ for any $\tilde{u}_0, \tilde{u}_1 \in {\cal U}$ such that $\tilde{u}_0 \preceq \tilde{u}_1$,
\end{center}
and we also have 
\begin{center}
$g(x,x',x'',x''',c,\tilde{u})\succeq y \succ g(x^{*},x^{**},x^{***},x^{****},c,\tilde{u})$ and $g(x,x',x'',x''',c,\tilde{u})\prec y \preceq g(x^{*},x^{**},x^{***},x^{****},c,\tilde{u})$ for some $x,x', x'',x''',x^{*},x^{**},x^{***},x^{****} \in \Omega_X$ and $y \in \Omega_Y$ and  for any $\tilde{u} \in {\cal U}$.
\end{center}
This implies the negation of Assumption 5 $\mathbb{P}((Y_{x,{M}_{x'},{N}_{x'',{M}_{x'''}}}\prec y \preceq Y_{x^{*},{M}_{x^{**}},{N}_{x^{***},{M}_{x^{****}}}}|C=c)\ne 0$ and $\mathbb{P}(Y_{x^{*},{M}_{x^{**}},{N}_{x^{***},{M}_{x^{****}}}}\prec y \preceq (Y_{x,{M}_{x'},{N}_{x'',{M}_{x'''}}}|C=c)\ne 0$ for some $x,x', x'',x''',x^{*},x^{**},x^{***},x^{****}  \in \Omega_Y$ and $y \in \Omega_Y$ since $g(x,x',x'',x''',c,\tilde{u})\succeq y \succ g(x^{*},x^{**},x^{***},x^{****},c,\tilde{u}) \Leftrightarrow Y_{x',M_{x}}(c,\tilde{u})\succeq y \succ Y_{x''',M_{x''}}(c,\tilde{u})$ and $g(x,x',x'',x''',c,\tilde{u})\prec y \preceq g(x^{*},x^{**},x^{***},x^{****},c,\tilde{u}) \Leftrightarrow Y_{x^{*},{M}_{x^{**}},{N}_{x^{***},{M}_{x^{****}}}}(c,\tilde{u})\succeq y \succ Y_{x,{M}_{x'},{N}_{x'',{M}_{x'''}}}(c,\tilde{u})$. \vspace{0.2cm}

\noindent (Assumption \ref{AS1}' $\Rightarrow$ Assumption \ref{AS1}.)
For any $c \in \Omega_C$, we denote $\tilde{u}_{sup}=\sup\{\tilde{u}:g(x,x',x'',x''',c,\tilde{u})\preceq y\}$. 
We consider the situations ``the function $g(x,x',x'',x''',c,\tilde{U})$ is monotonic increasing on $\tilde{U}$'' and ``the function $g(x,x',x'',x''',c,\tilde{U})$ is monotonic decreasing on $\tilde{U}$'', separately. \vspace{0.2cm}

\noindent {\bf (1).}
If the function $g(x,x',x'',x''',c,\tilde{U})$ is {\bf monotonic increasing} on $\tilde{U}$
for all $x \in \Omega_X$ almost surely w.r.t. $\mathbb{P}_{\tilde{U}}$, we have
\begin{equation}
\begin{aligned}
&g(x,x',x'',x''',c,\tilde{u}_{sup}) \preceq g(x,x',x'',x''',c,\tilde{u})\\
&\text{ and }g(x^{*},x^{**},x^{***},x^{****},c,\tilde{u}_{sup}) \preceq g(x^{*},x^{**},x^{***},x^{****},c,\tilde{u})
\end{aligned}
\end{equation}
for $\mathbb{P}_{\tilde{U}}$-almost every $\tilde{u} \in \Omega_{\tilde{U}}$ such that $\tilde{u}\succeq \tilde{u}_{sup}$.
We have the following statements:
\begin{enumerate}
\item Supposing $g(x,x',x'',x''',c,\tilde{u}_{sup})\succ g(x^{*},x^{**},x^{***},x^{****},c,\tilde{u}_{sup})$, 
we have $y= g(x,x',x'',x''',c,\tilde{u}_{sup}) \succ g(x^{*},x^{**},x^{***},x^{****},c,\tilde{u}_{sup})\succeq g(x^{*},x^{**},x^{***},x^{****},c,\tilde{u})=Y_{x''',M_{x''}}(c,\tilde{u})$ for $\mathbb{P}_{\tilde{U}}$-almost every $\tilde{u} \in \Omega_{\tilde{U}}$ such that $g(x,x',x'',x''',c,\tilde{u})\prec y$.
It means $Y_{x,{M}_{x'},{N}_{x'',{M}_{x'''}}}(c,\tilde{u})\prec y \Rightarrow Y_{x^{*},{M}_{x^{**}},{N}_{x^{***},{M}_{x^{****}}}}(c,\tilde{u}) \prec y$ for $\mathbb{P}_{\tilde{U}}$-almost every $\tilde{u} \in \Omega_{\tilde{U}}$ and  $\mathbb{P}(Y_{x,{M}_{x'},{N}_{x'',{M}_{x'''}}}\prec y \preceq Y_{x^{*},{M}_{x^{**}},{N}_{x^{***},{M}_{x^{****}}}}|C=c)=0$.
\item Supposing $g(x,x',x'',x''',c,\tilde{u}_{sup})\preceq g(x^{*},x^{**},x^{***},x^{****},c,\tilde{u}_{sup})$, we have $g(x^{*},x^{**},x^{***},x^{****},c,\tilde{u}) \succeq g(x^{*},x^{**},x^{***},x^{****},c,\tilde{u}_{sup}) \succeq g(x,x',x'',x''',c,\tilde{u}_{sup}) =y$ for $\mathbb{P}_{\tilde{U}}$-almost every $\tilde{u} \in \Omega_{\tilde{U}}$ such that $g(x,x',x'',x''',c,\tilde{u})\succeq y$.
It means $Y_{x,{M}_{x'},{N}_{x'',{M}_{x'''}}}(c,\tilde{u})\succeq y \Rightarrow Y_{x''',M_{x''}}(c,\tilde{u}) \succeq y$ for $\mathbb{P}_{\tilde{U}}$-almost every $\tilde{u} \in \Omega_{\tilde{U}}$ and  $\mathbb{P}(Y_{x^{*},{M}_{x^{**}},{N}_{x^{***},{M}_{x^{****}}}}\prec y \preceq Y_{x,{M}_{x'},{N}_{x'',{M}_{x'''}}}|C=c)=0$.
\end{enumerate}
Then, these results imply Assumption \ref{AS1}'. \vspace{0.2cm}

\noindent {\bf (2).}
If the function $g(x,x',x'',x''',c,\tilde{U})$ is {\bf monotonic decreasing} on $\tilde{U}$
for all $x \in \Omega_X$ almost surely w.r.t. $\mathbb{P}_{\tilde{U}}$, we have
\begin{equation}
\begin{aligned}
&g(x,x',x'',x''',c,\tilde{u}_{sup}) \succeq g(x,x',x'',x''',c,\tilde{u})\\
&\text{ and }g(x^{*},x^{**},x^{***},x^{****},c,\tilde{u}_{sup}) \succeq g(x^{*},x^{**},x^{***},x^{****},c,\tilde{u})
\end{aligned}
\end{equation}
for $\mathbb{P}_{\tilde{U}}$-almost every $\tilde{u} \in \Omega_{\tilde{U}}$ such that $\tilde{u}\succeq \tilde{u}_{sup}$.
We have the following statements:
\begin{enumerate}
    \item Supposing $g(x,x',x'',x''',c,\tilde{u}_{sup})\preceq g(x^{*},x^{**},x^{***},x^{****},c,\tilde{u}_{sup})$, 
we have $y= g(x,x',x'',x''',c,\tilde{u}_{sup}) \preceq g(x^{*},x^{**},x^{***},x^{****},c,\tilde{u}_{sup})\preceq g(x^{*},x^{**},x^{***},x^{****},c,\tilde{u})=Y_{x''',M_{x''}}(c,\tilde{u})$ for $\mathbb{P}_{\tilde{U}}$-almost every $\tilde{u} \in \Omega_{\tilde{U}}$ such that $g(x,x',x'',x''',c,\tilde{u})\succeq y$.
It means $Y_{x',M_{x}}(c,\tilde{u})\succeq y \Rightarrow Y_{x''',M_{x''}}(\tilde{u}) \succeq y$ for $\mathbb{P}_{\tilde{U}}$-almost every $\tilde{u} \in \Omega_{\tilde{U}}$ and  $\mathbb{P}(Y_{x^{*},{M}_{x^{**}},{N}_{x^{***},{M}_{x^{****}}}}\prec y \preceq Y_{x,{M}_{x'},{N}_{x'',{M}_{x'''}}}|C=c)=0$.
\item Supposing $g(x,x',x'',x''',c,\tilde{u}_{sup})\succ g(x^{*},x^{**},x^{***},x^{****},c,\tilde{u}_{sup})$, we have $g(x^{*},x^{**},x^{***},x^{****},c,\tilde{u}) \prec g(x^{*},x^{**},x^{***},x^{****},c,\tilde{u}_{sup}) \preceq g(x,x',x'',x''',c,\tilde{u}_{sup}) =y$ for $\mathbb{P}_{\tilde{U}}$-almost every $\tilde{u} \in \Omega_{\tilde{U}}$ such that $g(x,x',x'',x''',c,\tilde{u})\prec y$.
It means $Y_{x',M_{x}}(c,\tilde{u})\prec y \Rightarrow Y_{x''',M_{x''}}(c,\tilde{u}) \prec y$ for $\mathbb{P}_{\tilde{U}}$-almost every $\tilde{u} \in \Omega_{\tilde{U}}$ and  $\mathbb{P}(Y_{x,{M}_{x'},{N}_{x'',{M}_{x'''}}}\prec y \preceq Y_{x^{*},{M}_{x^{**}},{N}_{x^{***},{M}_{x^{****}}}}|C=c)=0$.
\end{enumerate}
Then, these resluts imply Assumption \ref{AS1}'.
In conclusion, Assumption \ref{AS1}' implies Assumption \ref{AS1} under Assumption \ref{SUP1}.

\end{proof}

\section{Path-Specific PNS with Three Mediators}
\label{appC}

We consider the following SCM ${\cal M}_3$:
\begin{gather}
Y:=f_Y(X,M^1,M^2,M^3,C,U_Y),M_3:=f_{M^3}(X,M^1,M^2,C,U_{M^3}),\nonumber\\
M_2:=f_{M^2}(X,M^1,C,U_{M^2}), M^1:=f_{M^1}(X,C,U_{M^1}),X:=f_X(C,U_X),C:=f_C(U_C),
\end{gather}
where all variables can be vectors, 
and $U_X$, $U_C$, $U_Y$, $U_{M^1}$, $U_{M^2}$, and $U_{M^3}$ are latent exogenous variables.
Assume that the domains $\Omega_Y$ and $\Omega_{U_Y} \times \Omega_{U_{M^1}} \times \Omega_{U_{M^2}} \times \Omega_{U_{M^3}}$ are totally ordered sets with $\preceq$.
Three mediators are causally ordered, or $M^1$ is the cause of $M^2$ and $M^2$ is the cause of $M^3$.
We give the definitions of the path-specific PNS with three mediators, and they have five counterfactual conditions, respectively.

%\subsection{Definitons of natural path-specific PNS with three mediators}

\begin{definition}[Path-Specific PNS with Three Mediators]
%For each $x',x \in \Omega_X$, $y \in \Omega_Y$, and $c \in \Omega_C$, 
We define eight types of path-specific PNS with three mediators as follows:
\begin{align}
&\text{\normalfont PNS}^{X \rightarrow Y}(y;x',x,{\cal E},c)\defeq\mathbb{P}\Bigg(Y_{x'} \prec y \preceq Y_{x}, Y_{x',M^1_{x}} \prec y, Y_{x',M^1_{x},M^2_{x,M^1_{x}}} \prec y,\nonumber\\
&\hspace{8cm}Y_{x',M^1_{x},M^2_{x,M^1_{x}},M^3_{x,M^1_{x},M^2_{x,M^1_{x}}}} \prec y\Bigg|{\cal E},C=c\Bigg),
\end{align}
\begin{align}
&\text{\normalfont PNS}^{X \rightarrow M^3 \rightarrow  Y}(y;x',x,{\cal E},c)\defeq\mathbb{P}\Bigg(Y_{x'} \prec y \preceq Y_{x}, Y_{x',M^1_{x}} \prec y, Y_{x',M^1_{x},M^2_{x,M^1_{x}}} \prec y,\nonumber\\
&\hspace{8cm}y \preceq Y_{x',M^1_{x},M^2_{x,M^1_{x}},M^3_{x,M^1_{x},M^2_{x,M^1_{x}}}}\Bigg|{\cal E},C=c\Bigg),
\end{align}
\begin{align}
&\text{\normalfont PNS}^{X \rightarrow M^2 \rightarrow  Y}(y;x',x,{\cal E},c)\defeq \mathbb{P}\Bigg(Y_{x'} \prec y \preceq Y_{x}, Y_{x',M^1_{x}} \prec y, y \preceq  Y_{x',M^1_{x},M^2_{x,M^1_{x}}},\nonumber\\
&\hspace{8cm}Y_{x',M^1_{x},M^2_{x',M^1_{x}},M^3_{x',M^1_{x},M^2_{x,M^1_{x}}}} \prec y\Bigg|{\cal E},C=c\Bigg),
\end{align}
\begin{align}
&\text{\normalfont PNS}^{X \rightarrow M^2 \rightarrow M^3 \rightarrow  Y}(y;x',x,{\cal E},c)\defeq \mathbb{P}\Bigg(Y_{x'} \prec y \preceq Y_{x}, Y_{x',M^1_{x}} \prec y, y \preceq  Y_{x',M^1_{x},M^2_{x,M^1_{x}}},\nonumber\\
&\hspace{8cm}y \preceq Y_{x',M^1_{x},M^2_{x',M^1_{x}},M^3_{x',M^1_{x},M^2_{x,M^1_{x}}}}\Bigg|{\cal E},C=c\Bigg),
\end{align}
\begin{align}
&\text{\normalfont PNS}^{X \rightarrow M^1 \rightarrow M^2 \rightarrow  Y}(y;x',x,{\cal E},c)\defeq\mathbb{P}\Bigg(Y_{x'} \prec y \preceq Y_{x},y \preceq Y_{x',M^1_{x}},Y_{x',M^1_{x},M^2_{x',M^1_{x'}}} \prec y,\nonumber\\
&\hspace{8cm}Y_{x',M^1_{x},M^2_{x',M^1_{x'}},M^3_{x',M^1_{x},M^2_{x',M^1_{x}}}} \prec y\Bigg|{\cal E},C=c\Bigg),
\end{align}
\begin{align}
&\text{\normalfont PNS}^{X \rightarrow M^1 \rightarrow M^2 \rightarrow M^3 \rightarrow  Y}(y;x',x,{\cal E},c)\defeq\mathbb{P}\Bigg(Y_{x'} \prec y \preceq Y_{x},y \preceq Y_{x',M^1_{x}},Y_{x',M^1_{x},M^2_{x',M^1_{x'}}} \prec y,\nonumber\\
&\hspace{8cm}y \preceq Y_{x',M^1_{x},M^2_{x',M^1_{x'}},M^3_{x',M^1_{x},M^2_{x',M^1_{x}}}}\Bigg|{\cal E},C=c\Bigg),
\end{align}
\begin{align}
&\text{\normalfont PNS}^{X \rightarrow M^1  \rightarrow  Y}(y;x',x,{\cal E},c)\defeq\mathbb{P}\Bigg(Y_{x'} \prec y \preceq Y_{x},y \preceq Y_{x',M^1_{x}},y \preceq Y_{x',M^1_{x},M^2_{x',M^1_{x'}}},\nonumber\\
&\hspace{8cm}Y_{x',M^1_{x'},M^2_{x',M^1_{x'}},M^3_{x',M^1_{x},M^2_{x',M^1_{x'}}}} \prec y\Bigg|{\cal E},C=c\Bigg),
\end{align}
\begin{align}
&\text{\normalfont PNS}^{X \rightarrow M^1  \rightarrow M^3 \rightarrow  Y}(y;x',x,{\cal E},c)\defeq\mathbb{P}\Bigg(Y_{x'} \prec y \preceq Y_{x},y \preceq Y_{x',M^1_{x}},y \preceq Y_{x',M^1_{x},M^2_{x',M^1_{x'}}},\nonumber\\
&\hspace{8cm}y \preceq Y_{x',M^1_{x'},M^2_{x',M^1_{x'}},M^3_{x',M^1_{x},M^2_{x',M^1_{x'}}}}\Bigg|{\cal E},C=c\Bigg),
\end{align}
where
${\cal E}\defeq(X=x^*, Y\in {\cal I}_Y)$,
and ${\cal I}_Y$ is a half-open interval $[y^l,y^u)$ or a closed interval $[y^l,y^u]$ w.r.t. $\prec$. 
\end{definition}

We have the following seven decompositions:
\begin{align}
&\text{\normalfont PNS}^{(M^1,M^2); X \rightarrow Y}(y;x',x,{\cal E},c)=\text{\normalfont PNS}^{X \rightarrow Y}(y;x',x,{\cal E},c)+\text{\normalfont PNS}^{X \rightarrow M^3 \rightarrow  Y}(y;x',x,{\cal E},c)\\
&\text{\normalfont PNS}^{(M^1,M^2); X \rightarrow {M^2} \rightarrow  Y}(y;x',x,{\cal E},c)=\text{\normalfont PNS}^{X \rightarrow M^2 \rightarrow  Y}(y;x',x,{\cal E},c)+\text{\normalfont PNS}^{X \rightarrow M^2 \rightarrow M^3 \rightarrow  Y}(y;x',x,{\cal E},c)\\
&\text{\normalfont PNS}^{(M^1,M^2); X \rightarrow {M^1} \rightarrow {M^1} \rightarrow  Y}(y;x',x,{\cal E},c)=\text{\normalfont PNS}^{X \rightarrow M^1 \rightarrow M^2 \rightarrow  Y}(y;x',x,{\cal E},c)+\text{\normalfont PNS}^{X \rightarrow M^1 \rightarrow M^2 \rightarrow M^3 \rightarrow  Y}(y;x',x,{\cal E},c)\\
&\text{\normalfont PNS}^{(M^1,M^2); X \rightarrow {M^1}  \rightarrow  Y}(y;x',x,{\cal E},c)=\text{\normalfont PNS}^{X \rightarrow M^1  \rightarrow  Y}(y;x',x,{\cal E},c)+\text{\normalfont PNS}^{X \rightarrow M^1  \rightarrow M^3 \rightarrow  Y}(y;x',x,{\cal E},c),\\
&\text{\normalfont ND-PNS}^{{M^1}}(y;x',x,{\cal E},c)=\text{\normalfont PNS}^{(M^1,M^2); X \rightarrow Y}(y;x',x,{\cal E},c)+\text{\normalfont PNS}^{(M^1,M^2); X \rightarrow {M^2} \rightarrow  Y}(y;x',x,{\cal E},c)\\
&\text{\normalfont NI-PNS}^{{M^1}}(y;x',x,{\cal E},c)=\text{\normalfont PNS}^{(M^1,M^2); X \rightarrow {M^1} \rightarrow {M^1} \rightarrow  Y}(y;x',x,{\cal E},c)+\text{\normalfont PNS}^{(M^1,M^2); X \rightarrow {M^1}  \rightarrow  Y}(y;x',x,{\cal E},c)\\
&\text{\normalfont T-PNS}(y;x',x,{\cal E},c)=\text{\normalfont ND-PNS}^{{M^1}}(y;x',x,{\cal E},c)+\text{\normalfont NI-PNS}^{{M^1}}(y;x',x,{\cal E},c),
\end{align}
where
\begin{align}
&\text{\normalfont PNS}^{(M^1,M^2); X \rightarrow Y}(y;x',x,{\cal E},c)\defeq\mathbb{P}(Y_{x'} \prec y \preceq Y_{x}, Y_{x',{M^1}_{x}} \prec y,
 Y_{x',{M^1}_{x},{M^2}_{x,{M^1}_{x}}} \prec y|{\cal E},C=c),\\
&\text{\normalfont PNS}^{(M^1,M^2); X \rightarrow {M^2} \rightarrow  Y}(y;x',x,{\cal E},c)\defeq\mathbb{P}(Y_{x'} \prec y \preceq Y_{x}, Y_{x',{M^1}_{x}} \prec y, y \preceq  Y_{x',{M^1}_{x},{M^2}_{x,{M^1}_{x}}}|{\cal E},C=c),\\
&\text{\normalfont PNS}^{(M^1,M^2); X \rightarrow {M^1} \rightarrow {M^1} \rightarrow  Y}(y;x',x,{\cal E},c)\defeq\mathbb{P}(Y_{x'} \prec y \preceq Y_{x},y \preceq Y_{x',{M^1}_{x}},Y_{x',{M^1}_{x},{M^2}_{x',{M^1}_{x'}}} \prec y|{\cal E},C=c),\\
&\text{\normalfont PNS}^{(M^1,M^2); X \rightarrow {M^1}  \rightarrow  Y}(y;x',x,{\cal E},c)\defeq\mathbb{P}(Y_{x'} \prec y \preceq Y_{x},y \preceq Y_{x',{M^1}_{x}},y \preceq Y_{x',{M^1}_{x},{M^2}_{x',{M^1}_{x'}}}|{\cal E},C=c),\\
&\text{\normalfont ND-PNS}^{{M^1}}(y;x',x,{\cal E},c)\defeq\mathbb{P}(Y_{x'} \prec y \preceq Y_{x}, Y_{x',{M^1}_{x}} \prec y|{\cal E},C=c),\\
&\text{\normalfont NI-PNS}^{{M^1}}(y;x',x,{\cal E},c)\defeq\mathbb{P}(Y_{x'} \prec y \preceq Y_{x},y \preceq Y_{x',{M^1}_{x}}|{\cal E},C=c).
\end{align}

%\section{Bounding Path-Specific PNS with Two Mediators}
%\label{appboun}
%\input{A3_2Bound}

\section{Additional Numerical Experiments}
\label{appD}

\subsection{Special Cases}
\label{appD1}

We provide three additional experiments under (1) no effect between ${M}$ and ${N}$, (2) no effect between $\{{M},{N}\}$ and $Y$, (3) only effect through $X \rightarrow {M} \rightarrow {N} \rightarrow Y$.

{\bf (1). No effect between ${M}$ and ${N}$.}
We consider the situation where there is no effect between ${M}$ and ${N}$.

{\bf Setting.}
We consider the following linear SCM:
\begin{equation}
\begin{gathered}
Y:=X+{M}+ {N}+ C+U_Y,
%{N}:=X+ {M}+ C+U_{{N}},\\
{N}:=X+ C+U_{{N}},
{M}:=X+C+U_{{M}}, X:=C+U_X, C:=U_C,
\end{gathered}
\end{equation}
where $U_C\sim {\cal N}(0,1)$, $U_X\sim {\cal N}(0,1)$, $U_Y\sim {\cal N}(0,1)$, $U_{{M}} \sim {\cal N}(0,1)$, $U_{{N}} \sim {\cal N}(0,1)$ and they are mutually independent normal distributions.
This SCM satisfies Assumptions \ref{SCAS}, \ref{ASM}, \ref{SUP1}, \ref{AS1}, and \ref{AS1}'.
We let $x'=0$, $x=1$, $y=0$, $c=0$, and ${\cal E}=\emptyset$.
We simulate 1000 times with the sample size $N=20$, $N=100$, and $N=10000$.

{\bf Results.}
The ground truth of $\text{\normalfont T-PNS}$ is $0.458$.
The ground truth of $\text{\normalfont PNS}^{X \rightarrow Y}$ is $0.082$ and the estimates are
\begin{center}
\textbf{$N=20$}:\, \, \, \, $0.083$ (95\%CI: $[0.003,0.213]$),\\\vspace{0.1cm}
\textbf{$N=100$}:\, \, \,  $0.081$ (95\%CI: $[0.044,0.127]$),\\\vspace{0.1cm}
\textbf{$N=10000$}: $0.082$ (95\%CI: $[0.078,0.086]$).
\end{center}
The ground truth of $\text{\normalfont PNS}^{X \rightarrow {N}  \rightarrow Y}$ is $0.158$ and the estimates are
\begin{center}
\textbf{$N=20$}:\, \, \, \, $0.158$ (95\%CI: $[0.030,0.350]$),\\\vspace{0.1cm}
\textbf{$N=100$}:\, \, \,  $0.158$ (95\%CI: $[0.102,0.221]$),\\\vspace{0.1cm}
\textbf{$N=10000$}: $0.158$ (95\%CI: $[0.151,0.164]$).
\end{center}
The ground truth of $\text{\normalfont PNS}^{X \rightarrow {M} \rightarrow {N}  \rightarrow Y}$ is $0.000$ and the estimates are
\begin{center}
\textbf{$N=20$}:\, \, \, \, $0.018$ (95\%CI: $[0.000,0.090]$),\\\vspace{0.1cm}
\textbf{$N=100$}:\, \, \,  $0.008$ (95\%CI: $[0.000,0.038]$),\\\vspace{0.1cm}
\textbf{$N=10000$}: $0.001$ (95\%CI: $[0.000,0.004]$).
\end{center}
The ground truth of $\text{\normalfont PNS}^{X \rightarrow {M}  \rightarrow Y}$ is $0.218$ and the estimates are
\begin{center}
\textbf{$N=20$}:\, \, \, \, $0.199$ (95\%CI: $[0.057,0.353]$),\\\vspace{0.1cm}
\textbf{$N=100$}:\, \, \,  $0.211$ (95\%CI: $[0.154,0.265]$),\\\vspace{0.1cm}
\textbf{$N=10000$}: $0.217$ (95\%CI: $[0.212,0.223]$).
\end{center}
All means of the estimators are close to the ground truth. 
However, estimators for small sample sizes have large 95 $\%$ CIs.

{\bf (2). No effect between $\{{M},{N}\}$ and $Y$.}
We consider the situation where there is no effect between $\{{M},{N}\}$ and $Y$.

{\bf Setting.}
We consider the following linear SCM:
\begin{equation}
\begin{gathered}
Y:=X+ C+U_Y,
{N}:=X+ {M}+ C+U_{{N}},
{M}:=X+C+U_{{M}}, X:=C+U_X, C:=U_C,
\end{gathered}
\end{equation}
where $U_C\sim {\cal N}(0,1)$, $U_X\sim {\cal N}(0,1)$, $U_Y\sim {\cal N}(0,1)$, $U_{{M}} \sim {\cal N}(0,1)$, $U_{{N}} \sim {\cal N}(0,1)$ and they are mutually independent normal distributions.
This SCM satisfies Assumptions \ref{SCAS}, \ref{ASM}, \ref{SUP1}, \ref{AS1}, and \ref{AS1}'.
We let $x'=0$, $x=1$, $y=0$, $c=0$, and ${\cal E}=\emptyset$.
We simulate 1000 times with the sample size $N=20$, $N=100$, and $N=10000$.

{\bf Results.}
The ground truth of $\text{\normalfont T-PNS}$ is $0.346$.
The ground truth of $\text{\normalfont PNS}^{X \rightarrow Y}$ is $0.346$ and the estimates are
\begin{center}
\textbf{$N=20$}:\, \, \, \, $0.282$ (95\%CI: $[0.037,0.481]$),\\\vspace{0.1cm}
\textbf{$N=100$}:\, \, \,  $0.314$ (95\%CI: $[0.207,0.400]$),\\\vspace{0.1cm}
\textbf{$N=10000$}: $0.339$ (95\%CI: $[0.328,0.347]$).
\end{center}
The ground truth of $\text{\normalfont PNS}^{X \rightarrow {N}  \rightarrow Y}$ is $0.000$ and the estimates are
\begin{center}
\textbf{$N=20$}:\, \, \, \, $0.027$ (95\%CI: $[0.000,0.185]$),\\\vspace{0.1cm}
\textbf{$N=100$}:\, \, \,  $0.011$ (95\%CI: $[0.000,0.074]$),\\\vspace{0.1cm}
\textbf{$N=10000$}: $0.001$ (95\%CI: $[0.000,0.007]$).
\end{center}
The ground truth of $\text{\normalfont PNS}^{X \rightarrow {M} \rightarrow {N}  \rightarrow Y}$ is $0.000$ and the estimates are
\begin{center}
\textbf{$N=20$}:\, \, \, \, $0.011$ (95\%CI: $[0.000,0.103]$),\\\vspace{0.1cm}
\textbf{$N=100$}:\, \, \,  $0.005$ (95\%CI: $[0.000,0.042]$),\\\vspace{0.1cm}
\textbf{$N=10000$}: $0.000$ (95\%CI: $[0.000,0.004]$).
\end{center}
The ground truth of $\text{\normalfont PNS}^{X \rightarrow {M}  \rightarrow Y}$ is $0.000$ and the estimates are
\begin{center}
\textbf{$N=20$}:\, \, \, \, $0.025$ (95\%CI: $[0.000,0.168]$),\\\vspace{0.1cm}
\textbf{$N=100$}:\, \, \,  $0.012$ (95\%CI: $[0.000,0.071]$),\\\vspace{0.1cm}
\textbf{$N=10000$}: $0.001$ (95\%CI: $[0.000,0.007]$).
\end{center}
All means of the estimators are close to the ground truth. 
However, estimators for small sample sizes have large 95 $\%$ CIs.

{\bf (3). Only effect through $X \rightarrow {M} \rightarrow {N} \rightarrow Y$.}
We consider the situation where there is only effect through $X \rightarrow {M} \rightarrow {N} \rightarrow Y$.

{\bf Setting.}
We consider the following linear SCM:
\begin{equation}
\begin{gathered}
Y:= {N}+ C+U_Y,
{N}:={M}+ C+U_{{N}},
{M}:=X+C+U_{{M}}, X:=C+U_X, C:=U_C,
\end{gathered}
\end{equation}
where $U_C\sim {\cal N}(0,1)$, $U_X\sim {\cal N}(0,1)$, $U_Y\sim {\cal N}(0,1)$, $U_{{M}} \sim {\cal N}(0,1)$, $U_{{N}} \sim {\cal N}(0,1)$ and they are mutually independent normal distributions.
This SCM satisfies Assumptions \ref{SCAS}, \ref{ASM}, \ref{SUP1}, \ref{AS1}, and \ref{AS1}'.
We let $x'=0$, $x=1$, $y=0$, $c=0$, and ${\cal E}=\emptyset$.
We simulate 1000 times with the sample size $N=20$, $N=100$, and $N=10000$.

{\bf Results.}
The ground truth of $\text{\normalfont T-PNS}$ is $0.219$.
The ground truth of $\text{\normalfont PNS}^{X \rightarrow Y}$ is $0.000$ and the estimates are
\begin{center}
\textbf{$N=20$}:\, \, \, \, $0.017$ (95\%CI: $[0.000,0.124]$),\\\vspace{0.1cm}
\textbf{$N=100$}:\, \, \,  $0.008$ (95\%CI: $[0.000,0.049]$),\\\vspace{0.1cm}
\textbf{$N=10000$}: $0.001$ (95\%CI: $[0.000,0.004]$).
\end{center}
The ground truth of $\text{\normalfont PNS}^{X \rightarrow {N}  \rightarrow Y}$ is $0.000$ and the estimates are
\begin{center}
\textbf{$N=20$}:\, \, \, \, $0.021$ (95\%CI: $[0.000,0.144]$),\\\vspace{0.1cm}
\textbf{$N=100$}:\, \, \,  $0.008$ (95\%CI: $[0.000,0.050]$),\\\vspace{0.1cm}
\textbf{$N=10000$}: $0.001$ (95\%CI: $[0.000,0.005]$).
\end{center}
The ground truth of $\text{\normalfont PNS}^{X \rightarrow {M} \rightarrow {N}  \rightarrow Y}$ is $0.219$ and the estimates are
\begin{center}
\textbf{$N=20$}:\, \, \, \, $0.146$ (95\%CI: $[0.000,0.293]$),\\\vspace{0.1cm}
\textbf{$N=100$}:\, \, \,  $0.189$ (95\%CI: $[0.123,0.247]$),\\\vspace{0.1cm}
\textbf{$N=10000$}: $0.215$ (95\%CI: $[0.208,0.222]$).
\end{center}
The ground truth of $\text{\normalfont PNS}^{X \rightarrow {M}  \rightarrow Y}$ is $0.000$ and the estimates are
\begin{center}
\textbf{$N=20$}:\, \, \, \, $0.034$ (95\%CI: $[0.000,0.189]$),\\\vspace{0.1cm}
\textbf{$N=100$}:\, \, \,  $0.014$ (95\%CI: $[0.000,0.068]$),\\\vspace{0.1cm}
\textbf{$N=10000$}: $0.001$ (95\%CI: $[0.000,0.007]$).
\end{center}
All means of the estimators are close to the ground truth. 
However, estimators for small sample sizes have large 95 $\%$ CIs.

\subsection{Sensitivity Analysis for Violation of Monotonicity}
\label{appD2}

We conduct a sensitivity analysis to assess the impact of violations of the monotonicity assumption.

{\bf Setting.}
We consider the following SCM:
\begin{equation}
\begin{gathered}
Y:=X+{M}+ {N}+ C+\alpha U_Y +(1-\alpha) U_Y^4,
{N}:=X+ {M}+ C+U_{{N}},
{M}:=X+C+U_{{M}}, X:=C+U_X, C:=U_C,
\end{gathered}
\end{equation}
where $U_C\sim {\cal N}(0,1)$, $U_X\sim {\cal N}(0,1)$, $U_Y\sim {\cal N}(0,1)$, $U_{{M}} \sim {\cal N}(0,1)$, $U_{{N}} \sim {\cal N}(0,1)$, which are mutually independent normal distributions.
%This SCM satisfies Assumptions \ref{SCAS}, \ref{ASM}, \ref{SUP1}, \ref{AS1}, and 4.3'.
This SCM violates the monotonicities.
We let $x'=0$, $x=1$, $y=0$, $c=0$, and ${\cal E}=\emptyset$.
We simulate 1000 times with the sample size $N=20$, $N=100$, and $N=10000$.
We examine the cases $\alpha = 0.5$ and $\alpha = 0$, which correspond to a moderate violation and a strong violation of monotonicity, respectively.
The case where $\alpha = 1$ corresponds to the setting described in Section 5.

{\bf Results ($\alpha=0.5$; moderate violation).}
The ground truth of $\text{\normalfont T-PNS}$ is $0.365$.
The ground truth of $\text{\normalfont PNS}^{X \rightarrow Y}$ is $0.039$ and the estimates are
\begin{center}
\textbf{$N=20$}:\, \, \, \, $0.048$ (95\%CI: $[0.000,0.188]$),\\\vspace{0.1cm}
\textbf{$N=100$}:\, \, \,  $0.048$ (95\%CI: $[0.000,0.115]$),\\\vspace{0.1cm}
\textbf{$N=10000$}: $0.048$ (95\%CI: $[0.040,0.057]$).
\end{center}
The ground truth of $\text{\normalfont PNS}^{X \rightarrow {N}  \rightarrow Y}$ is $0.073$ and the estimates are
\begin{center}
\textbf{$N=20$}:\, \, \, \, $0.053$ (95\%CI: $[0.000,0.176]$),\\\vspace{0.1cm}
\textbf{$N=100$}:\, \, \,  $0.056$ (95\%CI: $[0.000,0.104]$),\\\vspace{0.1cm}
\textbf{$N=10000$}: $0.056$ (95\%CI: $[0.050,0.062]$).
\end{center}
The ground truth of $\text{\normalfont PNS}^{X \rightarrow {M} \rightarrow {N}  \rightarrow Y}$ is $0.111$, and the estimates are
\begin{center}
\textbf{$N=20$}:\, \, \, \, $0.078$ (95\%CI: $[0.000,0.214]$),\\\vspace{0.1cm}
\textbf{$N=100$}:\, \, \,  $0.070$ (95\%CI: $[0.000,0.135]$),\\\vspace{0.1cm}
\textbf{$N=10000$}: $0.063$ (95\%CI: $[0.055,0.071]$).
\end{center}
The ground truth of $\text{\normalfont PNS}^{X \rightarrow {M}  \rightarrow Y}$ is $0.142$ and the estimates are
\begin{center}
\textbf{$N=20$}:\, \, \, \, $0.101$ (95\%CI: $[0.000,0.253]$),\\\vspace{0.1cm}
\textbf{$N=100$}:\, \, \,  $0.080$ (95\%CI: $[0.000,0.163]$),\\\vspace{0.1cm}
\textbf{$N=10000$}: $0.069$ (95\%CI: $[0.057,0.081]$).
\end{center}

{\bf Results ($\alpha=0$; strong violation).}
The ground truth of $\text{\normalfont T-PNS}$ is $0.330$.
The ground truth of $\text{\normalfont PNS}^{X \rightarrow Y}$ is $0.035$ and the estimates are
\begin{center}
\textbf{$N=20$}:\, \, \, \, $0.048$ (95\%CI: $[0.000,0.201]$),\\\vspace{0.1cm}
\textbf{$N=100$}:\, \, \,  $0.041$ (95\%CI: $[0.000,0.125]$),\\\vspace{0.1cm}
\textbf{$N=10000$}: $0.032$ (95\%CI: $[0.022,0.043]$).
\end{center}
The ground truth of $\text{\normalfont PNS}^{X \rightarrow {N}  \rightarrow Y}$ is $0.064$ and the estimates are
\begin{center}
\textbf{$N=20$}:\, \, \, \, $0.038$ (95\%CI: $[0.000,0.149]$),\\\vspace{0.1cm}
\textbf{$N=100$}:\, \, \,  $0.034$ (95\%CI: $[0.000,0.091]$),\\\vspace{0.1cm}
\textbf{$N=10000$}: $0.034$ (95\%CI: $[0.027,0.042]$).
\end{center}
The ground truth of $\text{\normalfont PNS}^{X \rightarrow {M} \rightarrow {N}  \rightarrow Y}$ is $0.101$, and the estimates are
\begin{center}
\textbf{$N=20$}:\, \, \, \, $0.048$ (95\%CI: $[0.000,0.184]$),\\\vspace{0.1cm}
\textbf{$N=100$}:\, \, \,  $0.040$ (95\%CI: $[0.000,0.106]$),\\\vspace{0.1cm}
\textbf{$N=10000$}: $0.044$ (95\%CI: $[0.028,0.044]$).
\end{center}
The ground truth of $\text{\normalfont PNS}^{X \rightarrow {M}  \rightarrow Y}$ is $0.130$ and the estimates are
\begin{center}
\textbf{$N=20$}:\, \, \, \, $0.069$ (95\%CI: $[0.000,0.232]$),\\\vspace{0.1cm}
\textbf{$N=100$}:\, \, \,  $0.049$ (95\%CI: $[0.000,0.137]$),\\\vspace{0.1cm}
\textbf{$N=10000$}: $0.038$ (95\%CI: $[0.026,0.048]$).
\end{center}

The bias becomes large under strong violations of monotonicity.

\subsection{Binary Outcome}
\label{appD3}

We conduct additional experiments using a logistic model for binary outcomes.

{\bf Setting.}
We consider the following SCM:
$Y$ is randomly chosen from $\{0,1\}$ with the probability
\begin{align}
\mathbb{P}(Y=1)=\frac{1}{1+\exp(-10(X+{M}+ {N}+ C))},
\end{align}
and
\begin{equation}
\begin{gathered}
{N}:=X+ {M}+ C+U_{{N}},
{M}:=X+C+U_{{M}}, X:=C+U_X, C:=U_C,
\end{gathered}
\end{equation}
where $U_C\sim {\cal N}(0,1)$, $U_X\sim {\cal N}(0,1)$, $U_{{M}} \sim {\cal N}(0,1)$, and $U_{{N}} \sim {\cal N}(0,1)$, which are mutually independent normal distributions.
This SCM satisfies Assumptions \ref{SCAS}, \ref{ASM}, \ref{SUP1}, \ref{AS1}, and 4.3'.
We estimate the model parameters using logistic regression.
We let $x'=0$, $x=1$, $y=0$, $c=0$, and ${\cal E}=\emptyset$.
We simulate 1000 times with the sample size $N=20$, $N=100$, and $N=10000$.

{\bf Results.}
The ground truth of $\text{\normalfont T-PNS}$ is $0.466$.
The ground truth of $\text{\normalfont PNS}^{X \rightarrow Y}$ is $0.054$ and the estimates are
\begin{center}
\textbf{$N=20$}:\, \, \, \, $0.092$ (95\%CI: $[0.000,0.950]$),\\\vspace{0.1cm}
\textbf{$N=100$}:\, \, \,  $0.054$ (95\%CI: $[0.000,0.159]$),\\\vspace{0.1cm}
\textbf{$N=10000$}: $0.053$ (95\%CI: $[0.048,0.058]$).
\end{center}
The ground truth of $\text{\normalfont PNS}^{X \rightarrow {N}  \rightarrow Y}$ is $0.098$ and the estimates are
\begin{center}
\textbf{$N=20$}:\, \, \, \, $0.080$ (95\%CI: $[0.000,0.359]$),\\\vspace{0.1cm}
\textbf{$N=100$}:\, \, \,  $0.096$ (95\%CI: $[0.013,0.202]$),\\\vspace{0.1cm}
\textbf{$N=10000$}: $0.096$ (95\%CI: $[0.089,0.103]$).
\end{center}
The ground truth of $\text{\normalfont PNS}^{X \rightarrow {M} \rightarrow {N}  \rightarrow Y}$ is $0.141$, and the estimates are
\begin{center}
\textbf{$N=20$}:\, \, \, \, $0.110$ (95\%CI: $[0.000,0.306]$),\\\vspace{0.1cm}
\textbf{$N=100$}:\, \, \,  $0.130$ (95\%CI: $[0.000,0.232]$),\\\vspace{0.1cm}
\textbf{$N=10000$}: $0.141$ (95\%CI: $[0.137,0.146]$).
\end{center}
The ground truth of $\text{\normalfont PNS}^{X \rightarrow {M}  \rightarrow Y}$ is $0.173$ and the estimates are
\begin{center}
\textbf{$N=20$}:\, \, \, \, $0.149$ (95\%CI: $[0.000,0.439]$),\\\vspace{0.1cm}
\textbf{$N=100$}:\, \, \,  $0.182$ (95\%CI: $[0.031,0.329]$),\\\vspace{0.1cm}
\textbf{$N=10000$}: $0.178$ (95\%CI: $[0.167,0.190]$).
\end{center}
The estimates obtained from logistic regression are reliable when the sample size is large.

\section{Additional Information about the Application to Real-World}
\label{appE}
Let the evidence be ${\cal E}=(X=0,10<Y\leq 15)$, and the other settings are the same as in the body of the paper.
The estimates at $(y;x',x,{\cal E},c)$
%of $\text{\normalfont T-PNS}$, $\text{\normalfont PNS}^{X \rightarrow Y}$, $\text{\normalfont PNS}^{X \rightarrow {N} \rightarrow Y}$, $\text{\normalfont PNS}^{X \rightarrow {M} \rightarrow {N} \rightarrow Y}$, and $\text{\normalfont PNS}^{X \rightarrow {M} \rightarrow Y}$ given $C=c_1$
are 
\begin{align}
&\text{\normalfont T-PNS}: &23.950 \% (\text{CI}: [0.000\%,62.123\%]),\nonumber\\ 
&\text{\normalfont ND-PNS}^{{M}}: &2.430 \% (\text{CI}: [0.000\%,18.587\%]),\nonumber\\
&\text{\normalfont NI-PNS}^{{M}}: &21.520 \% (\text{CI}: [0.000\%,52.851\%]),\nonumber\\
&\text{\normalfont PNS}^{X \rightarrow Y}: &0.354 \% (\text{CI}: [0.000\%,4.998\%]),\nonumber\\
&\text{\normalfont PNS}^{X \rightarrow {N} \rightarrow Y}: &2.075 \% (\text{CI}: [0.000\%,16.048\%]),\nonumber\\
&\text{\normalfont PNS}^{X \rightarrow {M} \rightarrow {N} \rightarrow Y}:  &0.000 \% (\text{CI}: [0.000\%,0.000\%]),\nonumber\\
&\text{\normalfont PNS}^{X \rightarrow {M} \rightarrow Y}:  &21.520 \% (\text{CI}: [0.000\%,52.851\%]).\nonumber
\end{align}

%\yuta{
%We provide the estimates of path-specific causal effects for the binarized outcome $\mathbb{I}(Y<10)$ based on the method proposed by \citep{Daniel2015}.
%The estimates of path-specific causal effects by [Daniel et al., 2015] in the application are as follows:
%\begin{align}
%&\text{Total Effect}: & 0.082 (CI: [-0.016,0.222]),\\
%&\text{Path-specific Effect of} X \rightarrow Y:& -0.047 (CI: [-0.113,0.019]),\\
%&\text{Path-specific Effect of} X \rightarrow {N} \rightarrow Y: & 0.027 (CI: [-0.022,0.082]),\\
%&\text{Path-specific Effect of} X \rightarrow {M} \rightarrow {N} \rightarrow Y: &0.000 (CI: [0.000,0.000]),\\
%&\text{Path-specific Effect of} X \rightarrow {M} \rightarrow Y:& 0.102 (CI: [0.023,0.204]).
%\end{align}
%}

\end{document}